%% file: main.tex
\documentclass[11pt,a4paper]{report} 

\input{preamble.tex} 

\newcommand{\welchethesis}{Master} 
\newcommand{\welcherstudiengang}{Informatik - Software und System Engineering} 
\newcommand{\thesisofwas}{of Science}
\newcommand{\titel}{Introspective Uncertainty Estimation for LLM-Based Code Generation}
\newcommand{\kurztitel}{Introspektive Unsicherheitsabsch\"atzung f\"ur LLM-basierte Codegenerierung}
\newcommand{\autor}{Thomas Klassert}
\newcommand{\datum}{1. Mai 2026} 

\newcommand{\referent}{Prof.\ Dr.\ Adrian Ulges}
\newcommand{\korreferent}{M.Sc.\ Viola Campos}

\begin{document}
\include{chapters/vorspann} 
\pagenumbering{roman}
\pagestyle{frontstyle}

\include{chapters/abstract}

\include{chapters/zusammenfassung}

\phantomsection
\addcontentsline{toc}{chapter}{Contents}
\tableofcontents
\cleardoublepage
\pagenumbering{arabic}
\pagestyle{mainstyle}

\include{chapters/intro}

\include{chapters/background}

\include{chapters/related_work}

\include{chapters/methodology}

\include{chapters/experiments}

\include{chapters/discussion}

\include{chapters/conclusion}

\pagestyle{frontstyle}

\bibliographystyle{alpha}
\addcontentsline{toc}{chapter}{Bibliography}
\bibliography{thesis,online}

\appendix
\include{chapters/appendix}

\end{document}

%% file: preamble.tex
\usepackage[ngerman,english]{babel} 
\usepackage[utf8]{inputenc} 
\usepackage[T1]{fontenc} 
\usepackage{textcomp} 
\usepackage[hyphens]{url}
\usepackage{amssymb} 
\usepackage{emptypage} 
\usepackage{titling}

\usepackage{booktabs} 
\usepackage{longtable} 
\usepackage{graphicx} 
\usepackage{xstring} 
\usepackage{lmodern} 
\usepackage{fix-cm}
\usepackage{fancyhdr}
\usepackage{color}
\usepackage[table]{xcolor}
\usepackage{enumitem}
\usepackage{url}
\usepackage{acronym}
\usepackage[ddmmyyyy,hhmmss]{datetime}

\usepackage{tabu} 
\usepackage{multirow}
\usepackage{mathtools}
\usepackage{array} 
\usepackage{rotating} 
\usepackage{caption} 
\usepackage{algorithm}
\usepackage{algpseudocode}
\usepackage{amsmath}
\usepackage{mathpazo} 
\usepackage[scaled=.95]{helvet}
\usepackage{courier}

\usepackage{microtype}

\usepackage{graphicx} 
\usepackage{subfig} 
\usepackage{wrapfig} 
\usepackage{listings} 
\usepackage{float}
\newfloat{listing}{htbp}{scl}[chapter]
\floatname{listing}{Listing}

\usepackage[paper=a4paper,width=14cm,left=35mm,height=22cm]{geometry}
\usepackage{setspace}
\newcommand{\phv}{\fontfamily{phv}\fontseries{m}\fontsize{9}{11}\selectfont}
\usepackage{fancyhdr}

\newcommand{\mynameblock}{%
  \phv\raisebox{-.24\height}{\includegraphics[height=2.5ex]{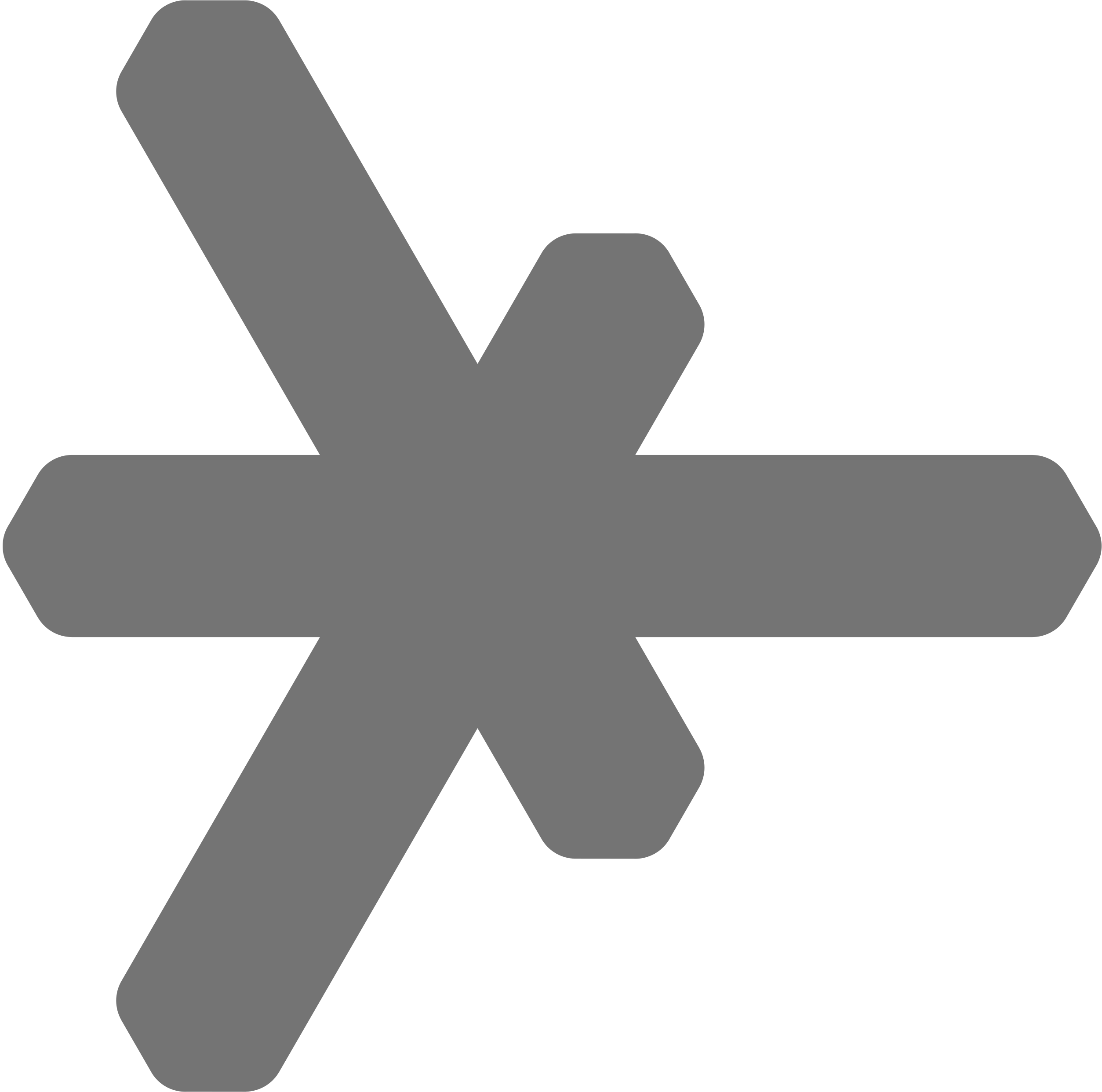}}%
  \,\ Thomas \textsc{Klassert}%
}

\renewcommand{\chaptermark}[1]{%
  \markboth{%
    \ifnum\value{chapter}>0
      \chaptername\enspace\thechapter\enspace\textbar\enspace
    \fi
    #1
  }{}%
}

\fancypagestyle{frontstyle}{%
  \fancyhf{}

  \fancyfoot[C]{\thepage}
}

\fancypagestyle{mainstyle}{%
  \fancyhf{}
  
  \fancyhead[C]{\small\selectfont\nouppercase{\leftmark}}
  \fancyfoot[L]{\mynameblock}
  \fancyfoot[C]{\thepage}
}

\usepackage{hyperref}
\usepackage{cleveref}

\usepackage{epigraph}
\usepackage{blindtext} 

\usepackage{soul}

\usepackage{enumitem}

%% file: chapters/vorspann.tex
\begin{titlepage}
  \begin{center}
    \vspace*{0.2cm}
    \includegraphics[width=0.55\textwidth]{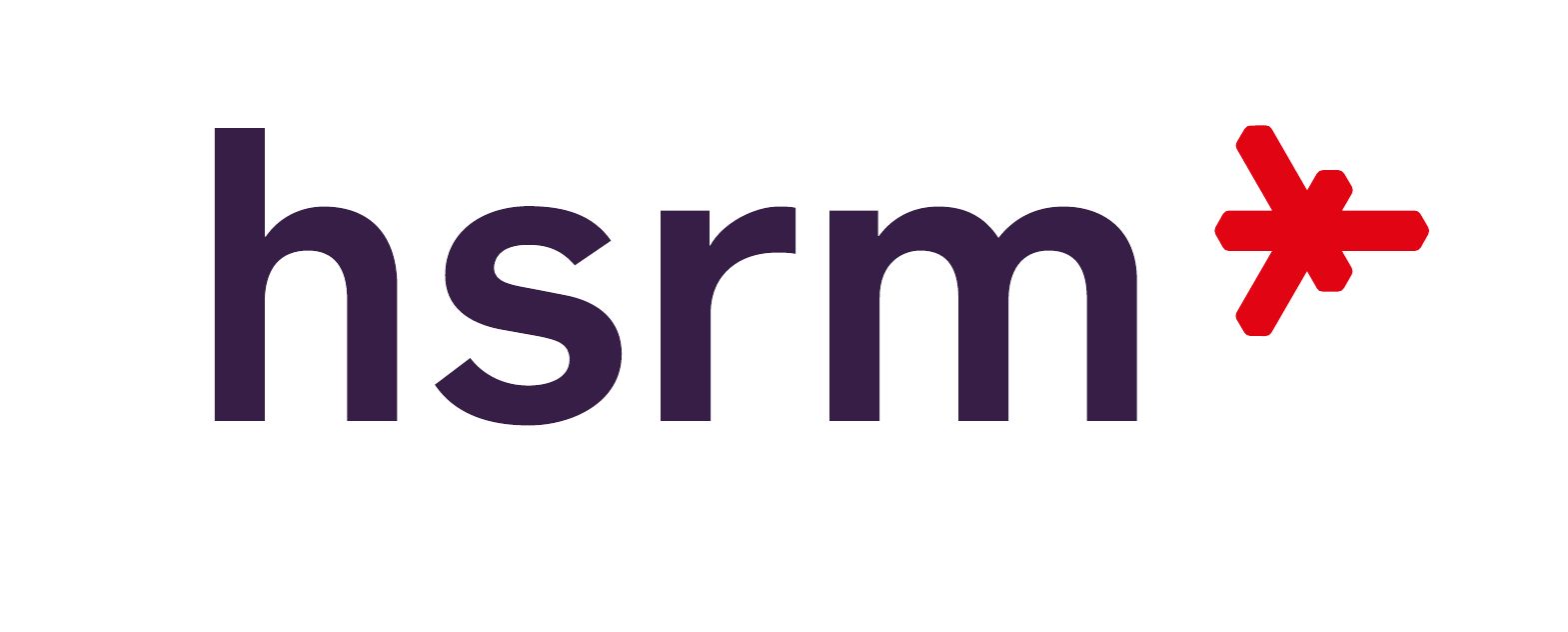}\\[0.5cm]

    \rule{0.95\textwidth}{1.6pt}\\[1.0cm]
    {\begin{spacing}{1.1}\huge\bfseries \titel\\[0.5cm]\end{spacing}}
    {\LARGE\normalfont \kurztitel}\\[0.85cm]
    \rule{0.95\textwidth}{1.6pt}\\[0.5cm]

    {\LARGE\bfseries \autor}\\[0.33cm]
    {\large \welchethesis-Thesis}\\[0.15cm]
    {\large zur Erlangung des akademischen Grades}\\[0.45cm]
    {\LARGE \welchethesis\ \thesisofwas}\\[0.1cm]
    {\Large \textit{(M.Sc.)}}\\[0.7cm]
    {\large im Studiengang}\\[0.1cm]
    {\large \welcherstudiengang}\\[0.7cm]

    {\large eingereicht am \datum ~am}\\
    {\large Fachbereich Design Informatik Medien der}\\
    {\large Hochschule RheinMain}\\[1.0cm]

    \begin{tabular}{ll}
      \textbf{Referent} & \referent \\
      \textbf{Korreferentin} & \korreferent
    \end{tabular}

    \vfill
  \end{center}
\end{titlepage}
\cleardoublepage

%% file: chapters/abstract.tex
\chapter*{Abstract}
\label{cha:abstract}
\addcontentsline{toc}{chapter}{Abstract}
\thispagestyle{frontstyle}
Large Language Models (LLMs) are increasingly used for code generation but can produce fluent yet functionally incorrect outputs, which limits trust in their usage for practical software engineering workflows~\cite{bui_correctness_2025, sharma_assessing_2025, huang_risk_2025}. This thesis investigates whether Introspective Uncertainty Estimation (IUE), based on internal hidden-state representations of LLMs, can reliably indicate correctness at the response and line levels for code generation tasks. The objective is to determine the extent to which hidden states encode information about functional code correctness and how this can be leveraged for practical risk assessment and fault localization.
Methodologically, this thesis combines response-level evaluation on LiveCodeBench (LCB)~\cite{jain_livecodebench_2024} and BigCodeBench (BCB)~\cite{zhuo_bigcodebench_2025} with an augmentation pipeline that derives token- and line-level labels from incorrect programs. In this setup, it compares static and dynamic response-level features, evaluates generalization across tasks, programming domains, and token positions, and studies line-level fault localization.

The results show that hidden states contain a strong response-level correctness signal. Static single-token probes perform best, reaching $0.90$ AUROC and $0.96$ F1 on LCB in the best settings, generally surpassing the thresholds of previously reported static probe baselines~\cite{snyder_early_2024, bui_correctness_2025} for IUE. More elaborate dynamic token-selection and sequence-modeling strategies yield no consistent gains. While generalization across tasks, domains, and token positions is feasible, setting-dependent degradation largely remains for real-world software projects. At a fine granularity, line-level prediction in mixed-program settings is substantially harder than response-level estimation. However, in a conditional localization setup with known-incorrect programs, Top-$K$ point-of-failure ranking remains effective, achieving a Top-3 hit rate of $81\%$ in the best setting.
Overall, the findings suggest that hidden states are a robust and informative resource for estimating functional code correctness and localizing faults, supporting a two-stage workflow that combines response-level risk screening with targeted line-level prioritization, and motivate further research on IUE for LLM-based code generation.
\cleardoublepage

%% file: chapters/zusammenfassung.tex
\begin{otherlanguage}{ngerman}
\chapter*{Zusammenfassung}
\label{cha:zusammenfassung}
\addcontentsline{toc}{chapter}{Zusammenfassung}
\thispagestyle{frontstyle}
Large Language Models (LLMs) werden zunehmend zur Codegenerierung eingesetzt, können jedoch flüssige, aber funktional fehlerhafte Ausgaben produzieren, was das Vertrauen in ihren Einsatz in praktischen Softwareentwicklungs-Workflows einschränkt~\cite{bui_correctness_2025, sharma_assessing_2025, huang_risk_2025}.  In dieser Thesis wird untersucht, ob Introspective Uncertainty Estimation (IUE), das auf internen Hidden-State-Repräsentationen von LLMs basiert, Korrektheit auf Antwort- und Zeilenebene von Codegenerierungsaufgaben zuverlässig messen kann.
Methodisch kombiniert die Arbeit die Bewertung auf Antwortebene in LiveCodeBench (LCB)~\cite{jain_livecodebench_2024} und BigCodeBench (BCB)~\cite{zhuo_bigcodebench_2025} mit einer Augmentations-Pipeline, die Token- und Zeilen-Labels aus fehlerhaften Programmen ableitet. In dieser Konfiguration vergleicht sie statische und dynamische Merkmale auf Antwortebene, bewertet die Generalisierung über Aufgaben, Programmierdomänen und Tokenpositionen hinweg und untersucht die Fehlerlokalisierung auf Zeilenebene.

Die Ergebnisse zeigen, dass die Merkmale ein starkes Signal für Codekorrektheit auf Antwortebene enthalten. Statische Single-Token-Probes erzielen die besten Ergebnisse. Unter den besten Bedingungen erreichen sie $0.90$ AUROC und $0.96$ F1 auf LCB und übertreffen damit im Allgemeinen die Schwellwerte zuvor veröffentlichter statischer Baselines~\cite{snyder_early_2024, bui_correctness_2025}. Aufwändigere Strategien zur dynamischen Token-Auswahl und Sequenzmodellierung führen zu keinen konsistenten Verbesserungen. Während eine Generalisierung über Aufgaben, Domänen und Token-Positionen hinweg möglich ist, bleibt eine kontextabhängige Verschlechterung bei Softwareprojekten in der Praxis überwiegend bestehen. Die Vorhersage auf Zeilenebene in gemischten Programmkonfigurationen ist wesentlich schwieriger als die Schätzung auf Antwortebene. In einem Setup zur bedingten Lokalisierung mit nachweislich falschen Programmen bleibt ein Top-$K$-Ranking der Fehlerquellen jedoch effektiv und erreicht in der besten Konfiguration eine Top-3-Trefferquote von $81\%$.
Insgesamt deuten die Ergebnisse darauf hin, dass Hidden States eine robuste und informative Ressource zur Einschätzung der funktionalen Korrektheit von Code und zur Lokalisierung von Fehlerquellen darstellen. Sie regen zu weiteren Forschungen im Bereich der IUE für LLM-basierte Codegenerierung an.
\cleardoublepage
\end{otherlanguage}

%% file: chapters/intro.tex
\chapter{Introduction} \label{cha:intro}

Large Language Models (LLMs) have undergone a rapid transition from research prototypes to practical tools. In the field of software engineering, these systems are now being utilized to assist with tasks that were previously exclusively performed by humans. These tasks include generating code from natural language prompts, repairing bugs, and refactoring legacy implementations~\cite{sharma_assessing_2025}. This progress signals a fundamental transformation in how software may be developed, reviewed, and maintained. However, this shift is constrained by a fundamental reliability problem: LLMs are capable of generating fluent, convincing, and still incorrect code~\cite{lee_hallucination_2025}.

The core motivation of this thesis is the tension between capability and trust. On the one hand, the productivity gains are substantial and increasingly difficult to ignore. However, hallucinations in code generation are still common enough to pose a real engineering risk~\cite{bui_correctness_2025, huang_risk_2025}. In the context of code generation, this thesis uses the term \textit{hallucinations} to refer to two failure modes: (i) the model produces code that is syntactically invalid or fails to execute, and (ii) the model produces code that executes but is semantically unfaithful to the specified functional requirements. These failures undermine the trustworthiness of the generated code and can introduce significant errors into software applications~\cite{lee_hallucination_2025}. Evidence suggests that LLM-generated code can contain more defects than human-written alternatives, increasing debugging overhead and reducing confidence in LLM-assisted workflows. As a result, developers must spend extra time validating generated code, which can offset a substantial proportion of the promised productivity gains~\cite{bui_correctness_2025}.

A common approach to mitigate these risks is post-hoc verification, which can involve static or dynamic code analysis, testing or fact-checking against specifications. While these methods are valuable, they often require expensive infrastructure or significant human effort. Furthermore, they depend on the existence of high-quality oracles, such as comprehensive test suites, formal specifications or knowledge sources. In real development environments, however, test suites may be incomplete, references may be unavailable, and requirements may be inadequately specified. Consequently, verification may confirm correctness while missing broader semantic errors~\cite{bui_correctness_2025, sharma_assessing_2025, huang_risk_2025}. In short, the above post-hoc methods detect some failures, but they do not solve the deeper problem of uncertainty during generation.

Uncertainty Estimation (UE) or Uncertainty Quantification (UQ) attempts to address this issue by estimating the likelihood of a model being incorrect. However, the most widely used methods are information-theoretic and assess uncertainty using the token probabilities provided by the model~\cite{shelmanov_acl_2025, sharma_assessing_2025}. The uncertainty signals derived from these output-level model probabilities are often poorly calibrated for tasks that estimate correctness downstream~\cite{spiess_calibration_2024, campos_multicalibration_2025}. A model can be confidently wrong, particularly in domains where superficial fluency is weakly correlated with functional correctness~\cite{zhang_sirens_2025}. For code generation, this miscalibration can be especially dangerous because high-confidence incorrect code can pass uncertainty inspection while still violating functional requirements. Therefore, while information-theoretic UE provides useful signals, it lacks the robustness required for software engineering.

This motivates a different approach: Introspective Uncertainty Estimation (IUE), which is based on internal model representations and is the focus of this thesis. Rather than relying solely on the model's outputs, introspective approaches examine the process by which the model generates them. Hidden states and attention weights encode intermediate computations that may contain richer reliability information than token probabilities alone~\cite{snyder_early_2024, shelmanov_acl_2025}. In Natural Language Generation (NLG), this approach has already shown promising results. Recent work demonstrates that internal states can encode signals related to factuality and truthfulness, and probe-based methods have achieved meaningful performance in hallucination detection~\cite{snyder_early_2024, ch-wang_androids_2024, preis_hallucination_2025}. These findings suggest that reliability information may be embedded in the model's internal representations throughout decoding.

At the same time, prior introspective work also highlights an unresolved challenge concerning generalization. Probe models trained in one task or domain may exhibit degradation when deployed in an alternative setting~\cite{ch-wang_androids_2024, preis_hallucination_2025}. This raises a critical scientific question: do internal signals capture general properties of model reliability, or are they predominantly task-specific artifacts?

Another intriguing aspect is the level of detail involved in predicting hallucinations. For developers, it is not enough to simply identify that a generated solution is incorrect. They need to know exactly where the error lies. Coarse, response-level hallucination detection leaves them with the costly task of verifying the entire output, whereas fine-grained detection could identify risk in specific lines, transforming uncertainty estimates into practical debugging support. This is particularly important for code, where defects are often concentrated in short sequences of characters~\cite{huang_risk_2025}. Therefore, line-level fault localization, which ranks lines most likely to contain bugs, bridges the gap between UE and practical developer assistance. This improves trust by mitigating potential hallucinations and accelerating the correction process.

This thesis is therefore driven by two related interests. The first is an engineering interest: to reduce the risk of hallucinations and to provide practical assistance to developers without assuming the existence of perfect external oracles. The second is a scientific interest: to determine whether and to what extent internal LLM representations encode reliability information in a domain with objective correctness signals.

Specifically, this work is based on the hypothesis that internal model states during code generation contain predictive information about the correctness of the code, and that this information can be learned. If this hypothesis is validated, it will establish IUE as a practical addition to verification pipelines, enabling more reliable LLM-based code generation. If this hypothesis is not validated, the result would clarify the boundaries of introspective methods and guide future UE research towards alternative mechanisms.

The motivation remains the same in both cases. LLMs are reshaping software engineering, but without reliable UE, their adoption will continue to be limited by risk. In order to establish trustworthy LLM-assisted programming, it is necessary to investigate new methods for reliability assessment that enable hallucination detection for code generation~\cite{bui_correctness_2025, sharma_assessing_2025, huang_risk_2025, ribeiro_llms_2025}.

\section{Scope} \label{sec:scope}

The scope of this thesis is intentionally limited to examine the behavior and limits of IUE for LLM-based code generation. Two decisions define the experimental space.

First, this work focuses exclusively on \textit{introspective} methods that estimate correctness based on internal model artifacts observed during an LLM's forward pass. Consequently, other UE paradigms are considered outside the scope of this work: information-theoretic methods~\cite{duan-etal-2024-shifting, campos_multicalibration_2025} based on output probabilities, consistency-based methods~\cite{manakul-etal-2023-selfcheckgpt} that utilize multiple decoding samples, as well as other black-box methods~\cite{vashurin_benchmarking_2025} and hybrid approaches~\cite{sharma_assessing_2025, liang_learning_2024}. This restriction is intentional, as it allows for a clear analysis of what internal representations alone can and cannot reveal about code correctness during generation.

Second, among internal artifacts, this thesis exclusively uses \textit{hidden states} as the feature representation for probing. Prior work indicates that hidden-state probes are strong baselines for capturing model knowledge signals relative to alternative internal representations such as attention outputs or multilayer perceptron (MLP) activations~\cite{liang_learning_2024}. Additionally, hidden states are more computationally practical for long sequences involving model reasoning and generated programs because their dimensionality remains constant per generated token. In contrast, attention-based representations scale with sequence length, resulting in a substantially larger memory footprint due to the quadratic complexity of self-attention~\cite{vaswani_attention_2023}. For this reason, this thesis does not explore attention-weight-based probing.

\section{Research Questions} \label{sec:rqs}

This thesis investigates whether IUE can be used to accurately predict uncertainty for various code generation tasks, whether it generalizes across them, and whether it is useful for localizing faults in programs. These tasks include code synthesis at a purely algorithmic level and the ability to accurately use library-provided tools to solve a coding task.

Accordingly, this work is guided by the following research questions:

\begin{enumerate}[label=(\roman*)]
\item \textbf{RQ1:} To what extent can internal artifacts produced by LLMs during the forward pass of code generation tasks be used to estimate code correctness? \label{item:rq1}
\item \textbf{RQ2:} To what extent do introspective uncertainty estimation methods generalize across code generation tasks? \label{item:rq2}
\item \textbf{RQ3:} Is fine-grained hallucination detection on code feasible using introspective uncertainty estimation methods? \label{item:rq3}
\end{enumerate}

\noindent \textbf{RQ1.}
Prior work has shown that hidden states from selected token positions can be used to estimate response-level correctness in code generation~\cite{bui_correctness_2025, ribeiro_llms_2025}. This thesis builds on this line of research by investigating whether richer representations of the generation process enhance predictive capabilities. Rather than relying solely on individual token states, it examines sequential hidden-state signals across multiple token positions, alternative architectural choices beyond standard MLP probes, and entropy-based filtering strategies to identify informative token positions. The underlying question is whether correctness estimation improves when introspection captures more of the sequential context of the generation process.

\noindent \textbf{RQ2.}
Generalization is a known weakness of probe-based introspective methods in NLG, where performance often degrades across domains and tasks~\cite{ch-wang_androids_2024, preis_hallucination_2025}. For code generation, evidence on this issue remains limited. To our knowledge, this thesis is among the first to conduct a targeted study of cross-task generalization behavior for IUE in a code generation setting. This study systematically evaluates IUE capabilities across diverse benchmarks, programming domains, and token extraction positions. This is a critical question because practical deployments require methods that remain reliable beyond the exact task distribution on which they were trained.

\noindent \textbf{RQ3.}
Previous code-focused introspective work mostly operates at the response level, yielding coarse predictions about whether an entire output is correct. This thesis therefore explores whether introspective signals can support fine-grained hallucination detection and identify likely faulty lines within generated code. While fine-grained detection has shown promise in grounded NLG tasks such as summarization~\cite{ch-wang_androids_2024}, there is a lack of evidence supporting its application for code generation. To our knowledge, this thesis is among the first to examine whether introspective methods can be used for line-level fault localization in generated code. Specifically, this thesis investigates the use of features from all token positions in code generation. It also explicitly utilizes program synthesis benchmarks for evaluation, contributing to a more comprehensive understanding of the practical utility of fine-grained detection methods in real-world software engineering contexts.

\section{Contributions} \label{sec:contributions}

To address the research questions in this thesis, this work contributes datasets and an empirical study of Introspective Uncertainty Estimation (IUE) for LLM-based code generation, spanning both response-level correctness prediction and fine-grained fault localization. Specifically, RQ1 is addressed by probing hidden states for response-level correctness estimation, RQ2 by evaluating transfer across benchmarks and domains, and RQ3 by extending introspective methods to line-level fault localization. The primary contributions are as follows:

\begin{itemize}
    \item \textbf{Comprehensive Code Dataset:} We curate a dataset of code generations across two state-of-the-art code generation benchmarks, produced with four distinct open-weight LLMs, and assign response-level correctness labels using benchmark-provided test suites.
    \item \textbf{High-Granularity Augmented Dataset:} We augment one benchmark for two models with automatically repaired program versions of each incorrect program. This augmented dataset provides fine-grained, token- and line-level labels, which are derived by computing diffs between the original and repaired code.
    \item \textbf{Response-Level Introspective Signals:} We study how hidden states collected during decoding encode response-level code correctness information, and evaluate probe designs and feature choices for learning this signal.
    \item \textbf{Cross-Task Generalization Analysis:} We evaluate the generalization capabilities of IUE methods in code generation. This analysis tests model performance across diverse benchmarks, domains, and token extraction positions to determine the practical robustness of introspective signals.
    \item \textbf{Line-Level Fault Localization:} We develop and evaluate a line-level IUE approach. Our method assists developers by highlighting the lines of code that are most likely to be flawed, thus connecting our research to practical developer assistance.
\end{itemize}

The remainder of this thesis is structured as follows. Chapter~\ref{cha:bg} introduces the technical background of this work. Chapter~\ref{cha:related_work} synthesizes related literature. Chapter~\ref{cha:methodology} presents the methodological framework of this thesis, including the design of probing methods and data collection pipelines. Next, Chapter~\ref{cha:experiments} describes the experimental setups aligned with the research questions. The results are thoroughly discussed in Chapter~\ref{cha:discussion}. Finally, Chapter~\ref{cha:conclusion} concludes the thesis by summarizing the implications of the findings and outlining directions for future work.

To ensure reproducibility and facilitate future research, all code and datasets used in this thesis are publicly available at:

\url{https://github.com/tomatsch87/iue-thesis}

%% file: chapters/background.tex
\chapter{Background} \label{cha:bg}

This chapter introduces the background concepts necessary for understanding the methodology of this thesis. First, it provides an overview of LLMs, including key capabilities and limitations. Next, it discusses hallucinations. Then, it briefly reviews the transformer architecture underlying LLMs. Finally, the chapter concludes with a discussion of the model architectures for Introspective Uncertainty Estimation (IUE) used in this thesis.

\section{Large Language Models} \label{sec:llms}

Prior to the emergence of LLMs, Natural Language Processing (NLP) primarily relied on statistical methods and small, task-specific neural networks. This paradigm shifted toward building larger, generalized models as research demonstrated that scaling both model size and training data yields substantial performance improvements. LLMs embody this approach, operating as sequence prediction systems trained on large, unstructured text corpora under a self-supervised objective to predict the next token in a sequence. By utilizing large parameter scales to capture complex statistical patterns, LLMs can store extensive knowledge, and comprehend and generate language~\cite{zhao_survey_2026}.

Scaling up these architectures has unlocked unprecedented capabilities. The term ``large'' reflects the sheer number of parameters and the volume of training data required. As models surpass certain parameter thresholds, they demonstrate what researchers refer to as ``emergent abilities''~\cite{wei_emergent_2022}. These behaviors are entirely absent in smaller models and appear suddenly and unpredictably at larger model scales. Examples include few-shot learning and multi-step reasoning, which allow LLMs to solve a wide range of tasks, from mathematical problem-solving to executing intricate instructions, without the need for explicit retraining. This capacity for broad generalization enables LLMs to effectively function as general-purpose task solvers~\cite{zhao_survey_2026}. In the context of software engineering, these capabilities empower models to perform sophisticated code generation based on natural language prompts.

Despite their impressive performance and flexibility, LLMs have significant shortcomings and risks. They suffer from knowledge-recency issues because their parametric knowledge is limited to a training-data cutoff date. Additionally, because LLMs learn from vast, largely unfiltered web corpora, they can memorize and perpetuate societal biases, offensive language, and harmful content~\cite{zhao_survey_2026}. When applied to source code, LLMs can learn from open-source code repositories containing bugs or security vulnerabilities, which may be reproduced in generated code. Furthermore, training and deploying these architectures also requires substantial computational and financial resources, which severely limits their accessibility~\cite{zhao_survey_2026}. Ultimately, the main challenge addressed in this thesis is hallucinations.

\section{Hallucinations} \label{sec:hallu}

Hallucinations occur when large language models generate content that is unverifiable or factually incorrect~\cite{zhao_survey_2026}. In the field of NLG, this phenomenon is defined more broadly as the generation of content that appears to be nonsensical or unfaithful to the provided source material~\cite{huang_survey_2025, zhang_sirens_2025}. These errors are commonly categorized into two types. The first type is intrinsic hallucinations, which directly conflict with the provided source context. The second type is extrinsic hallucinations, which cannot be verified against any external knowledge or source. Therefore, there is no evidence that supports or contradicts extrinsic hallucinations.~\cite{huang_survey_2025}.

A more granular taxonomy distinguishes between two specific types of hallucinations based on the model's objective. First, factuality hallucinations refer to discrepancies between the generated content and verifiable real-world facts. They typically manifest as factual contradictions, such as entity or relation errors, or factual fabrications, which are unverifiable claims that lack universal consensus~\cite{huang_survey_2025}. Second, faithfulness hallucinations occur when the output diverges from user instructions or lacks internal consistency. Subcategories of this type include instruction inconsistency (deviating from a user's directive), context inconsistency (contradicting the provided context), and logical inconsistency (internal contradictions within a reasoning chain)~\cite{huang_survey_2025}.

The origins of why these errors occur are multifaceted and can be traced back to model architecture as well as the various stages of their training and deployment. Key contributing factors include limitations in parametric memorization, overinflated self-confidence, misleading alignment, and generation-time risk~\cite{zhang_sirens_2025}. During pre-training, models internalize vast amounts of information into their parametric memory. Hallucinations often arise because models have either memorized false or outdated knowledge from their training corpora, or they simply lack the knowledge necessary to answer a specific query~\cite{zhang_sirens_2025, zhao_survey_2026}. Furthermore, LLMs often suffer from overconfidence, systematically overestimating their capabilities and the extent of their factual knowledge. This tendency can cause models to fabricate answers with unwarranted certainty rather than acknowledging uncertainty~\cite{zhang_sirens_2025}.

Alignment processes, such as Reinforcement Learning from Human Feedback (RLHF), can potentially mislead models and inadvertently encourage hallucinations~\cite{zhang_sirens_2025}. Under these conditions, models may exhibit sycophancy by generating responses that appeal to human evaluators rather than maintaining strict truthfulness. This often occurs when models are trained on instructions that exceed their knowledge boundaries, forcing them to guess or fabricate~\cite{zhang_sirens_2025, huang_survey_2025, zhao_survey_2026}. Finally, the chosen generation strategy may result in hallucinations. Sampling-based decoding methods such as top-$p$ or top-$k$ introduce randomness to enhance creativity~\cite{zhang_sirens_2025}. However, this randomness increases the risk of sampling infrequent or irrelevant tokens from the distribution's tail~\cite{huang_survey_2025, zhao_survey_2026}. Consequently, models may suffer from hallucination snowballing, where they over-commit to an initial error and then maintain internal self-consistency by making subsequent errors that perpetuate the initial one~\cite{zhang_sirens_2025}.

\noindent \textbf{Code generation.}
As introduced in Chapter~\ref{cha:intro}, in the domain of software engineering and LLM-based code generation, this thesis defines hallucinations by two distinct failure modes. Specifically, a hallucination occurs when the model either generates code that is syntactically invalid or fails to execute, or when it produces code that executes but is semantically unfaithful to the specified functional requirements~\cite{lee_hallucination_2025}.

The implications of hallucinations in software engineering are significant. Code hallucinations may manifest as subtle bugs, memory leaks, or critical security vulnerabilities. Because LLMs can generate fluent code, these errors are often difficult for developers to identify, especially if they only occur under specific execution circumstances. Ultimately, deploying such unfaithful code undermines the reliability of software applications and introduces substantial operational risks~\cite{lee_hallucination_2025}.

For the scope of this thesis, it is not critical to diagnose the specific types of hallucinations or trace their exact technical origins. The primary objective is to reliably identify when an error of either type occurs so it can be mitigated effectively, improving the trustworthiness and functional correctness of generated code. To this end, we focus on IUE, as a practical approach to estimate uncertainty based on the internal representations of LLMs. This requires an understanding of the transformer architecture.

\section{The Transformer Architecture} \label{sec:transformer}

The success of modern LLMs is deeply rooted in the transformer architecture introduced in ``Attention Is All You Need''~\cite{vaswani_attention_2023}. The transformer was designed to overcome the fundamental bottlenecks of sequential computation inherent in recurrent neural networks (RNNs) by dropping recurrence entirely. Instead, it relies on self-attention to model global dependencies between inputs and outputs. This architecture enables parallel processing of input tokens, which significantly improves training efficiency on modern hardware and makes it feasible to train models with very large parameter counts on vast amounts of data~\cite{vaswani_attention_2023, zhao_survey_2026}.

Since this thesis focuses on IUE for transformer-based LLMs, understanding the internal representations produced during their forward pass is essential because these representations serve as features for our estimation models. Specifically, we examine transformers used for auto-regressive language modeling (decoder models), which receive a sequence of input tokens and predict output tokens sequentially by conditioning on prior context~\cite{jm3}.

\begin{figure}[!ht]
    \centering
    \includegraphics[width=0.5\textwidth]{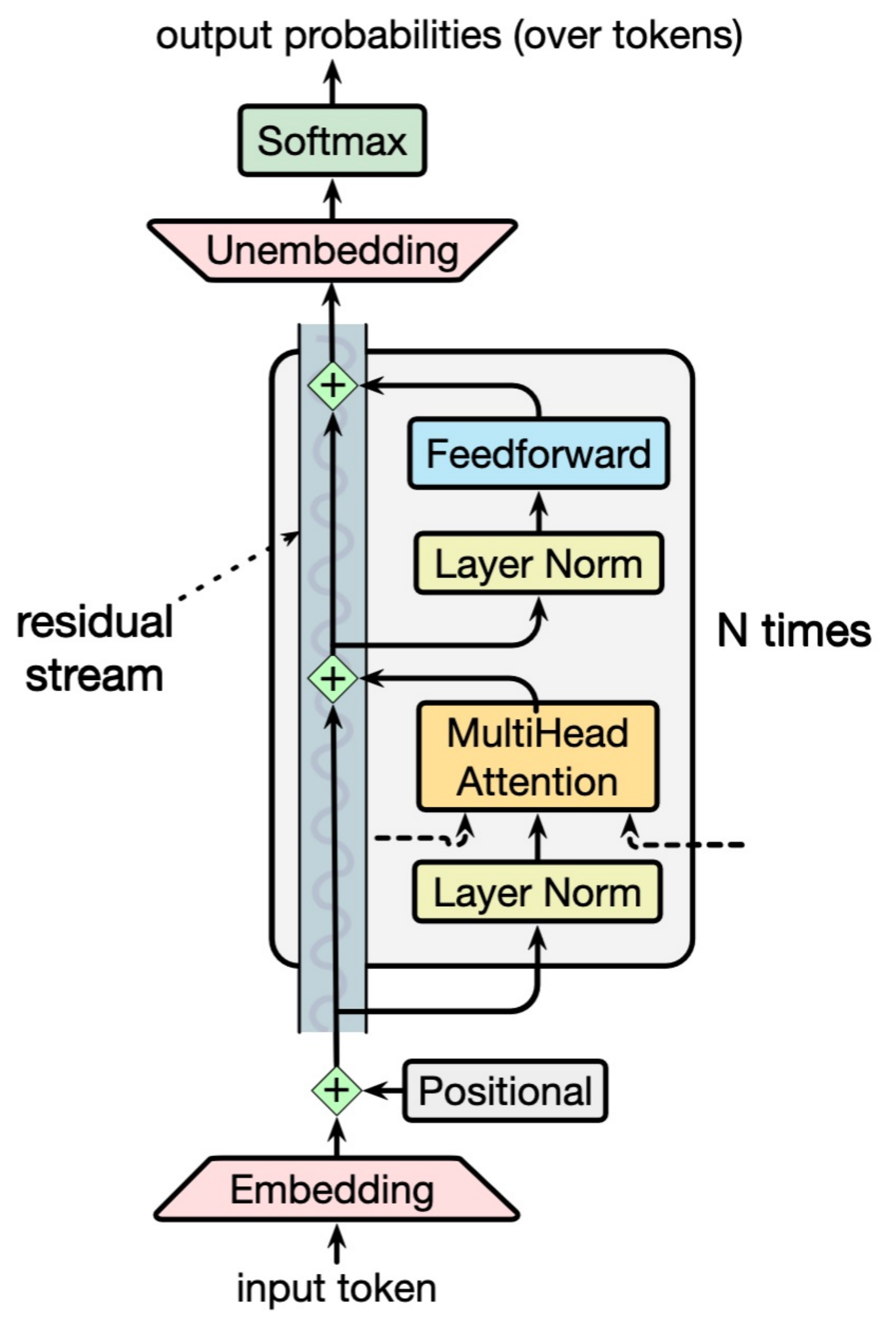}
    \caption{A visualization of the transformer showing the residual stream for processing an input token and outputting a distribution over next tokens.~\cite[Figure 8.1]{jm3}}
    \label{fig:transformer_decoder}
\end{figure}

When processing a prompt, the input text is decomposed into a sequence of discrete tokens. Each token is mapped to a high-dimensional embedding vector and combined with positional encodings~\cite{jm3}. This representation then flows through a stack of transformer blocks, as illustrated in Figure~\ref{fig:transformer_decoder}. Throughout this process, the evolving, embedding-based representation of each token is referred to as its hidden state. It is important to note that the dimensionality of this embedding remains static across the residual stream. In this thesis, we are particularly interested in the hidden state from the last layer of the model because this representation contains all internal computations before being processed to generate output probabilities for the next token through the ``unembedding'' step.

Within each block, the hidden state first enters a multi-head self-attention layer, which computes attention weights that determine the interaction strength between each token and all previous tokens in its context. Self-attention builds contextual representations by integrating information from preceding tokens into the embeddings, enabling the model to understand how tokens relate to each other over large spans of text~\cite{jm3}.

Although self-attention is effective at capturing long-range dependencies, it is computationally intensive, scaling quadratically with sequence length. As discussed in Section~\ref{sec:scope}, this quadratic complexity makes attention weights impractical as features, as their dimensionality increases quickly for long generation sequences~\cite{vaswani_attention_2023}.

After the self-attention mechanism, the hidden state passes through a feedforward layer~\cite{jm3}. The intermediate values computed by applying a non-linear activation function within this layer are referred to as MLP activations in this thesis. Across each layer and for each generated token, the model produces a rich set of intermediate artifacts, including attention weights, MLP activations, and updated hidden states that encode contextually enriched token representations. In this thesis, the primary features used for training IUE models are hidden states.

\section{Model Architectures} \label{sec:models}

In order to utilize the intermediate representations of LLMs, we train external classification models on these features with the intention of detecting hallucinations. This thesis explores three model architectures for this purpose: MLPs as a strong neural baseline, XGBoost~\cite{chen_xgboost_2016} as a regularized tree ensemble alternative, and long short-term memory (LSTM)~\cite{hochreiter_long_1997} networks, which can explicitly model long-range dependencies across sequences of token representations. The following subsections introduce these model architectures at a high level.

\subsection{Multilayer Perceptron (MLP)} \label{subsec:mlps}

In the field of explainable AI, researchers first introduced so-called probes as a diagnostic tool to address the black-box property of deep neural networks~\cite{alain_understanding_2018}. This non-invasive approach involves attaching an independent classifier, or probe, to specific internal layers of a frozen network. These probes act as ``thermometers,'' allowing researchers to measure the internal state of a model at many different locations simultaneously without interfering with its computations. Inserting a probe at a given layer queries internal information and helps researchers understand what the network has learned at different stages of its processing chain~\cite{alain_understanding_2018}.

In the context of IUE, probes are commonly used to predict hallucinatory behavior based on internal representations. The most frequently used model architecture for building these diagnostic probes is the multilayer perceptron (MLP)~\cite{preis_hallucination_2025, snyder_early_2024, ch-wang_androids_2024}. As simple feedforward neural networks~\cite{goodfellow_deep_2016}, MLPs provide an effective baseline for interpreting the high-dimensional features generated by transformers.

The primary objective of an MLP is to approximate a function that maps an input to a specific category or value. Its architecture is structured as a sequence of connected layers through which information flows strictly from the input to the output, with no feedback loops. Between the input and output layers are multiple hidden layers consisting of parallel units that apply linear transformations to the data via learned weights and biases. To capture complex, non-linear relationships, an activation function is applied to the output of these linear transformations. Through training, the model autonomously learns to represent the input data within these hidden layers to optimize its predictive accuracy~\cite{goodfellow_deep_2016}.

\subsection{XGBoost} \label{subsec:xgboost}

Gradient tree boosting is an ensemble method that builds a strong predictor by adding decision trees iteratively. Rather than fitting all trees in a single step, the model is trained additively. At each iteration, a new tree is selected to improve the current training objective based on the residual errors of the existing ensemble.~\cite{xgboost_introduction_2026, chen_xgboost_2016}.

XGBoost (Extreme Gradient Boosting) is a tree boosting system designed to deliver strong performance while remaining resource-efficient. A key feature of the system is its regularized learning objective, which minimizes a differentiable loss function while penalizing model complexity by taking into account the number of leaves and the $L_2$ norm of leaf weights~\cite{chen_xgboost_2016}. This smooths the learned weights and prevents overfitting. 

Its effectiveness also stems from several algorithmic improvements, such as cache-aware prefetching that reduces memory access overhead during split enumeration, and out-of-core computation utilizing block compression and sharding to process large datasets.~\cite{chen_xgboost_2016}. These design choices make XGBoost suitable for large-scale, high-dimensional supervised learning tasks.

Considering these properties, XGBoost offers a suitable probe architecture for IUE. It complements MLP-based probes with a different inductive bias and provides highly efficient training behavior.

\subsection{Long Short-Term Memory (LSTM)} \label{subsec:lstm}

\begin{figure}[!ht]
    \centering
    \includegraphics[width=0.6\textwidth]{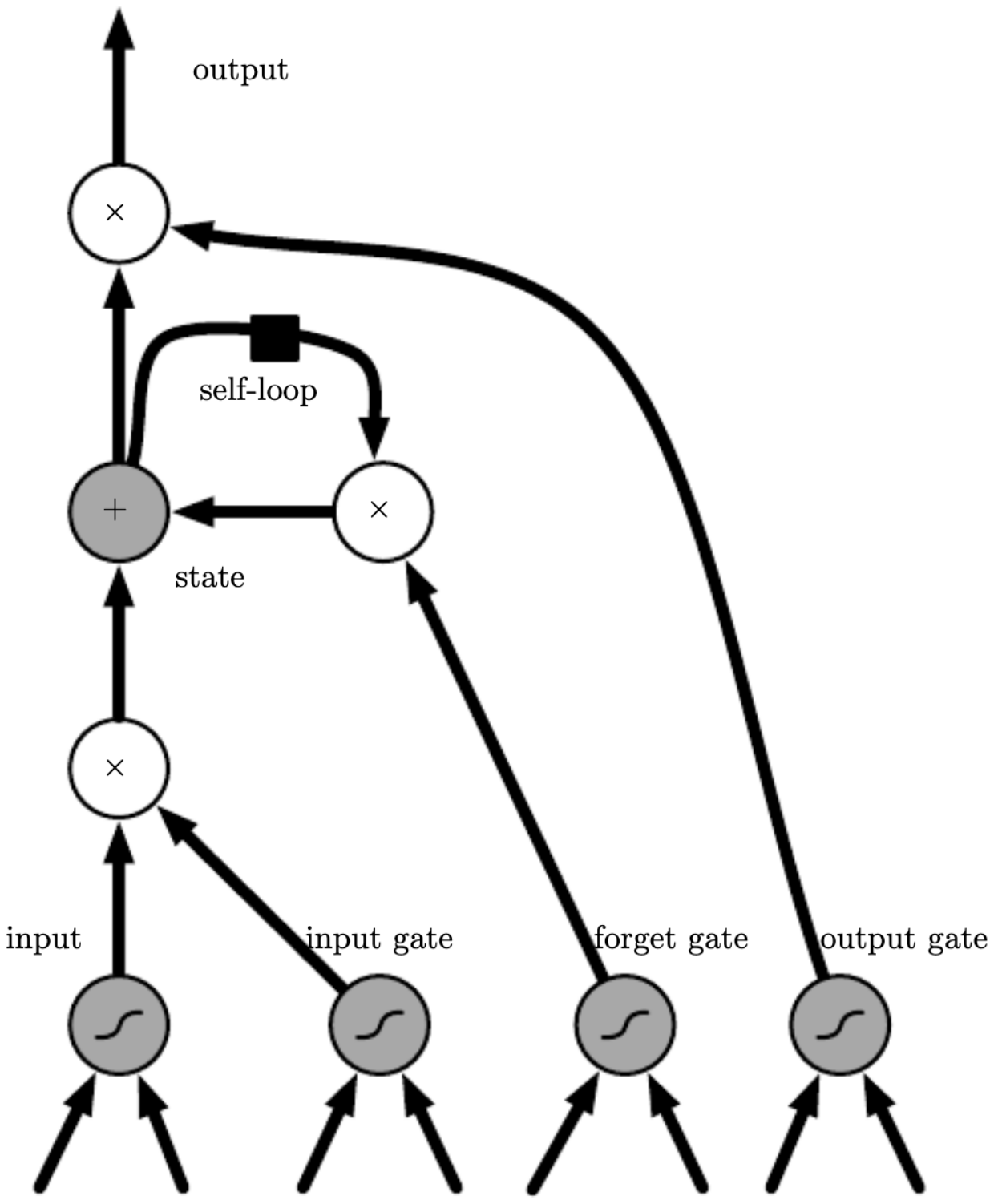}
    \caption{A block diagram of an LSTM cell. The cells are connected recurrently to each other, replacing the hidden units found in traditional recurrent networks. An input feature is computed using a standard artificial neuron. If the sigmoid input gate activates, its value can be accumulated into the state. The state unit has a linear self-loop whose weight is controlled by the forget gate. The cell's output can be shut off by the output gate. All the gating units have a sigmoid non-linearity. The state unit can also serve as an additional input for the gating units. The black square indicates a delay of one time step.~\cite[Figure 10.16]{goodfellow_deep_2016}}
    \label{fig:lstm_cell}
\end{figure}

LSTM~\cite{hochreiter_long_1997} networks are a specialized class of RNNs designed for processing sequential data. Unlike standard RNNs, which maintain a single hidden state per token, LSTMs use structured memory cells with gating mechanisms that control the flow of information over time, as illustrated in Figure~\ref{fig:lstm_cell}. The input, forget, and output gates determine which information is written, retained, and output at each time step.

This architecture is particularly relevant when dependencies span a larger range of input features. In conventional RNNs, gradients can vanish across many recurrent steps, making it difficult to learn relationships between distant elements in a sequence. LSTMs mitigate this effect with an additive memory path and gating-based updates. These features support more stable gradient flow and enable the model to preserve task-relevant context over longer time periods~\cite{goodfellow_deep_2016}.

At the same time, LSTMs have practical limitations. Their recurrent computation is inherently sequential, which reduces parallelism and can increase training time compared to alternative architectures. Additionally, while LSTMs mitigate vanishing-gradient issues, optimization can become unstable in steep regions of the loss landscape, so techniques such as gradient clipping are often necessary~\cite{goodfellow_deep_2016}.

For IUE, this trade-off makes LSTMs a meaningful complementary probe architecture because they can model the structure of token-wise generation using sequences of internal representations to provide the learning signal.

%% file: chapters/related_work.tex
\chapter{Related Work} \label{cha:related_work}

This chapter reviews previous research on detecting hallucinations and assessing the reliability of LLM outputs, focusing on contributions relevant to code generation. Following the taxonomy of \cite{huang_survey_2025}, we first discuss fact-checking approaches, including post-hoc verification techniques used to evaluate generated code. Next, we review UE methods that aim to predict hallucinations in oracle-free scenarios, before narrowing in on introspective approaches. Finally, we examine fine-grained hallucination detection methods that go beyond response-level predictions.

\section{Fact Checking} \label{sec:rw_fact_checking}

Fact checking aims to assess whether a model output is supported by reliable evidence. In the context of hallucination detection, this involves verifying generated claims against knowledge sources by either retrieving external evidence or checking a model's internal knowledge~\cite{huang_survey_2025, preis_hallucination_2025}. Although this literature was developed primarily for NLG, where verification focuses on factual ``claims'', the central idea maps directly to our definition of hallucinations in code generation (see Section~\ref{sec:hallu}). In this domain, verifying a claim is analogous to evaluating whether a generated program is syntactically valid and faithfully implements the specified functional requirements.

\noindent \textbf{External retrieval.}
External retrieval methods ground verification in a trusted text corpus. One example is FActScore~\cite{min_factscore_2023}, which evaluates long-form generations by decomposing them into atomic facts that can be verified independently. Each fact is checked against evidence retrieved from an external source using a pipeline that combines retrieval with an evaluation model to judge whether the evidence supports the claim. These decisions yield a fine-grained factuality score that is more informative than binary assessments at the response level. Retrieval-based fact checking provides a clear notion of what counts as evidence and offers interpretable failure cases. At the same time, its effectiveness depends on corpus coverage and retrieval quality.

\noindent \textbf{Internal checking.}
Internal checking methods try to reduce hallucinations without checking external data, using the model's ability to critique and revise its own outputs. Chain-of-Verification (CoVe)~\cite{dhuliawala_chain_2023} is an example of this approach. The model first produces an initial response, then generates verification questions for that response, answers those questions independently, and finally revises the original response based on the verification results. A key insight of this method is that LLMs can often answer short factoid questions correctly even when they have hallucinated that same information in a longer response. Such methods can mitigate certain hallucinations and improve consistency, but they are limited by the model's own knowledge boundaries.

\noindent \textbf{Post-hoc verification for generated code.}
The analogue of fact checking for code is post-hoc verification, which involves evaluating a generated program by analyzing or executing it against an external oracle, such as a test suite, a reference implementation or a formal specification with functional requirements~\cite{bui_correctness_2025}. Such approaches are crucial because LLM-generated code provides no inherent correctness guarantees.

However, post-hoc verification is resource-intensive and may miss context-specific issues~\cite{bui_correctness_2025}. Furthermore, it depends on the availability and quality of external oracles. In realistic development settings, oracles may be incomplete or absent, and even when present, they may be insufficiently precise~\cite{sharma_assessing_2025}. Several works address this limitation for research benchmarks.

For instance, EvalPlus~\cite{liu_chatgpt_2023} augments problems with substantially more generated tests, aiming to strengthen the oracle and reduce false positives where incorrect solutions pass weak test suites. Similarly, Turbulence~\cite{honarvar_turbulence_2025} evaluates robustness using parameterized ``question neighborhoods'' paired with oracle templates that include unit tests, reference solutions, and fuzzing generators, enabling systematic assessment of how models can fail under small semantic variations.

Code-specific hallucinations can also concern context, such as unavailable documentation. De-Hallucinator~\cite{eghbali_dehallucinator_2024} tackles project-specific Application Programming Interface (API) hallucinations due to a lack of local context. The approach uses an iterative grounding mechanism that leverages the model's initial, potentially incorrect code to retrieve relevant API references. The framework iteratively refines prompts with relevant context, nudging the LLM towards producing more reliable and factually grounded code that correctly refers to existing APIs.

Overall, fact checking and post-hoc verification provide principled ways to validate model outputs when reliable evidence sources or strong oracles exist. This thesis, however, focuses on practical oracle-free settings, where such external signals are unavailable. This motivates the use of UE methods that aim to predict the likelihood of incorrectness without depending on executing the program or retrieving external knowledge.

\section{Uncertainty Estimation (UE)} \label{sec:rw_ue}
UE, sometimes also referred to as Uncertainty Quantification (UQ), aims to predict how likely a model output is to be incorrect. For hallucination detection, UE applies the idea that hallucinations are closely linked to situations where the model is highly uncertain yet still produces confident responses~\cite{huang_survey_2025}. Therefore, UE provides a universal approach to hallucination detection~\cite{shelmanov_acl_2025}. In contrast to fact checking, UE is typically designed for oracle-free deployment since it does not rely on external knowledge retrieval or executing generated code at inference time. This makes UE appealing for practical settings where code is generated, but tests and specifications may be unavailable or incomplete~\cite{sharma_assessing_2025}.

UE methods can be categorized along several axes. Following the taxonomy of \cite{shelmanov_acl_2025}, the first distinction involves the learning signal. Some approaches are unsupervised and compute an uncertainty score using an intuitive statistic that reflects predictive uncertainty, whereas supervised approaches learn a predictor of correctness or hallucination from labeled data~\cite{shelmanov_acl_2025}. Notably, both types are considered oracle-free at inference time. However, supervised methods typically require labels during training, often obtained from benchmarks, human annotations, or augmentation techniques. 

A second distinction revolves around the assumed level of access to the underlying LLM. Black-box approaches only require the ability to query the model and observe its textual outputs, while white-box approaches assume access to internal model signals, such as token probabilities or intermediate representations~\cite{shelmanov_acl_2025}. The remainder of this section reviews representative UE methods in relation to this black-box/white-box distinction.

\subsection{Black-box Methods} \label{subsec:rw_black_box}
Black-box methods estimate uncertainty without access to the model's internal states or parameters, relying solely on its outputs and behavior. Such methods usually involve either prompting the model to express its confidence, or analyzing the variability of its responses, in order to infer uncertainty~\cite{shelmanov_acl_2025}.

\noindent \textbf{Verbalized uncertainty.}
The simplest black-box strategy is to instruct the model itself to report its level of confidence. After producing an answer, the model can be instructed to additionally output a probability as a calibrated score. This approach is appealing because it does not require any access to model internals and can be applied to all proprietary systems. Yet, the method relies on the model's ability to introspect, which can be impaired by systematic overconfidence~\cite{zhang_sirens_2025}. It also relies heavily on prompt formatting~\cite{shelmanov_acl_2025}. Previous work~\cite{tian-etal-2023-just} has shown that carefully designed elicitation strategies can improve calibration compared to the model's own probabilities in NLG settings. This is achieved by prompting the model to consider multiple hypotheses before assigning a confidence score. However, the sensitivity to prompt templates and dependence on model calibration generally limit the reliability of verbalized uncertainty in practice.

\noindent \textbf{Self-evaluation.}
Similarly, related approaches require the model to explicitly self-evaluate whether a candidate answer is correct. Depending on the available model access, this can be a black-box method or a white-box method if token probabilities of responses are used. Prior work~\cite{kadavath_language_2022} uses this premise as a white-box approach, asking the model to classify its response as true or false, and extracting the probability associated with the ``True'' token as an uncertainty measure. This leverages the model's internal confidence in its generated claim. The study demonstrates that, when prompted specifically to evaluate their own claims, LLMs can often predict their own correctness, suggesting that they possess an implicit sense of uncertainty regarding their own outputs. These findings support the hypothesis that models demonstrate a measurable capacity for self-evaluation, observing that LLMs ``(mostly) know what they know''~\cite{kadavath_language_2022}.

\noindent \textbf{Consistency-based uncertainty.}
Another prominent black-box approach measures uncertainty through disagreement across multiple generations. The guiding intuition is that diverse responses to the same prompt indicate high uncertainty, suggesting that the model is operating near its knowledge boundary and is more likely to hallucinate~\cite{shelmanov_acl_2025}. SelfCheckGPT~\cite{manakul-etal-2023-selfcheckgpt}, a zero-resource, black-box approach utilizes this concept by sampling multiple responses and measuring their consistency with the original response using similarity metrics such as Natural Language Inference (NLI). Their findings suggest that consistency-based metrics can outperform traditional uncertainty measures, such as token probabilities~\cite{manakul-etal-2023-selfcheckgpt}. While consistency-based UE is appealing when token probabilities are unavailable, it is limited by its relatively high computational cost. 

\noindent \textbf{Limitations.}
The limitations of black-box methods are evident. Methods involving verbalized uncertainty and self-evaluation suffer from overconfidence and are highly sensitive to prompting~\cite{tian-etal-2023-just}. A key limitation of consistency-based UE that remains unaddressed is that disagreement between generated samples may not be sufficient to indicate incorrectness. If a model repeatedly produces the same incorrect claim, the samples may appear consistent yet still be inaccurate. Conversely, ambiguity in a task can produce diverse yet entirely correct outputs, which is particularly true for software engineering~\cite{sharma_assessing_2025}. UE methods based on consistency also require multiple samples, which can be computationally expensive~\cite{manakul-etal-2023-selfcheckgpt}. This is particularly problematic for code generation, as generating multiple samples can be costly and time-consuming for developers. However, they remain a practical option in many settings as they do not require access to any internal model artifacts.

In summary, black-box UE offers broad applicability, but only limited access to the most informative uncertainty signals~\cite{snyder_early_2024, shelmanov_acl_2025}. This motivates white-box approaches, which leverage token probabilities and internal model representations, assuming they are available.

\subsection{White-box Methods} \label{subsec:rw_white_box}
White-box UE methods leverage access to internal model computations, such as token probabilities, hidden states, attention weights, or MLP activations. Therefore, they require stronger model access than black-box approaches. These methods analyze the aforementioned features to identify uncertainty signals that may indicate hallucinations~\cite{shelmanov_acl_2025, snyder_early_2024, liang_learning_2024}. It is important to note that \cite{shelmanov_acl_2025} does not consider MLP activations as signals of uncertainty. In contrast, \cite{snyder_early_2024, liang_learning_2024} study them and find that they also have significant predictive power for hallucination detection. Therefore, we include them in the definition of white-box methods.

While white-box methods technically encompass all internal signals, this subsection focuses exclusively on information-theoretic methods operating on output-level token probabilities, calibration approaches for code generation building on these probabilities, and hybrid methods combining consistency-based and information-theoretic approaches. The exploration of introspective approaches that exploit deeper internal representations, such as hidden states and attention weights, is reserved for Section~\ref{sec:rw_iue}.

\noindent \textbf{Information-theoretic uncertainty.}
The most common white-box UE signals treat the token probability distribution as a proxy for uncertainty~\cite{huang_survey_2025}. Typical metrics include the probability of the selected token, or the entropy of the next-token distribution. These signals can be computed at the token level and then aggregated to obtain uncertainty estimates at the sentence or response level. These measures are inexpensive to compute when token-level probabilities are available, and they have been widely adopted as baselines~\cite{shelmanov_acl_2025, fadeeva_fact-checking_2024}.

Despite their simplicity, the choice of aggregation is significant. For longer generations, uncertainty can be localized in only a small fraction of tokens while the majority of tokens carry little uncertainty. Therefore, simply averaging probabilities can obscure the signal at a crucial token position. Conversely, summing token-level scores tends to correlate with length, which can confuse uncertainty with verbosity~\cite{shelmanov_acl_2025}. These effects are especially relevant for code generation, where a single incorrect character can render an otherwise ``likely'' program incorrect~\cite{huang_risk_2025}.

However, information-theoretic UE may not always align correctly with the notion of correctness relevant to hallucination detection. Next-token probabilities are optimized for predicting likely continuations rather than for determining if the final generated statement is factual or if a program meets its functional requirements. As a result, models can be confidently wrong, and probability-based uncertainty can be systematically miscalibrated~\cite{zhang_sirens_2025, preis_hallucination_2025}.

Previous research~\cite{duan-etal-2024-shifting} refines information-theoretic UE by reweighting token contributions. They propose shifting attention toward relevant parts of the generation by calculating token relevance scores, which prioritize semantically critical components over redundant linguistic fillers during UE. They found that these irrelevant components disproportionately inflate traditional uncertainty measures. By shifting attention, their approach significantly outperforms traditional baselines. Nevertheless, the method introduces significant latency due to the computation of the attention shift.

\noindent \textbf{Calibration for code generation.}
In software engineering, the discrepancy between token-level likelihood and functional correctness is especially evident, since a syntactically valid program may still violate functional requirements~\cite{lee_hallucination_2025}. Prior work~\cite{spiess_calibration_2024} studies this mismatch and analyzes how well model probabilities predict functional correctness across coding tasks. The researchers observed that LLMs are not well-calibrated for code generation out of the box and that users would greatly benefit from a reliable indication of the confidence that the generated code is correct. The researchers demonstrate that standard post-hoc calibration methods can improve model calibration but also point out their limited robustness across coding tasks.

Other research~\cite{campos_multicalibration_2025} extends this perspective by considering group-dependent calibration for code generation. Their multicalibration approach strives to correct miscalibration correlated with properties of the prompt or the generated program (e.g., task complexity or program length). This produces confidence estimates that more accurately reflect correctness within relevant subgroups. However, their approach was primarily tested on short-form function synthesis tasks, which may not accurately reflect real-world software engineering.

\noindent \textbf{Hybrid approaches.}
Finally, some methods combine information-theoretic uncertainty and consistency-based uncertainty. In this regard, \cite{sharma_assessing_2025} introduce a semantic uncertainty framework for code generation. Their method targets a key limitation of consistency-based uncertainty that was previously mentioned: drastically different implementations can be semantically equivalent while programs that look similar can behave entirely different. The researchers proposed symbolic clustering to adapt information-theoretic uncertainty measures to the functional requirements of code generation. They found that while traditional semantic clustering is ineffective for code, symbolic clustering correlates strongly with correctness, even when assuming a uniform token-probability distribution. One key shortcoming is its reliance on symbolic execution, which involves loop unrolling of generated programs. This can result in false semantic equivalence despite deeply-rooted behavioral differences.

\noindent \textbf{Limitations.}
White-box methods have limitations based on the fact that efficient token probability measures are fundamentally limited by model calibration~\cite{spiess_calibration_2024, campos_multicalibration_2025} and by the relationship between next-token likelihood and semantic correctness. However, white-box methods can provide more direct signals of uncertainty, and hybrid methods that utilize consistency and information-theoretic aspects tend to be more accurate than standalone approaches~\cite{shelmanov_acl_2025}.

This motivates UE that moves beyond the token probability distributions and instead exploits internal representations produced during decoding. These representations may encode richer information about reliability than token probabilities alone~\cite{snyder_early_2024, shelmanov_acl_2025}.

\section{Introspective Uncertainty Estimation (IUE)} \label{sec:rw_iue}
IUE methods analyze a model's internal computations. Rather than relying solely on output-level token probabilities, we shift our focus to approaches that examine artifacts produced during decoding that are deeply rooted inside the model itself, including hidden states, self-attention maps, and MLP activations. These methods are based on the intuition that hallucinated and factual generations often have different distributions in internal representations, even when final token probabilities appear similarly confident. There is strong evidence supporting this assumption, showing that these artifacts can outperform token probability-based signals for UE~\cite{snyder_early_2024, shelmanov_acl_2025}.

Following prior taxonomies, IUE literature can be organized along two dimensions: the learning signal and the representation type~\cite{shelmanov_acl_2025}. The learning signal distinguishes between unsupervised scoring methods and supervised probing methods, which are trained using labeled hallucination data~\cite{shelmanov_acl_2025}. The representation dimension distinguishes approaches based on hidden states, attention weights, and MLP activations, with some methods combining multiple artifacts~\cite{snyder_early_2024, liang_learning_2024}. Following this taxonomy, we discuss introspective methods, which are first categorized according to their learning signal and subsequently by other attributes, such as representation type, beginning with unsupervised methods.

\subsection{Unsupervised Methods} \label{subsec:rw_unsup}
Unsupervised IUE methods compute uncertainty scores directly from internal representations, which makes them particularly attractive when labeled hallucination data is scarce or costly to obtain.

\noindent \textbf{Hidden states.}
INSIDE~\cite{chen_inside_2024}, introduces a method that combines introspection with consistency-based analysis. Given multiple responses to the same prompt, the method extracts sentence embeddings from internal states and computes an eigenvalue-based score (EigenScore) related to differential entropy in the embedding space. Intuitively, when the model is knowledge-consistent, the sampled responses remain semantically clustered, resulting in lower dispersion. However, when the model is uncertain, semantic spread increases, and the EigenScore reflects this internal behavior. To mitigate overconfident failure cases, the approach adds feature clipping at inference time, which truncates extreme activation values before scoring.

This research demonstrates that introspection-based metrics can outperform purely information-theoretic or consistency-based baselines across several NLG benchmarks~\cite{chen_inside_2024}. At the same time, INSIDE inherits the practical limitation of consistency-based approaches in that it requires multiple generations per prompt, thereby increasing computational cost.

\noindent \textbf{Multiple internal representations.}
Previous research~\cite{sriramanan_llm-check_2024} proposes LLM-Check, which is designed to detect hallucinations in a single response and avoid repeated sampling. The framework uses teacher forcing\footnote{Teacher forcing, also referred to as forced decoding or prefilling in the context of this thesis, is a strategy for training neural networks on sequential inputs. Instead of using model output from a prior time step, it always uses ground truth as input~\cite{lamb_professor_2016}.} on a fixed prompt-response pair to extract internal features in a single forward pass. They compute three features: (i) a hidden score from the hidden-state covariance structure, (ii) an attention score derived from the log determinants of self-attention maps, and (iii) output-level uncertainty features, such as perplexity and entropy, from the models' token probabilities.

Their methodology achieves substantial speedups compared to consistency-based baselines and remains competitive or superior on hallucination detection benchmarks~\cite{sriramanan_llm-check_2024}. Their results also reinforce a recurring pattern in IUE that the middle-to-late layers of the model often carry the strongest reliability signal. However, the approach necessitates white-box access, and performance can vary significantly across layers, underscoring the importance of layer selection for optimal performance.

Ultimately, unsupervised introspective methods show that internal representations alone provide useful signals of uncertainty. Despite this, they tend to be less accurate than supervised methods~\cite{shelmanov_acl_2025}.

\subsection{Supervised Methods} \label{subsec:rw_sup}
Supervised IUE methods use labeled data to train a predictor of correctness or hallucination from internal representations. While these methods typically outperform unsupervised approaches on in-domain tasks, they depend on labeled hallucination data, which may be expensive to obtain depending on the domain or task~\cite{shelmanov_acl_2025}.

\noindent \textbf{Hidden states.}
Azaria and Mitchell~\cite{azaria_internal_2023} present one of the key early studies in this area. Their SAPLMA framework trains a lightweight MLP classifier on the LLM's internal hidden states, which are collected while the model processes true and false statements. To reduce overfitting on artifacts from specific topics, they evaluate held-out topics to measure whether the probes capture a broader internal truthfulness signal. They report significant improvements over self-evaluation and information-theoretic baselines, with the strongest layers typically found in the middle to late-middle regions of the model.

The study set an important methodological precedent by demonstrating that simple MLP probes can extract meaningful truthfulness signals from a model’s internal states~\cite{azaria_internal_2023}. Yet, the approach has clear limitations, as it requires a curated true/false statement dataset. Additionally, the authors reported that performance varies with domain coverage during training, foreshadowing a broader generalization challenge.

\noindent \textbf{Multiple internal representations.}
Snyder et al.~\cite{snyder_early_2024} expand the analysis of representations by comparing four types of internal features: integrated gradients, softmax probabilities, self-attention scores, and fully connected (MLP) activations. They ask whether hallucination risk can be identified early, before the full answer is produced. Their results show that internal representations, particularly self-attention scores and MLP activations, are more predictive than token probabilities across several datasets and models. In many settings, these internal probes achieve AUROC scores above 0.70, with peak performance around 0.81.

Their work yielded several findings that are particularly relevant to this thesis. First, probe effectiveness increases with layer depth. Early layers provide little information, whereas middle and later layers more effectively distinguish hallucinated from non-hallucinated outputs. Next, detectable differences emerge as early as the first generated token, even before a complete answer has been produced. Finally, the authors find that simple MLP classifiers are sufficient for their objective, as deeper probe architectures did not consistently improve performance~\cite{snyder_early_2024}.

\noindent \textbf{Architectural scaling.}
In contrast, Prei{\ss}~\cite{preis_hallucination_2025} investigates whether more complex model architectures improve hallucination detection performance using hidden states. Their approach uses a shared MLP feature extractor across layers, followed by layer comparison and aggregation modules. Their best configuration outperforms a last-layer baseline, albeit modestly (by roughly 3--4\% F1 in reported settings), prompting the question of whether increased architectural complexity is practical.

The work confirms two broader points in the literature. First, middle-to-late layers are the most informative. Second, cross-benchmark generalization remains weak, and performance observed on one dataset transfers poorly to others. The study also explores token-level hallucination localization and reports difficulty under severe class imbalance, where non-hallucinated tokens dominate~\cite{preis_hallucination_2025}. This result indicates that transitioning from response-level to token-level predictions poses a considerable challenge with regard to supervised IUE.

\noindent \textbf{Hallucination mitigation.}
Other research~\cite{liang_learning_2024} uses introspection not only for detection but as a component in hallucination mitigation pipelines. The researchers propose a hybrid framework in which linear probes estimate the uncertainty of a model by examining hidden states, attention outputs, and MLP outputs. These probe predictions are then used to generate training signals for a reward model that guides reinforcement learning via Reinforcement Learning from Knowledge Feedback (RLKF).

Their findings align with earlier results, showing that probe performance increases from the early to middle layers and remains high in the later layers. This indicates that several layers of computation are necessary before knowledge-state information becomes explicit. They also demonstrate that hidden states generally provide the strongest baseline for capturing model knowledge relative to attention or MLP outputs~\cite{liang_learning_2024}.

Overall, supervised IUE shows strong in-domain predictive capabilities based on internal representations, with hidden states and middle-to-late layers of the model often providing the best results. However, there remains a persistent gap between in-domain benchmark performance and robust cross-domain behavior.

\subsection{Code Generation} \label{sec:rw_code_gen}
Recent work has expanded the scope of IUE from NLG to software engineering, where correctness is measured by program behavior. This research is the focal point of the thesis. The shift is significant because it tests whether introspective signals remain informative when the target is functional program correctness rather than textual factuality. Several works have already indicated that hidden states carry predictive information for code correctness~\cite{bui_correctness_2025, ribeiro_llms_2025, huang_risk_2025}.

\noindent \textbf{Hallucination detection.}
Bui et al.~\cite{bui_correctness_2025} propose OPENIA, a supervised framework that predicts code correctness directly from hidden representations, thereby reducing reliance on post-hoc verification for hallucination detection. They use popular code generation benchmarks to generate programs and derive labels from the benchmark test suites. Then, they train lightweight MLP probes on features extracted from different layers and token positions.

Their analysis reveals a practical interaction between layer depth and token selection. Shallow layers are sensitive to token position, while deeper layers remain stable across positions. The study found that deep-layer representations of the last code token often achieved the strongest F1 scores~\cite{bui_correctness_2025}. This aligns with causal decoding, in which the last token has contextualized the full generated context. While OPENIA experiments with four distinct static token positions (first token, first code token, last code token, and last token of the response) and observes significant performance differences among them, their reliance on individual, static hidden states risks missing dynamic reliability signals that may emerge at various points throughout the code generation process.

This limitation directly motivates the first research question \hyperref[item:rq1]{(RQ1)} of this thesis, which investigates whether capturing richer representations, such as sequential hidden-state signals across multiple tokens, or dynamically selecting relevant tokens via entropy-based filtering, can enhance predictive capabilities for code correctness over single-token baselines.

\noindent \textbf{Hallucination mitigation.}
Ribeiro et al.~\cite{ribeiro_llms_2025} focus on hallucination mitigation rather than pure detection. They do this by ranking candidate programs with an internal correctness score. They use representation-engineering techniques to extract a correctness axis from hidden states, usually at the final token position. First, they encode contrastive pairs of correct and incorrect programs. Then, they build layer-wise difference vectors. Finally, they use a principal direction to score new candidates by projection. The ranking surpasses verbalized and token probability measures and is more efficient than test-intensive reranking pipelines. They also recognize that their labels, which are based on benchmark-provided test suites, do not capture the non-functional qualities of the code. This highlights the limitations of using test suites to generate correctness labels for programs.

The study also explores out-of-distribution (OOD) behavior and notes a fundamental trade-off between specialization and generalization. The researchers found that fitting their model to in-distribution data achieves higher accuracy by exploiting dataset-specific nuances. However, this specialization leads to instability when the model is tested on unseen data. Conversely, fitting on OOD data captures more general features, albeit at the cost of lower overall accuracy~\cite{ribeiro_llms_2025}.

This motivates the second research question \hyperref[item:rq2]{(RQ2)} of this thesis. Specifically, it expands upon these initial OOD observations by systematically evaluating the generalization capabilities of IUE methods for code generation across diverse benchmarks, programming domains, and token extraction positions to determine the practical robustness of introspective signals.

\subsection{Limitations} \label{sec:rw_iue_limits}
Despite substantial progress, current IUE methods have several limitations in both NLG and code generation. Some of these limitations directly motivated this thesis.

All IUE methods require direct access to internal model artifacts, which limits their applicability to proprietary models~\cite{preis_hallucination_2025}. Generalization remains fragile, as probes trained on one benchmark often perform poorly on unseen benchmarks or across tasks and domains. These results suggest a generalization gap and indicate that introspective signals may be partly task-specific~\cite{preis_hallucination_2025, ribeiro_llms_2025}.

Performance is sensitive to model design choices and depends strongly on layer selection and token position choices. Results from NLG and code generation indicate that hidden states and middle-to-late layers are often the most informative. Yet, optimal settings vary by task and model~\cite{azaria_internal_2023, snyder_early_2024, liang_learning_2024, bui_correctness_2025}.

Ultimately, the quality of supervised learning is limited by the quality of the labels. In code generation, benchmark test suites are generally used to generate program labels, and probes inherit the strengths and weaknesses of these tests. This approach can present challenges, as test suites may be incomplete or unreliable, and their results may not fully capture the non-functional quality requirements of the code~\cite{spiess_calibration_2024, ribeiro_llms_2025}.

In summary, IUE demonstrates that internal computations carry strong reliability information, including for code generation tasks. These findings necessitate transitioning from response-level evaluation to a fine-grained hallucination detection approach. The latter aims to identify the specific tokens, sentences or lines of code that are likely to be incorrect rather than providing a global correctness assessment. This level of detail is particularly important in the context of code generation, where the ability to precisely identify potential fault locations can offer developers more practical support than binary response-level estimates~\cite{huang_risk_2025}.

\section{Fine-grained Hallucination Detection} \label{sec:rw_fghd}
As previously discussed, IUE shows that internal computations carry strong reliability information during code generation. However, a global correctness assessment at the response level burdens developers with the task of manually identifying errors or regenerating the program when it is deemed incorrect. In software engineering, shifting from response-level to line-level predictions is valuable because identifying the exact location of potential errors, such as specific tokens, sentences, or lines of code, streamlines the debugging and correction process~\cite{huang_risk_2025}.

\subsection{Natural Language Generation (NLG)} \label{subsec:rw_nlg}
Fine-grained hallucination detection is more established in the domain of NLG, where researchers have explored both attention-based and hidden-state-based methods to identify ungrounded or incorrect spans of text.

\noindent \textbf{Attention weights.}
One approach to span-level detection leverages attention weights. Lookback Lens~\cite{chuang-etal-2024-lookback} is a method designed to detect and mitigate contextual hallucinations. It does so by analyzing how much a model ``looks back'' at the provided context versus its own previously generated tokens. The core concept is that attention weights strongly correlate with hallucinations, and a lack of focus on the source context can be strong indicator of non-factual generation. To measure this, the authors calculate a ``lookback ratio'' using internal attention weights. They then train a simple span-based linear classifier on these features to predict the truthfulness of text segments. 

When implemented via a sliding window approach and used for classifier-guided decoding, the Lookback Lens successfully reduced hallucinations by 9.6\% in summarization tasks. Furthermore, the researchers observed the method's impressive ability to transfer across different model sizes without requiring retraining~\cite{chuang-etal-2024-lookback}.

Despite its successes, the Lookback Lens has several limitations. First, it is a supervised method that relies on annotated examples to train the detector. Second, the sliding window strategy can result in mixed-content segments containing both factual and hallucinated information, which complicates the binary classification task. Most importantly, the method exhibits strong feature dependency, requiring the entire set of attention heads across all layers for optimal detection performance~\cite{chuang-etal-2024-lookback}. While feasible for short summarization tasks, extracting and processing full self-attention maps is impractical for longer generation tasks, such as code synthesis, due to the quadratic complexity of self-attention~\cite{vaswani_attention_2023}.

\noindent \textbf{Hidden states.}
Alternatively, researchers have explored fine-grained detection using hidden states. In their work~\cite{ch-wang_androids_2024}, they trained linear probes at the token level to detect specific ungrounded spans during evaluation while simultaneously also providing a coarse response-level assessment. These probes were trained on a high-quality dataset containing both ``organic'' (sampled) and ``synthetic'' (edited) hallucinations across tasks such as summarization and dialogue. 

A key finding of their study is that hidden states are highly predictive of hallucinations, with probes surpassing expert human annotators in span-level F1 scores on the summarization task. However, the researchers also observed a significant drop in probe accuracy when training and testing across different tasks, indicating a generalization gap. Furthermore, probes generally performed better when trained and tested on the same type of hallucinations. Models trained on synthetic data were less accurate when evaluated on organic data, and vice versa~\cite{ch-wang_androids_2024}. This suggests that synthetic hallucinations may not accurately replicate the internal signals from organic hallucinations.

Additionally, their approach suffers from the difficulty of obtaining high-quality labeled data. Annotating hallucinations at the span level is difficult, even for human annotators, and significant disagreement between annotators regarding exact span boundaries can introduce noise into the training data~\cite{ch-wang_androids_2024}.

Perhaps most notably, the study also underscores the issue of signal localization and timing~\cite{ch-wang_androids_2024}. The internal signal indicating a hallucination may not align precisely with the generated token itself. Instead, a relevant signal could be present in preceding tokens or only manifest later in the generation process. Current model architectures may fail to capture these complex, sequential context relationships.

This limitation further motivates the first research question \hyperref[item:rq1]{(RQ1)} of this thesis, which explores whether capturing richer representations, such as sequential hidden-state signals across multiple token positions, or dynamic token selection strategies, can improve performance in estimating code correctness.

\subsection{Code Generation}  \label{subsec:rw_fg_codegen}
The transition of fine-grained hallucination detection from NLG to code generation introduces unique challenges and opportunities. The correctness of code is determined by its functional behavior rather than by factuality, and a single incorrect token can render an entire program non-functional~\cite{huang_risk_2025}. Thus, precise error localization is even more critical in code generation than in NLG.

\noindent \textbf{Line-level hallucination detection.}
Huang et al.~\cite{huang_risk_2025} proposed PtTrust, a framework intended to provide fine-grained, line-level risk assessment for code LLMs. Their approach rests on the assumption that while internal states contain information capable of identifying risks in specific lines of code, it is difficult to train on these high-dimensional representations directly. 

To address this issue, PtTrust first uses a Sparse Autoencoder (SAE) for unsupervised representation learning. The authors train using simple mutation techniques, such as swapping or deleting single lines of code, to deliberately generate faulty snippets. The SAE then compresses these high-dimensional hidden states into sparse, low-dimensional latent variables. Next, they perform the ``semantic binding'' step, using benchmark samples with annotated labels to map these latent representations to functional correctness. This enables the system to learn to localize the riskiest lines of code~\cite{huang_risk_2025}.

A notable aspect of their methodology is the fine-grained label collection process. Recognizing the time-consuming nature of manually annotating diverse errors, the authors combined automated diff tools with LLM feedback to identify incorrect code lines. They then validated these labels by asking the model to provide explicit repair solutions. In cases where the LLM fails to suggest a solution, the authors perform a manual review~\cite{huang_risk_2025}.

Ultimately, their downstream classifier achieved strong performance in their evaluation process, surpassing \cite{chuang-etal-2024-lookback}, and demonstrated cross-language generalization capabilities. It correctly detected errors in Java code despite being trained solely on Python data~\cite{huang_risk_2025}.

\noindent \textbf{Limitations.}
Despite these strong results, the PtTrust framework has notable  limitations that highlight gaps in the existing literature. First, their feature selection strategy differs from well-established findings. They extract hidden states exclusively from the early layers of the model, claiming that these layers control high-level decision-making processes. However, this conflicts with the broader consensus in the IUE literature that middle-to-late layers are the most predictive of hallucinations~\cite{azaria_internal_2023, snyder_early_2024, liang_learning_2024}. Furthermore, their reliance on simple code mutation techniques for pre-training the SAE may fail to capture the patterns of real-world software defects.

Another critical implementation choice is their exclusive focus on the newline token (\texttt{\textbackslash n}) at the end of every code line. They discard the hidden states of all other tokens, which significantly reduces the computational processing costs of model training~\cite{huang_risk_2025}. However, this strict filtering discards the vast majority of the generation context. Since crucial reliability signals can emerge at various points during the generation of a line of code, relying exclusively on the newline token could result in the loss of significant predictive power.

Finally, although Huang et al. demonstrate successful line-level detection, their evaluation primarily focuses on generation tasks such as code repair, translation, and editing. This leaves a significant gap in understanding the effectiveness of fine-grained introspective methods for program synthesis tasks, which are considerably more complex and realistic scenarios for LLM-assisted software development~\cite{ribeiro_llms_2025, huang_risk_2025}.

These limitations motivate the third research question \hyperref[item:rq3]{(RQ3)} of this thesis, which investigates whether leveraging full hidden state information across all token positions from the code generation can improve fine-grained hallucination detection performance. Additionally, shifting the evaluation focus to complex program synthesis tasks can provide a more comprehensive understanding of the practical utility of fine-grained detection methods in real-world software engineering contexts.

\noindent \textbf{Summary.}
In conclusion, while fine-grained hallucination detection has demonstrated considerable promise in isolating errors at the span and line levels, existing methods face significant challenges. Attention-based models are limited by prohibitive memory requirements for long sequences, while hidden-state probes can struggle with cross-task generalization.

In the realm of code generation, state-of-the-art approaches rely on aggressive feature pruning, such as exclusively training on newline tokens, and have yet to be thoroughly validated on complex program synthesis tasks. These challenges set the stage for the methodology of this thesis, which strives to systematically address the identified research gaps.

%% file: chapters/methodology.tex
\chapter{Methodology} \label{cha:methodology}

This chapter outlines the overarching methodological framework of this thesis. First, Section~\ref{sec:tasks} formalizes code generation tasks and introduces the benchmarks used in this thesis. Section~\ref{sec:datasets} outlines the data curation pipelines and describes how LLM responses are generated, how test suites are used to assign response-level correctness labels, and how internal hidden states are extracted to create features for probing. To enable fine-grained hallucination detection, Section~\ref{sec:augmentation} subsequently describes the data augmentation strategy used to derive token- and line-level correctness labels. Finally, Section~\ref{sec:model_design} establishes the model designs, feature selections, training objectives, and evaluation metrics used in this thesis.

\section{Code Generation Tasks} \label{sec:tasks}

In this thesis, we consider code generation to be an interaction with an auto-regressive language model (see Section \ref{sec:transformer}). Given a prompt that describes a problem or a set of functional requirements, the model generates a response that is expected to contain a code segment. For this code segment to be considered correct, it must be syntactically valid and executable, and it must successfully pass a set of predefined test cases. Depending on the task, the prompt may provide a high-level problem description, a partial code snippet, or a buggy implementation. The expected output may contain a complete program, code completion, bug fix, refactoring step, or translation between programming languages~\cite{sharma_assessing_2025}.

While the spectrum of code generation tasks is broad, this thesis specifically focuses on code synthesis. Code synthesis requires the model to generate a complete, executable program based only on an initial problem using natural language. This challenging task serves as a highly relevant testbed for our research questions because it demands deep reasoning and generation of complete program structures without leveraging existing code.

To rigorously evaluate Introspective Uncertainty Estimation (IUE) across diverse synthesis settings, we employ two established code generation benchmarks: LiveCodeBench (LCB)~\cite{jain_livecodebench_2024} and BigCodeBench (BCB)~\cite{zhuo_bigcodebench_2025}. These benchmarks assess various aspects of programming capability, ranging from algorithmic reasoning to practical tool usage. 

\subsection{LiveCodeBench (LCB)} \label{sec:lcb}

LCB focuses on algorithmic code synthesis. The benchmark evaluates a model's ability to implement complex algorithms using standard Python data structures, emphasizing logical reasoning and problem solving~\cite{jain_livecodebench_2024}. 

A key motivation behind LCB is to provide a contamination-free evaluation framework. Here, contamination is defined as the use of data in evaluation that has been seen by the model during training. LCB achieves this by continuously collecting newly released problems from competitive programming platforms, enabling researchers to evaluate models on data generated after their respective training cutoff dates~\cite{jain_livecodebench_2024}. The primary goal of this thesis is not rigorous performance benchmarking, but rather the analysis of internal model representations during generation. Consequently, the methodology prioritizes compiling a large, diverse dataset of problems and responses over strict contamination control. Since this thesis uses models that were published after the benchmark's cutoff date, we recognize the possible impact of data contamination on model performance.

For dataset curation, this thesis relies exclusively on LCB's code synthesis scenario, despite the fact that the benchmark also supports self-repair and code execution tasks. Instruction-tuned models are prompted using the standard zero-shot prompt provided by the benchmark. This prompt includes multiple input and output examples for each problem. A complete example of the LCB prompt format is provided in Appendix~\ref{sec:app_prompt_lcb}. The format requires the model to read from standard input and write to standard output. 

LCB curates high-quality metadata and classifies tasks by difficulty based on platform ratings. The distribution of the 1055 tasks utilized from LCB's problem set is detailed in Table~\ref{tab:lcb_difficulty}. Furthermore, each problem is paired with a robust suite of test cases, averaging 17 tests per problem, which allows for rigorous verification of functional correctness based on dynamic execution. Correctness is measured using the \textit{Pass@1} metric, which we also adopt to derive binary correctness labels.

\begin{table}[htpb]
\centering
\begin{tabular}{lcc}
\toprule
\textbf{Difficulty} & \textbf{Count} & \textbf{Percentage} \\
\midrule
Easy & 322 & 30.5\% \\
Medium & 383 & 36.3\% \\
Hard & 350 & 33.2\% \\
\bottomrule
\end{tabular}
\caption{Distribution of problem difficulties in the LiveCodeBench dataset.}
\label{tab:lcb_difficulty}
\end{table}

\subsection{BigCodeBench (BCB)} \label{sec:bcb}

While LCB focuses on self-contained algorithmic reasoning, BigCodeBench (BCB) applies code synthesis to practical software engineering domains. BCB is a benchmark designed to test a model's ability to utilize diverse function calls from popular Python libraries and follow complex instructions~\cite{zhuo_bigcodebench_2025}. 

\begin{figure}[!ht]
    \centering
    \includegraphics[width=0.95\textwidth]{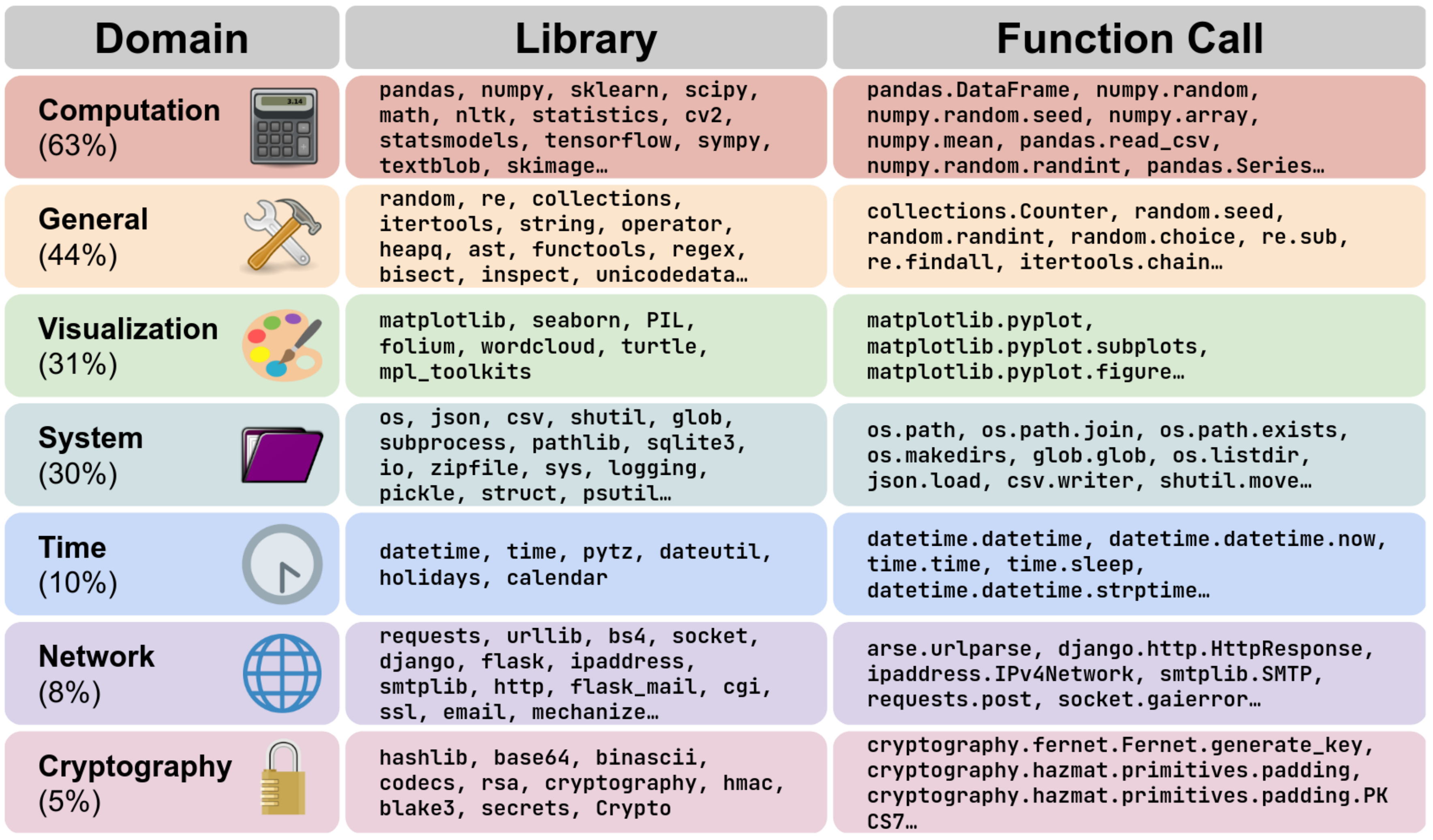}
    \caption{Examples of tool usage in BigCodeBench (BCB). Each function call belongs to a domain-specific library. The distribution of each domain is computed based on the frequency of domain-specific libraries appearing per task. For example, ``63\%'' in ``Computation'' means that there are 63\% tasks in BCB using at least one computation library.~\cite[Figure 4]{zhuo_bigcodebench_2025}}
    \label{fig:bcb_domains}
\end{figure}

In the real world, software development rarely involves writing standalone logic from scratch. Rather, it requires integrating external libraries and APIs. BCB reflects this by including 1,140 tasks that require compositional reasoning and the use of over 100 popular Python libraries~\cite{zhuo_bigcodebench_2025}. These libraries span various software domains, as shown in Figure~\ref{fig:bcb_domains}. It should be noted that tasks frequently require invoking functions from multiple libraries, causing the total frequency percentages to exceed 100\%. These multi-library problems make BCB a highly realistic yet challenging environment for hallucination detection.

The benchmark categorizes task instructions into two formats: the so-called Complete version, which uses structured PEP-257 docstrings that include relevant examples, and the Instruct version, which summarizes the requirements into concise, natural-language directives~\cite{zhuo_bigcodebench_2025}. Importantly, neither prompt variant offers additional context about the necessary libraries for the task. In this thesis, we primarily use the Instruct version (see full prompt example in Appendix~\ref{sec:app_prompt_bcb_instruct}) provided by the benchmark to reflect how a human developer might informally ask an LLM a question.

Notably, the BCB authors observed a degradation in model performance when using the Instruct prompts compared to the Complete versions. In their evaluation, models experienced an average performance drop of 8.5\% when transitioning to these shorter directives~\cite{zhuo_bigcodebench_2025}. The authors attribute this discrepancy to the models' sensitivity to verbose instructions. Without the detailed parameter definitions, exception handling guidance, and interactive input/output examples found in the Complete docstring prompts, LLMs may have difficulty inferring the necessary implementation details.

Correspondingly, during preliminary experiments for this thesis, a similar discrepancy in performance was observed for code correctness estimation. Probing accuracy was significantly higher for LCB than for BCB when using the Instruct format. Hypothesizing that this variance could partially stem from the difference in prompt structures, specifically the presence of multiple input/output examples in LCB, we also designed a Fusion prompt version for BCB (see full prompt example in Appendix~\ref{sec:app_prompt_bcb_fusion}). This fusion strategy combines the high-level intent of the Instruct version with the detailed execution examples of the Complete version using the LCB prompt template. To this end this thesis explores the impact of this prompt engineering choice on the quality of extracted internal representations.

Similar to LCB, BCB provides comprehensive test suites to evaluate functional correctness via the \textit{Pass@1} metric.

\subsection{Alignment with Research Questions} \label{sec:alignment_rqs}

Although both LCB and BCB fall under the broad category of code synthesis, their problem formulations, reasoning contexts, and required knowledge are fundamentally different. LCB requires self-contained algorithmic logic, while BCB demands compositional reasoning and the integration of function calls from multiple libraries. Given these clear differences, this thesis treats evaluating on LCB and BCB as distinct code generation tasks.

This distinction is essential to addressing \hyperref[item:rq2]{RQ2}. First, comparing models in these two environments allows us to systematically measure cross-task (cross-benchmark) generalization capabilities. Second, BCB's rich metadata  gives us the ability to investigate cross-domain generalization within the same benchmark. This enables us to assess how well probe performance transfers when testing on a held-out domain (e.g., visualization) after training on the remaining domains.

Furthermore, the complexity and diversity of these code synthesis benchmarks provide an ideal foundation for investigating fine-grained hallucination detection, thereby tackling \hyperref[item:rq3]{RQ3}. As identified in Chapter~\ref{cha:related_work}, existing fine-grained introspective methods for code generation primarily focus on constrained generation tasks such as code editing or repair. Leveraging datasets that demand complete program synthesis from scratch mirrors the complexity of real-world software development and addresses this critical research gap.

In summary, the code synthesis tasks defined by LCB and BCB offer a comprehensive and challenging testing environment for evaluating the capabilities of IUE. Treating these benchmarks as distinct tasks allows us to systematically analyze cross-task and cross-domain generalization while exploring the potential for fine-grained hallucination detection in complex code synthesis settings. The next section discusses using these benchmarks to curate the datasets necessary for training and evaluating IUE models.

\section{Datasets} \label{sec:datasets}

This thesis trains supervised probes to predict code correctness from internal LLM representations. As discussed in Chapter~\ref{cha:related_work}, supervised introspective methods tend to be more accurate than unsupervised alternatives when suitable labels are available~\cite{snyder_early_2024, shelmanov_acl_2025}. To facilitate this approach, we have curated datasets from LCB and BCB, both of which offer executable test suites that can serve as a correctness oracle. In line with the scope defined in Section~\ref{sec:scope}, we focus exclusively on hidden states as features.

\begin{figure}[htp]
\centering
\includegraphics[width=0.85\textwidth]{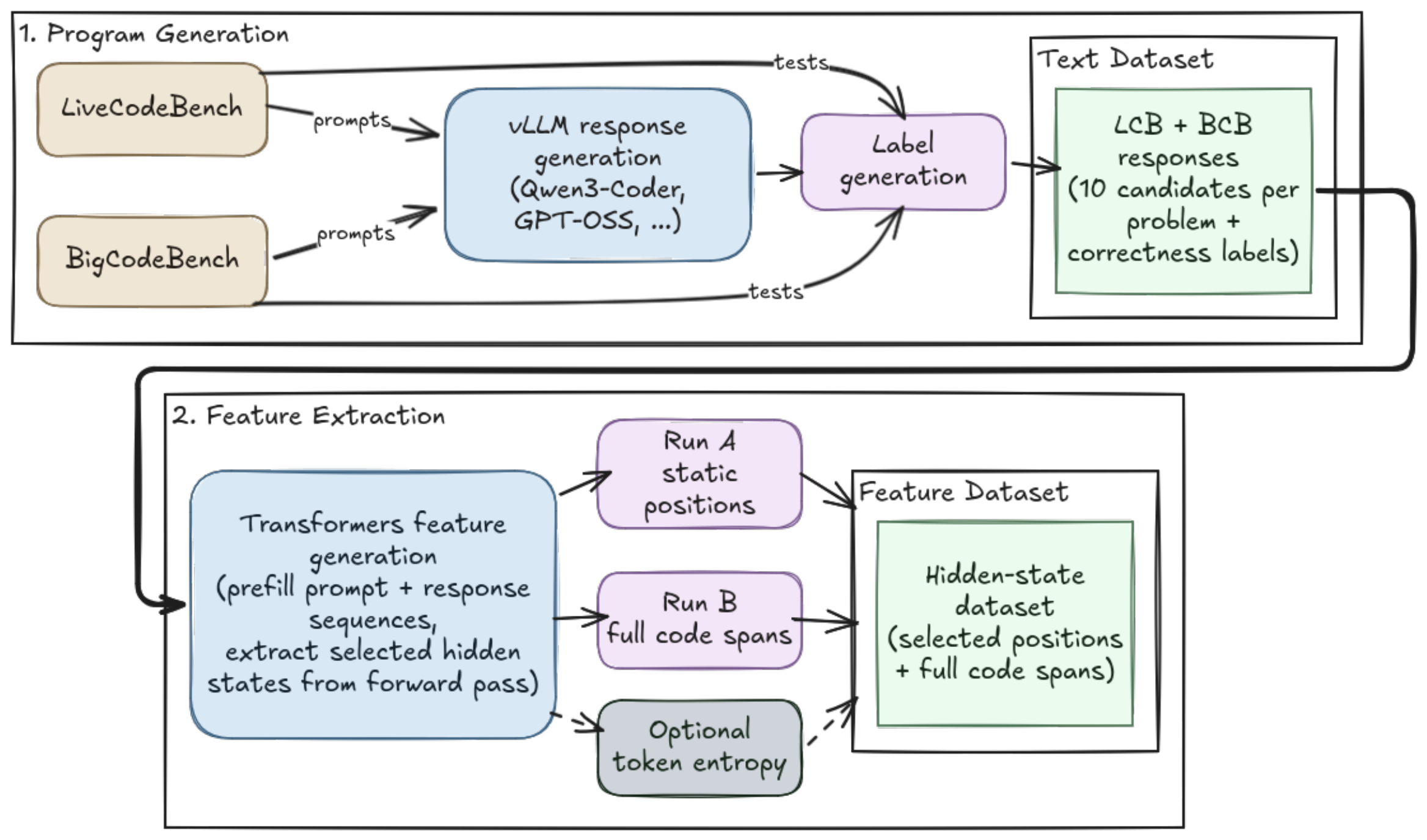}
\caption{Response-level dataset generation pipeline: In Phase 1, benchmark prompts are used to generate candidate programs for each LLM, and correctness labels are assigned by executing benchmark test suites. In Phase 2, each saved prompt-response pair is replayed using prefilling with the same LLM to extract hidden-state features. Feature extraction can be executed in multiple runs to target various feature positions.}
\label{fig:response-level-pipeline}
\end{figure}

Data curation follows a response-level pipeline with two stages: (i) program and label generation from benchmark prompts and test suites, and (ii) extraction of hidden-state features via ``prefilling'' (teacher forcing). Figure~\ref{fig:response-level-pipeline} summarizes this process. We apply this pipeline to LCB and both BCB prompt variants, Instruct and Fusion.

The following subsections provide a detailed description of both phases, beginning with program generation and label construction and concluding with feature extraction.

\subsection{Program Generation} \label{sec:program_generation}
The first phase generates candidate programs for each benchmark task and derives response-level labels from test outcomes. This establishes the text-based foundation of our dataset and determines the learning signal used by our response-level probes.

\noindent \textbf{Model selection.}
Four open-weight, instruction-tuned LLMs are used: Qwen3-Coder~\cite{yang_qwen3_2025}, GPT-OSS-20B~\cite{openai_gpt-oss-120b_2025}, NVIDIA-Nemotron-3-Nano~\cite{nvidia_nvidia_2025}, and Olmo 3.1 Instruct (32B)~\cite{olmo_olmo_2025}. Table~\ref{tab:open_weight_llms} summarizes release metadata and architecture dimensions. The selection is based on three criteria: openness to enable introspection, recency to reflect state-of-the-art model behavior, and feasibility under hardware constraints for generation and feature extraction.

\begin{table}[htpb]
\centering
\footnotesize
\setlength{\tabcolsep}{3.5pt}
\renewcommand{\arraystretch}{0.95}
\begin{tabular}{p{5.5cm}p{2.0cm}p{2.0cm}p{1.0cm}p{1.4cm}}
\toprule
\textbf{Model} & \textbf{Release Date} & \textbf{Knowledge Cutoff} & \textbf{Layers} & \textbf{Hidden Sizes} \\
\midrule
\href{https://huggingface.co/Qwen/Qwen3-Coder-30B-A3B-Instruct}{Qwen3-Coder}~\cite{yang_qwen3_2025} & Jul 31, 2025 & Nov 20, 2024 & 48 & 2048 \\
\href{https://huggingface.co/openai/gpt-oss-20b}{GPT-OSS-20B}~\cite{openai_gpt-oss-120b_2025} & Aug 5, 2025 & Jun 1, 2024 & 24 & 2880 \\
\href{https://huggingface.co/nvidia/NVIDIA-Nemotron-3-Nano-30B-A3B-BF16}{NVIDIA-Nemotron-3-Nano}~\cite{nvidia_nvidia_2025} & Dec 15, 2025 & Nov 25, 2025 & 52 & 2688 \\
\href{https://huggingface.co/allenai/Olmo-3.1-32B-Instruct}{Olmo 3.1 Instruct (32B)}~\cite{olmo_olmo_2025} & Nov 20, 2025 & Dec 1, 2024 & 64 & 5120 \\
\bottomrule
\end{tabular}
\caption{Overview of the open-weight LLMs that were used for code generation to create the datasets.}
\label{tab:open_weight_llms}
\end{table}

\noindent \textbf{Generation strategy.}
The pipeline follows a two-phase workflow. In the first phase, responses are generated using vLLM~\cite{kwon_efficient_2023} and stored as text. In the second phase, hidden states are extracted from saved prompt-response pairs using the Transformers~\cite{wolf_huggingfaces_2020} library. This decoupled approach is widely used in related work~\cite{sriramanan_llm-check_2024, ch-wang_androids_2024, ribeiro_llms_2025, huang_risk_2025} and provides an important practical benefit: generation throughput remains high with vLLM, while feature extraction remains flexible. After generating the response, Transformers~\cite{wolf_huggingfaces_2020} can be used to perform multiple feature extractions targeting different layers or token positions for the same generation, enabling reproducibility.

For the generation process, we sample 10 responses per problem and LLM. We keep the model defaults for decoding controls (temperature, top-$p$, top-$k$) and set \textit{max\_tokens} to 4000. After this limit is reached, the response is truncated. This value was chosen as a compromise. Relative to shorter limits (e.g., 2000), it reduces truncation and increases the probability of obtaining code samples from the generations. Compared to longer limits (e.g., 32000), it keeps prefilling-based hidden-state extraction feasible under memory constraints. Per model, this yields 10,550 generations for LCB and 11,400 generations for each BCB variant.

\noindent \textbf{Failure cases.}
During dataset curation, we distinguish between successful generations and ``\textit{no\_program}'' outcomes. The latter includes truncation before a  code segment is produced, malformed fenced code blocks, and explicit refusal outputs. Some model-specific patterns were observed: GPT-OSS occasionally produced refusal responses, while Nemotron and Olmo more often exhausted the token budget on long reasoning traces, failing to produce a program. 

These patterns affect the label balance of the datasets, which can impact model training. They reveal an inherent conflict in our pipeline: larger generation budgets enhance task completion rates but also increase the memory requirements of full-sequence feature extraction. In our setup, this trade-off is a key constraint at the systems level.

\noindent \textbf{Label generation.}
Labels are assigned by executing the benchmark-provided tests. A generated response is labeled \textit{correct} if it passes all test cases, and \textit{tests\_failed} if it fails at least one test case. Responses from which no fenced code block could be extracted are considered invalid programs and labeled \textit{no\_program}. We filter all \textit{no\_program} data points from the datasets and only use \textit{correct} and \textit{tests\_failed} data for the experiments in this thesis.

This labeling strategy is scalable and reproducible, but it inherits known limitations of test-based supervision. As discussed in Section~\ref{sec:rw_iue_limits}, incomplete tests can miss behavioral defects, and passing all tests does not guarantee broader code quality properties such as readability or maintainability~\cite{spiess_calibration_2024, ribeiro_llms_2025}. Addressing these limitations could be a valuable direction for future work.

\noindent \textbf{Text dataset statistics.}
Table~\ref{tab:text_dataset_stats} summarizes the response-level label counts for all models across the three dataset variants based on the definitions established above.

\begin{table}[htpb]
\centering
\footnotesize
\setlength{\tabcolsep}{4.5pt}
\renewcommand{\arraystretch}{0.95}
\begin{tabular}{lccc}
\toprule
\textbf{Model} & \textbf{LCB} & \textbf{BCB Instruct} & \textbf{BCB Fusion} \\
\midrule
Qwen3-Coder & 5561 / 3963 / 1026 & 5348 / 6052 / 0 & 6743 / 4650 / 7 \\
GPT-OSS-20B & 6488 / 1504 / 2558 & 5148 / 6195 / 57 & 6493 / 4847 / 60 \\
NVIDIA-Nemotron-3-Nano & 4641 / 663 / 5246 & 4949 / 6379 / 72 & 6217 / 5056 / 127 \\
Olmo 3.1 Instruct (32B) & 4551 / 3381 / 2618 & 4122 / 6732 / 546 & 5252 / 5869 / 279 \\
\bottomrule
\end{tabular}
\caption{Response-level label statistics across dataset variants. Each cell reports \textit{correct / tests\_failed / no\_program} counts.}
\label{tab:text_dataset_stats}
\end{table}

Two observations stand out from the statistics. First, LCB yields substantially more \textit{no\_program} responses than both BCB variants. The \textit{no\_program} rates on LCB range from 10\% to nearly 50\% for certain models. In contrast, BCB Instruct remains below 5\% for all models, and BCB Fusion below 3\%. These results demonstrate noticeable label imbalances in the training data, particularly for some LCB-based model.

Second, the BCB Fusion format increases the number of correct programs for all models (Qwen3-Coder: 5348 $\rightarrow$ 6743, GPT-OSS-20B: 5148 $\rightarrow$ 6493, NVIDIA-Nemotron-3-Nano: 4949 $\rightarrow$ 6217, Olmo 3.1 Instruct (32B): 4122 $\rightarrow$ 5252). This is consistent with the intuition that concrete execution examples reduce ambiguity in instruction interpretation. A domain-wise illustration for BCB Instruct and Fusion labels that supports this observation on a domain level is presented Appendix~\ref{sec:app_bcb_domains}.

\subsection{Feature Extraction} \label{sec:feature_extraction}

The second phase of the pipeline involves extracting hidden states from generated prompt-response pairs. This process is referred to as ``prefilling'' in this thesis.

\noindent \textbf{Prefilling.}
Feature extraction is performed the using the Transformers~\cite{wolf_huggingfaces_2020} library. Specifically, we replay each saved prompt-response pair in a single forward pass using the original LLM and decoding controls to record internal activations at response token positions. We use this approach because vLLM~\cite{kwon_efficient_2023} does not expose any hidden-state representations, and prefilling enables consistent and efficient extraction across many samples.

However, this choice introduces an approximation. Extracted representations are obtained from replayed texts rather than being saved during the original generation process. This thesis investigated this effect and found small but significant differences in token probabilities between vLLM and Transformers. This shows that differences in implementation between inference engines can cause discrepancies that affect the model's internal computations. While this approximation is a limitation of this thesis because it may affect the performance of our probes, it is a well-established paradigm in recent introspective research~\cite{sriramanan_llm-check_2024, ch-wang_androids_2024, ribeiro_llms_2025, huang_risk_2025} and represents a necessary trade-off to enable large-scale experimentation.

\noindent \textbf{Hidden state extraction.}
Hidden states are obtained using the Transformers inference flag  ``\textit{output\_hidden\_states=True}''. For each sample, this returns a tuple of length \textit{num\_layers + 1}, where each element is a tensor with shape \textit{(batch\_size, sequence\_length, hidden\_size)}. The first tensor in each tuple corresponds to the static input embeddings before they are processed by the first layer of the transformer model. To avoid padding the prompt-response sequences, we extract features from one sample at a time (batch size 1), albeit at the cost of reduced throughput.

This thesis primarily uses hidden states from the final layer of the models. Previous work consistently reports strong uncertainty signals in middle-to-late layers, including last-layer features for code correctness prediction~\cite{azaria_internal_2023, snyder_early_2024, liang_learning_2024, bui_correctness_2025}. Instead of performing exhaustive layer sweeps, we fix the layer and focus on token position and feature design.

We run multiple feature extraction pipelines that target different token positions of the model response. The first follows the static token-position setup used in prior code-focused work: first token, first code token, last code token, and last token~\cite{bui_correctness_2025}. The second pipeline extracts dense sequences across all code token positions, from the beginning of the code segment to the end of the response. This enables experiments using sequential features for \hyperref[item:rq1]{RQ1} and fine-grained token- and line-level analyses for \hyperref[item:rq3]{RQ3}, utilizing all available feature positions. Optionally, we store token probability distributions and compute per-token entropy~\cite{shelmanov_acl_2025} to serve as an auxiliary uncertainty signal for filtering our main features.

\noindent \textbf{Feature visualization.}
To conduct an initial qualitative analysis, t-SNE~\cite{maaten_visual_2008} was used to visualize select hidden states in the LCB data for the Qwen3-Coder model in Figure~\ref{fig:qwen3_tsne_combined}. The same illustration is provided for all models in Appendix~\ref{sec:app_feature_vis}. Correct and incorrect samples show partial separation, supporting the hypothesis that hidden states encode usable correctness information. The next section builds on this response-level dataset and introduces the augmentation pipeline used to derive token- and line-level labels.

\begin{figure}[H]
    \centering
    \includegraphics[width=0.8\textwidth]{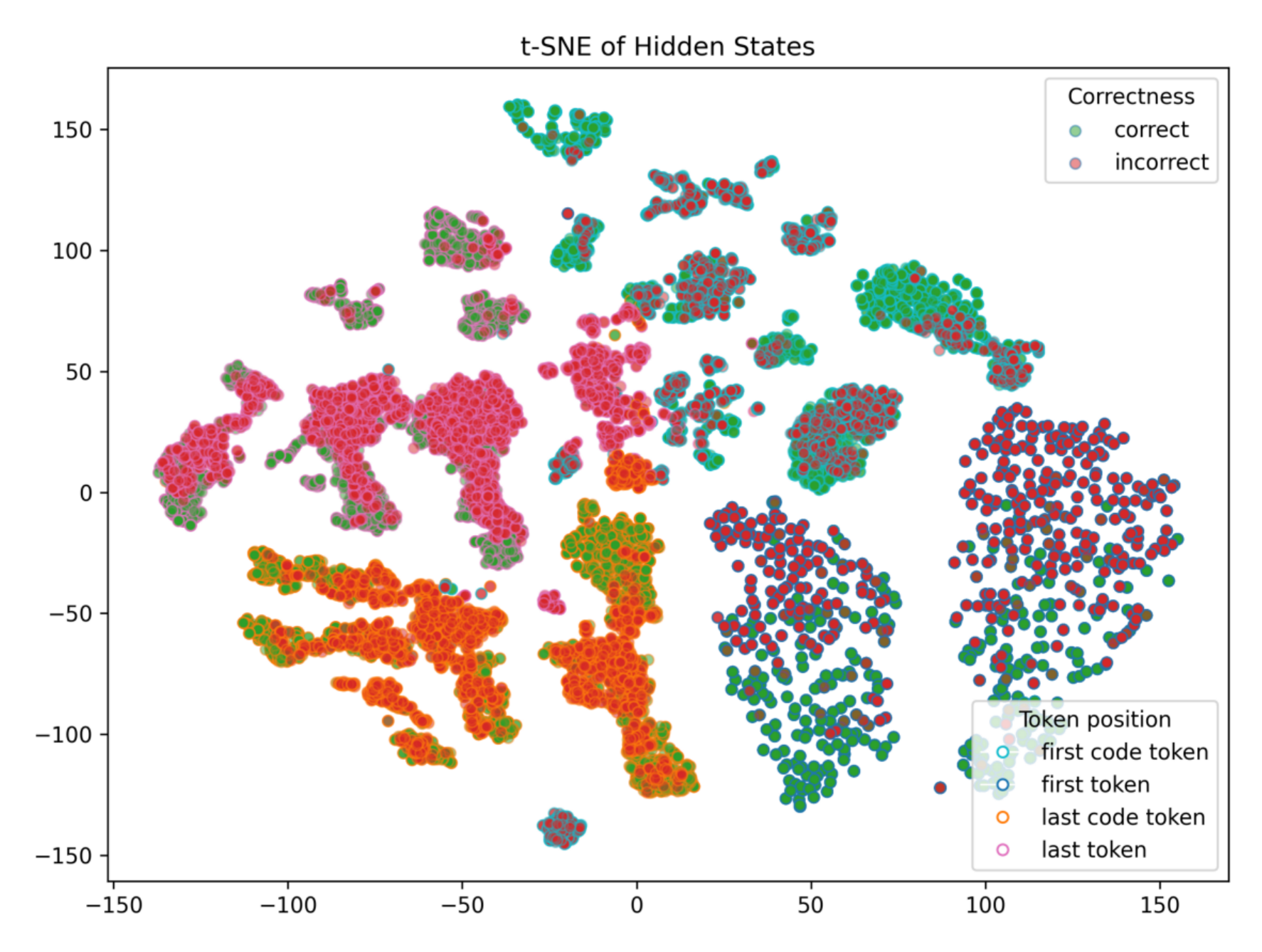}
    \caption{2D t-SNE distributions of select hidden-state features extracted from the last layer of Qwen3-Coder on LCB. The distributions differ between correct v.s. incorrect code generations and show varying degrees of separation across token positions.}
    \label{fig:qwen3_tsne_combined}
\end{figure}

\section{Augmentation} \label{sec:augmentation}

The objective of dataset augmentation is to convert response-level correctness labels into token- and line-level correctness labels. This provides the necessary training data to address \hyperref[item:rq3]{RQ3}. At the response level, a sample is either correct or incorrect. However, we have already established that this level of detail is insufficient for code generation because it does not provide practical developer assistance. In code generation, small local defects can invalidate an otherwise plausible program~\cite{huang_risk_2025}. Consequently, the goal of augmentation is to identify likely fault regions in incorrect samples of our program datasets. To this end, we examine the tokens and lines of these programs.

The design of this augmentation stage is driven by two constraints. First, this thesis focuses on realistic synthesis settings, in which errors should reflect the actual behavior of the generation rather than artificial perturbations. Second, token-level labels are expensive to obtain manually at scale. Therefore, we require an automated, reproducible pipeline aligned with functional correctness, as defined by benchmark test suites.

\noindent \textbf{Alternative approaches.}
Three alternatives were evaluated before the final augmentation setup was chosen:

First, replacing the augmentation with an external, human-annotated bug dataset would likely improve the precision of the fine-grained labels, but it would also introduce a domain and distribution shift relative to our other work. Given the scope and resource constraints of this thesis, this approach was not pursued.

Second, mutation-based augmentation, such as deleting or swapping lines, is simple and scalable, and has been explored in related work~\cite{huang_risk_2025}. However, synthetic code edits differ from ``organic'' model failures, reducing the likelihood that probes learn transferable uncertainty signals.

Third, fully synthetic generation of buggy code from a correct dataset sample was considered. However, previous research in the NLG domain suggests that these synthetic hallucinations may not match the internal feature patterns of genuine hallucinations~\cite{ch-wang_androids_2024}. This risk motivated a conservative strategy focused on organic model failures.

\noindent \textbf{Our approach.}
Based on these considerations, we adopt a ``fix-then-diff'' approach: starting from organically incorrect programs, we employ an auxiliary LLM as an oracle to generate test-passing, minimal-edit fixes for each candidate. We then derive fine-grained labels from the discrepancies between the buggy and fixed programs, and optionally filter the data to focus on a specific localization task. 

\begin{figure}[htp]
\centering
\includegraphics[width=0.9\textwidth]{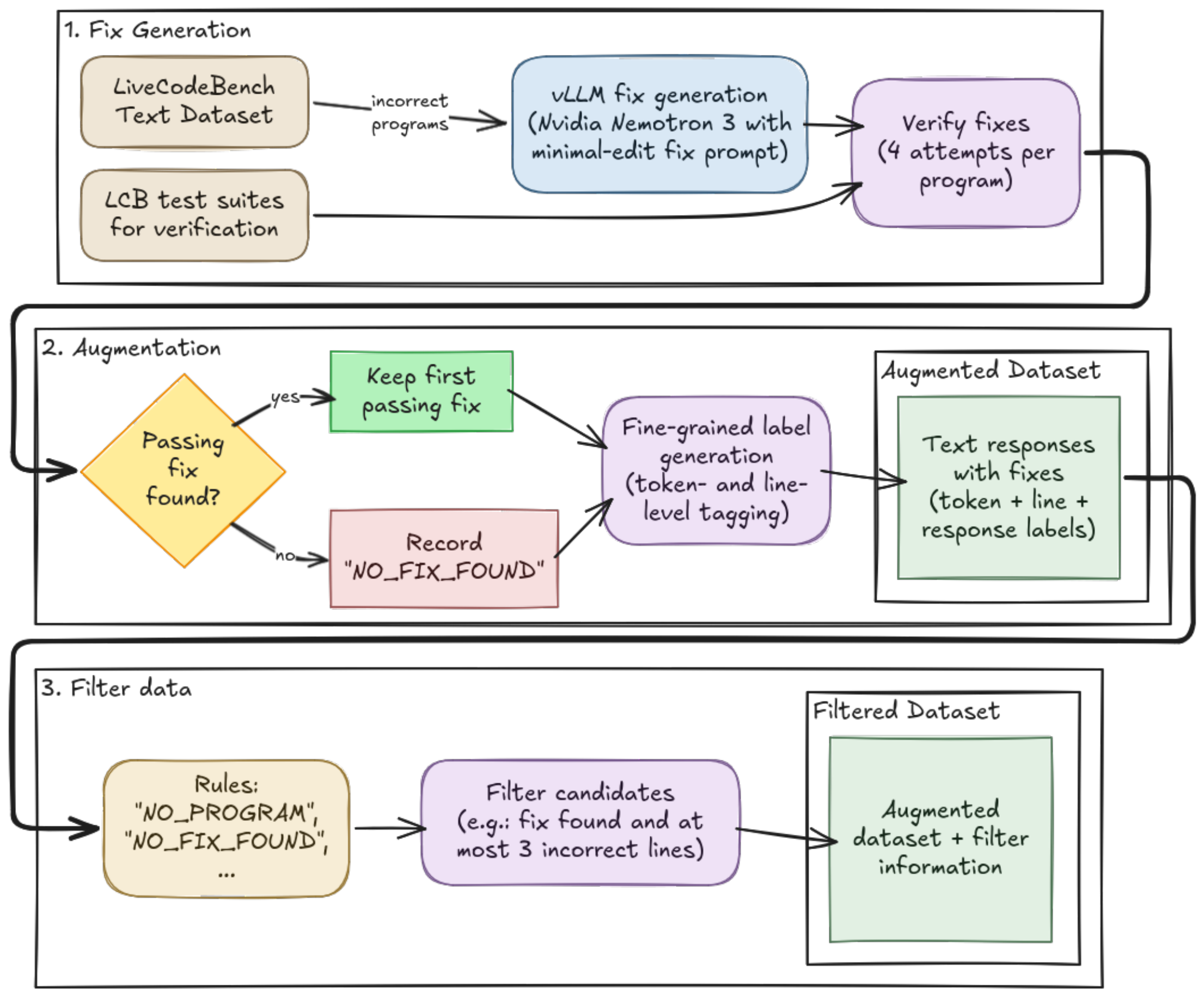}
\caption{Token- and line-level dataset augmentation pipeline: In Phase 1, incorrect programs are repaired by an auxiliary model and validated with benchmark test suites. In Phase 2, buggy and test-passing repaired programs are used to derive token- and line-level correctness labels. In Phase 3, augmented samples can be filtered based on rules stemming from the augmentation data.}
\label{fig:line-level-pipeline}
\end{figure}

It is important to acknowledge that this approach assumes the computed diff accurately isolates the true underlying fault. While the auxiliary model is explicitly instructed to perform minimal edits, it may occasionally rewrite larger portions of the code rather than making a localized fix. This can introduce noise into the fine-grained labels. We aim to mitigate this risk through our data filtering strategies. Figure~\ref{fig:line-level-pipeline} summarizes this pipeline.

\subsection{Fix Generation}
We only augment LiveCodeBench (LCB) because evaluating fixed programs is faster on LCB than on BigCodeBench (BCB). The main difference is that BCB requires the sanitization of code segments and a containerized evaluation protocol, both of which cause significant runtime overhead. Due to compute constraints, augmentation is further limited to two models, Qwen3-Coder and GPT-OSS-20B. We include Qwen3-Coder because it performed best in response-level IUE experiments. This enables us to analyze whether response-level performance translates to fine-grained tasks.

For incorrect samples, an auxiliary repair model receives the original prompt and buggy program, and is instructed to apply minimal edits to generate a fix (see full prompt in Appendix~\ref{sec:app_augmentation_prompt}). In this thesis, we used NVIDIA-Nemotron-3-Nano for fix generation. Up to four repair attempts are generated per sample. After each attempt, the benchmark test suite is executed, and the first test-passing attempt is accepted as the reference fix. If all attempts fail, the sample is flagged with ``\textit{no\_fix\_found}''. Preliminary validation of this setup on 1000 incorrect programs from LCB yielded an 80\% fix rate, motivating further usage for large-scale augmentation.

For already correct programs, no repair is needed. Their response-level label is expanded to token level by marking all code tokens as correct.

\subsection{Fine-grained Label Generation}
Given an original buggy program and a test-passing repaired program, token-level labels are produced by alignment and differentiation:
\begin{enumerate}
    \item Both versions are tokenized with the original tokenizer that generated the buggy program, and each token is mapped to its source line.
    \item Token-level alignment and differentiation is computed with \\ \texttt{difflib.SequenceMatcher}~\cite{difflib_sequence_2026}, operating on token IDs.
    \item Tokens belonging to aligned matching spans are labeled as correct and unmatched tokens are labeled as incorrect.
\end{enumerate}

In practice, naive token matching is noisy. Whitespace, newline, and comment tokens can change without influencing program behavior. Conversely, inserted tokens may alter behavior even when surrounding tokens still match. To reduce this noise, we apply two adjustments.

First, whitespace, newline, and comment tokens are explicitly tracked and excluded from mismatch counting. Second, insertion-aware line handling is implemented. If a sequence of tokens is inserted by a fixed program, the nearest corresponding line after the insertion is flagged as incorrect in the buggy program. This prevents insert-only repairs from being incorrectly labeled as fully correct due to matching context (see example in Appendix~\ref{sec:app_augmentation_insertions}).

Then, token labels are promoted to the line level. A line is marked as incorrect if at least one non-whitespace, newline, or comment token on that line is incorrect. Otherwise, it is marked correct. This line-level projection intentionally sacrifices some token-level precision to obtain a more robust signal for practical fault localization.

\subsection{Data Filters}
The augmentation pipeline stores three filter flags per sample, which can be used to optionally filter the data for a specific localization task:
\begin{enumerate}
    \item \textit{no\_program}: no fenced code segment was extracted from the original response.
    \item \textit{no\_fix\_found}: no test-passing repair was generated within the four attempts.
    \item \textit{line\_filter}: the augmented sample contains more than three incorrect lines of code. This constraint is used to identify samples in which the fix altered significant portions of the code.
\end{enumerate}

Table~\ref{tab:aug_filter_stats} summarizes candidate retention and filtering outcomes for the base augmented datasets of Qwen3-Coder and GPT-OSS-20B.

\begin{table}[htpb]
\centering
\footnotesize
\setlength{\tabcolsep}{2.8pt}
\renewcommand{\arraystretch}{0.95}
\begin{tabular}{lp{2.0cm}p{2.0cm}p{2.2cm}p{2.0cm}}
\toprule
\textbf{Model} & \textbf{Kept (n, \%)} & \multicolumn{3}{c}{\textbf{Dropped by Filter (n, \%)}} \\
\cmidrule(lr){3-5}
& & \textbf{no\_program} & \textbf{no\_fix\_found} & \textbf{line\_filter} \\
\midrule
Qwen3-Coder & 6329 (59.99\%) & 1134 (10.75\%) & 1928 (18.27\%) & 1159 (10.99\%) \\
GPT-OSS-20B & 6756 (64.04\%) & 2672 (25.33\%) & 893 (8.46\%) & 229 (2.17\%) \\
\bottomrule
\end{tabular}
\caption{Filtering statistics for augmented LCB datasets. Kept (candidate was retained) / Dropped (candidate was filtered). Percentages are relative to all candidates ($n=10550$ per model).}
\label{tab:aug_filter_stats}
\end{table}

Two different patterns emerge from the filtered data. For Qwen3-Coder, most of the discarded samples are flagged as \textit{no\_fix\_found}, indicating that a significant portion of its faulty programs are too complex to repair within the minimal-edit constraints. Conversely, GPT-OSS-20B's discarded samples are primarily driven by \textit{no\_program} errors, originating from extraction failures from the initial response-level generation. We highlight this distinction because it differentiates dropped samples caused by an inability to synthesize a fix from dropped samples caused by a failure to produce executable code in the first place.

Based on these filters, we construct three distinct dataset variants per model to support different evaluation objectives. \textbf{V1} enforces all three filters, yielding a dataset strictly focused on the localization of small bugs. \textbf{V2} relaxes this constraint by omitting the \textit{line\_filter}, thereby broadening the dataset to include programs with larger, more complex fault regions. Finally, \textbf{V3} builds upon V2 by additionally removing all samples that were already correct at the response level. This variant is specifically designed to emulate a realistic downstream scenario: performing fine-grained fault localization on generated code that is already known to be defective from response-level evaluation.

The resulting token- and line-level label distributions for all three variants across both models are presented in Table~\ref{tab:aug_summary_combined}. The ``Tokens T/F'' and ``Lines T/F'' columns provide the absolute label counts alongside the percentage of incorrect (false) labels (\%F), deliberately excluding whitespace, newline, and comment tokens. The ``Tok.\ $\mu$'' and ``Line $\mu$'' columns report the average sequence lengths in tokens and lines per program. Ultimately, the ``Diff. E/M/H'' column illustrates the distribution of problems across \textit{Easy}, \textit{Medium}, and \textit{Hard} difficulty levels.

\begin{table}[htpb]
\centering
\footnotesize
\setlength{\tabcolsep}{4.5pt}
\renewcommand{\arraystretch}{0.95}
\begin{tabular}{lccrrp{2.5cm}}
\toprule
\textbf{Var.} & \textbf{Tokens T/F (\%F)} & \textbf{Lines T/F (\%F)} & \textbf{Tok. $\mu$} & \textbf{Line $\mu$} & \textbf{Diff. E/M/H (\%)} \\
\midrule
\multicolumn{6}{c}{\textbf{Qwen3-Coder}} \\
\midrule
V1 & 756482/11431 (1.49\%) & 105056/1290 (1.21\%) & 121.3 & 16.8 & 49.3/39.1/11.6 \\
V2 & 833748/137425 (14.15\%) & 116387/18194 (13.52\%) & 133.2 & 18.5 & 43.3/39.6/17.1 \\
V3 & 182887/137425 (42.90\%) & 25575/18194 (41.57\%) & 185.0 & 25.3 & 11.0/46.9/42.1 \\
\midrule
\multicolumn{6}{c}{\textbf{GPT-OSS-20B}} \\
\midrule
V1 & 971108/3386 (0.35\%) & 143270/393 (0.27\%) & 144.2 & 21.3 & 46.7/38.1/15.2 \\
V2 & 986560/15424 (1.54\%) & 145551/2070 (1.40\%) & 145.2 & 21.4 & 45.9/38.3/15.8 \\
V3 & 68850/15424 (18.30\%) & 10005/2070 (17.14\%) & 204.1 & 29.2 & 12.1/49.9/38.0 \\
\bottomrule
\end{tabular}
\caption{Augmentation summary for Qwen3-Coder and GPT-OSS-20B (LCB). Variant definitions: V1 = small bugs, V2 = complete data, V3 = known bugs. (\%F) denotes the percentage of incorrect (false) labels.}
\label{tab:aug_summary_combined}
\end{table}

Analyzing the three variants reveals a clear trade-off regarding class imbalance. \textbf{V1} offers the cleanest localization signal, but it suffers from extreme class imbalance due to its exceptionally low rate of false labels. \textbf{V2} maintains broader dataset coverage, but introduces noisier and more extensive fault regions. Finally, \textbf{V3} achieves a significantly more balanced distribution by exclusively targeting responses that are already known to be incorrect. However, removing trivially correct generations in V3 also skews the problem difficulty distribution toward medium and hard tasks, which require more tokens to solve.

\noindent \textbf{Evaluation.}
A manual audit of 100 line-level annotated samples was conducted to assess labeling reliability. The resulting labels aligned with the observed bug locations in nearly 90\% of the cases. Most residual errors stem from edge cases involving insertions that slightly modify surrounding token sequences, triggering a ``replace'' operation-code that does not match the original snippet (see example in Appendix~\ref{sec:app_augmentation_insertions}).

\noindent \textbf{Limitations.}
Several limitations remain. First, it is important to acknowledge that our approach is based on the assumption that the computed diff accurately isolates the true underlying fault. In practice, minimal-edit prompting does not strictly enforce minimal repairs. Accepted fixes may unnecessarily rename identifiers, modify imports, or restructure code. Second, even with a whitespace- and insertion-aware implementation, alignment on token IDs is inherently fragile amidst the refactoring of code. Third, manual quality estimation is subjective and only partially indicative of the label quality across the entire dataset. Despite these constraints, the augmentation pipeline provides scalable, fine-grained label generation that is sufficient for evaluating line-level IUE behavior in this thesis.

With the datasets now established, the focus of this methodology shifts from data curation to the architecture and training objectives of our introspection models. The following section describes the design decisions adopted for these probes.

\section{Model Design} \label{sec:model_design}

This section specifies how the architectures introduced in Section~\ref{sec:models} are used as probes to investigate the research questions (see Section~\ref{sec:rqs}) of this thesis. The central design goal is to distinguish the quality of the internal representations we study from the capabilities of the model architectures used for classification. We therefore keep the hidden-state source fixed at the final layer of the LLMs, as explained in Subsection~\ref{sec:feature_extraction}, and vary (i) which token positions are encoded as features and (ii) which lightweight probe maps these features to correctness signals.

Formally, for a generated code segment with $T$ code tokens, we denote the final-layer hidden-state sequence as
\[
H = (h_1, \dots, h_T), \quad h_t \in \mathbb{R}^{d}.
\]
Where $d$ is the hidden size of the model (see Table~\ref{tab:open_weight_llms}). Each probe outputs a probability $\hat{p} \in [0,1]$ for the target correctness label at response, token, or line level, depending on the task setup.

\subsection{Probes} \label{subsec:probes}

Three probe families are introduced that primarily differ in the dimensionality and structure of their input features. The first two types use static feature sizes, and the third type is used for variable feature sizes.

\noindent \textbf{Static single-token probes.}
The first probe family follows prior introspective work on code correctness prediction~\cite{bui_correctness_2025}: one hidden state from a fixed token position in the response is used as input, $x=h_{t_*}$. We evaluate the four static positions introduced in Subsection~\ref{sec:feature_extraction} (first token, first code token, last code token, last token), since they represent different stages of causal decoding during code generation. Two classifier architectures are considered for these probes:
\begin{itemize}
    \item \textbf{MLP:} a compact feed-forward classifier, consistent with common probing practice in hallucination detection~\cite{snyder_early_2024, ch-wang_androids_2024, bui_correctness_2025}.
    \item \textbf{XGBoost:} a tree-ensemble alternative~\cite{chen_xgboost_2016} that offers a useful non-neural baseline with strong training efficiency and inference speed.
\end{itemize}
This probe family builds the baseline of this thesis. Its design allows us to directly evaluate the predictive power of hidden states for code correctness estimation at the response level, while isolating the impact of the model architecture from the impact of the four different feature representations.

\noindent \textbf{Static multi-token probes.}
To test whether features at different token positions carry complementary information, we introduce a second probe design that concatenates hidden states from multiple static token positions,
\[
x = h_{t_1} \mathbin\Vert h_{t_2} \mathbin\Vert \dots \mathbin\Vert h_{t_k} \in \mathbb{R}^{kd}.
\]
This approach allows us to investigate whether combining signals from different phases of the generation process improves the predictive performance for code correctness. Although it increases the overall feature dimension ($kd$) compared to single-token probes, the input size remains fixed for each model. This allows us to continue using MLP and XGBoost probes for classification.

\noindent \textbf{Sequential probes.}
Since static probes cannot model the sequential interactions of hidden states, we introduce sequential probes, which can process dynamic, variable-length features and capture the full context of code generation. As introduced in Section~\ref{sec:models}, LSTMs are well-suited for this task because they can effectively capture sequential feature dependencies. These probes consume hidden states from all code token positions until the end of the response as sequence input,
\[
X = (h_1, \dots, h_T) \in \mathbb{R}^{T \times d},
\]
and encode them using an LSTM-based classifier:
\[
z_t = \mathrm{LSTM}(h_t, z_{t-1}), \qquad
\hat{p} = f_{\theta}(z_T).
\]
Here, $z_T$ denotes the final LSTM state after processing the complete token sequence, and $f_{\theta}$ is a classifier head that maps $z_T$ to a probability in $[0,1]$. Within this thesis, $f_{\theta}$ is instantiated with varying MLP configurations to evaluate how the capacity of the classification head affects performance. This architecture is used to evaluate whether response-level IUE performance can be improved by using sequential dynamics from code token hidden states, and whether line-level fault localization can be learned from line-level hidden-state sequences.

The probe architectures are linked to our research questions as follows:
\begin{itemize}
    \item \hyperref[item:rq1]{\textbf{RQ1:}} We compare all three probe families (single-token, multi-token, and sequential) to quantify how predictive performance evolves as we transition from individual features to full sequences.
    \item \hyperref[item:rq2]{\textbf{RQ2:}} To evaluate generalization capabilities across different synthesis tasks, domains, and token positions, we use static single- and multi-token probes. This approach isolates the impact of distribution shifts by maintaining a consistent classifier architecture while systematically altering the training and testing data.
    \item \hyperref[item:rq3]{\textbf{RQ3:}} We repurpose both the static single-token and sequential probe architectures for token- and line-level prediction tasks, using the augmented correctness labels established in Section~\ref{sec:augmentation}.
\end{itemize}

\subsection{Feature Selection} \label{subsec:feat_select}

The feature selection process exhibits a progression from local to global contexts and is intentionally aligned with the three research questions. For static probes, the feature selection is straightforward: we use the hidden state from one or more fixed token positions. As previously discussed, we evaluate four static positions mainly used in prior code generation work~\cite{bui_correctness_2025}: the first token, first code token, last code token, and last token.

\noindent \textbf{Static features.}
\textit{Single-token} probes, which take one of four static positions as input, are the most prevalent design in prior research~\cite{snyder_early_2024, bui_correctness_2025, ribeiro_llms_2025}. This design enables direct comparison of the predictive power of different token positions while maintaining a fixed probe architecture. \textit{Multi-token} probes, on the other hand, concatenate hidden states from multiple static positions. In this thesis, we consider two configurations: one that concatenates the first and last tokens, and another that concatenates the first and last code tokens. These configurations are motivated by the intuition that the beginning and end of the generation encapsulate distinct phases of the process, such as initial planning and final execution. By encapsulating either the entire response or specifically the code segment, these configurations provide a middle ground between highly local single-token features and full sequential modeling, while still maintaining a fixed-size input.

\noindent \textbf{Dynamic features.}
\textit{Full sequence} features contain the hidden states of all code tokens, thereby preserving token-order information and long-range dependencies. This is essential for two reasons. First, correctness-relevant signals may emerge dynamically at various token positions throughout the decoding process rather than at a single designated point. Second, fine-grained fault localization requires token- and line-based scoring, which cannot be expressed using global static features.

To improve upon this sequential design and overcome the limitation of equally weighting all positions, we introduce selective filtering strategies for dynamic hidden state sequences. Since not all token positions may be equally informative, we evaluate structure-aware filters (e.g., restricted to \textit{newline} tokens, inspired by line-based approaches such as PtTrust~\cite{huang_risk_2025}) and uncertainty-aware filters based on token-level \textit{entropy}~\cite{shelmanov_acl_2025}.

The reliance on entropy as an auxiliary filtering signal is motivated by recent findings identifying uncertainty dynamics in code generation. Akyash et al.~\cite{akyash_decortl_2025} find that while generated code naturally has low entropy, sharp spikes occur at critical syntactic and semantic positions. Structural tokens naturally yield minimal entropy, whereas meaningful variables and control flow operators introduce sampling diversity and higher entropy. Consequently, occurrences of high uncertainty can indicate fragile, error-prone code tokens.

Zhu et al.~\cite{zhu_uncertainty-guided_2025} expand on this dynamic behavior, highlighting that prediction difficulty is concentrated at the beginning of a new line of code because this position dictates the subsequent logical structure of the following line. They isolated challenging generation steps by calculating normalized entropy at these line transitions and used thresholding to intervene with chain-of-thought reasoning when necessary.

Drawing from these insights, we adopt normalized token entropy as an auxiliary signal for filtering sequential hidden-state features, under the premise that focusing on more uncertain tokens guides the probe toward the most structurally consequential and informative positions in the code sequence.

For a given token $t$, normalized entropy is computed as
\[
u_t = -\frac{1}{\log |V|}\sum_{v \in V} p_t(v)\log p_t(v),
\]
where $|V|$ is the vocabulary size. This normalization maps $u_t \in [0,1]$ to support threshold-based filtering. In this thesis, entropy strictly serves as an auxiliary selector for hidden-state features, supposedly identifying uncertain generation states~\cite{akyash_decortl_2025, zhu_uncertainty-guided_2025}.

\subsection{Training Objectives} \label{subsec:train_obj}

To systematically address our research questions, we adjust the granularity of the target labels and the data splitting protocols while preserving a fundamental classification approach.

\noindent \textbf{Response-level prediction.}
The objective for \hyperref[item:rq1]{RQ1} is response-level code correctness prediction. Probes are trained to classify functionally correct programs as positive instances and incorrect programs as negative instances.

To analyze generalization capabilities for \hyperref[item:rq2]{RQ2}, we maintain the response-level objective but introduce systematic data distribution shifts. First, we measure cross-task generalization by training on one benchmark (e.g., LCB) and evaluating on the other (e.g., BCB-Instruct). Second, we evaluate cross-domain generalization in BCB by training on a subset of domains and testing on a held-out domain. Finally, to test cross-token-position robustness, we train single-token probes using randomized token sampling drawn from variable-length suffix windows controlled by a ``random tail'' percentage. This evaluates whether the probes learn position-invariant representations of global correctness or overfit to a fixed generation position.

\noindent \textbf{Fine-grained fault localization.}
For \hyperref[item:rq3]{RQ3}, the training objective shifts to predicting the token- and line-level targets derived from our augmentation pipeline. Crucially, we invert the target labels so that the positive class corresponds to incorrect tokens or lines. Since correct code tokens make up the vast majority of the augmented datasets, resulting in severe class imbalances, this label inversion ensures that the optimization objective aligns with the class that matters: the fault regions of the code.

In this fine-grained configuration, single-token probes are trained to predict labels at the token level. Line-level evaluations are produced from these probes by aggregating the resulting token scores using max-pooling. This relies on the assumption that an entire line is risky if it contains at least one highly suspicious token. Alternatively, filtering can be used to only focus on the last code token for each line. In contrast, sequential probes are trained to predict line-level correctness targets directly from a line's full sequence of hidden states. This approach leverages a more holistic feature context rather than relying on localized signals.

\subsection{Evaluation Metrics} \label{subsec:metrics}

To compare probes across response-level and fine-grained tasks, we report complementary metrics that evaluate ranking quality, thresholded classifications, calibration, and practical debugging utility.

\noindent \textbf{Area Under the Receiver Operating Characteristic (AUROC).}
The primary metric is AUROC, which summarizes a model's ability to discriminate between positive and negative classes across all possible decision thresholds. AUROC is a standard metric in related introspective work~\cite{snyder_early_2024}, and is particularly suitable for our evaluation because it is robust with regard to class imbalance. AUROC is defined as the probability that a randomly chosen positive instance is ranked higher than a randomly chosen negative instance. Its range is $[0,1]$, where $0.5$ corresponds to a random ranking, and higher values indicate better discrimination.

\noindent \textbf{F1 score.}
To complement the threshold-independent AUROC, we report the F1 score, the harmonic mean of Precision and Recall~\cite{preis_hallucination_2025}. The F1 score is widely reported in hallucination detection literature~\cite{preis_hallucination_2025, ch-wang_androids_2024, bui_correctness_2025}, as it effectively captures the precision-recall trade-off required for practical model use. It is defined as:
\[
\mathrm{F1} = \frac{2\,\cdot\mathrm{Precision}\cdot\mathrm{Recall}}
{\mathrm{Precision}+\mathrm{Recall}}.
\]
The range of F1 scores is $[0,1]$, where a higher score is better. Unless stated otherwise, F1 is computed at a fixed decision threshold of $0.5$, without threshold optimization.

\noindent \textbf{Brier Skill Score (BSS).}
To evaluate the effectiveness of the introspective models in producing meaningful uncertainty estimates, we also report BSS as a calibration-oriented metric~\cite{campos_multicalibration_2025}. Let $\hat{p}_i \in [0,1]$ denote the predicted probability that sample $i$ is correct and $y_i \in \{0,1\}$ its observed correctness label. The Brier score is
\[
\mathcal{B} = \frac{1}{n}\sum_{i=1}^{n}(\hat{p}_i - y_i)^2.
\]
Considering a so-called unskilled predictor that always outputs the empirical base rate $p_r = \frac{1}{n}\sum_{i=1}^{n} y_i$, the corresponding reference Brier score is
\[
\mathcal{B}_{\mathrm{ref}} = p_r \cdot (1-p_r).
\]
The BSS then measures relative improvement over this baseline:
\[
\mathrm{BSS} = \frac{\mathcal{B}_{\mathrm{ref}} - \mathcal{B}}{\mathcal{B}_{\mathrm{ref}}}.
\]
Positive BSS values indicate improvement over the unskilled baseline, while negative values indicate deterioration. As in~\cite{campos_multicalibration_2025}, when $p_r \in \{0,1\}$ (so $\mathcal{B}_{\mathrm{ref}}=0$), the convention $\mathrm{BSS}=1$ if $\mathcal{B}=0$ and $\mathrm{BSS}=-\infty$ otherwise is used.

\noindent \textbf{Top-$K$ hit rate.}
For fine-grained line-level fault localization, we follow practices from related research~\cite{huang_risk_2025} and report the Top-$K$ hit rate. For each sample, lines are ranked by their predicted risk. A hit is counted if at least one truly incorrect line appears in the top $K$:
\[
\mathrm{Hit@}K = \frac{1}{N}\sum_{i=1}^{N}
\mathbf{1}\!\left[\exists\,\ell \in \mathrm{TopK}(i): \tilde{y}_{i,\ell}=1\right].
\]
\newpage
Here, $N$ denotes the total number of evaluated programs, and the indicator function $\mathbf{1}[\cdot]$ evaluates to $1$ if the condition within the brackets is met. For each program $i$, the model's $K$ highest-risk line predictions form the set $\mathrm{TopK}(i)$. A hit is recorded if there exists at least one predicted line $\ell$ within this set that is incorrect, meaning its ground-truth label is $\tilde{y}_{i,\ell}=1$. By averaging these hits across all $N$ programs, this metric directly reflects practical debugging utility, measuring how often a developer would be effectively guided to a faulty region after inspecting only a few high-risk lines. In our evaluations, we report $K \in \{1, 2, 3\}$.

In summary, the model design incorporates increasingly complex hidden-state representations, various probe families, task-specific learning signals, and uses different evaluation metrics across research questions. Now that the methodological foundation has been established, the focus shifts to the experiments and results, which are presented in the next chapter.

%% file: chapters/experiments.tex
\chapter{Experiments} \label{cha:experiments}

In this chapter, we present the results of our experiments with introspective uncertainty estimation (IUE) models for LLM-based code generation. We evaluate response- and line-level probes across multiple LLMs and dataset variants. First, we describe our experimental setup. Next, we present the results of our response-level experiments, followed by the results of our line-level experiments.

\section{Experimental Setup} \label{sec:setup}
This section summarizes the implementation and evaluation protocols used for all IUE models that were trained in this thesis. First, we describe the software stack. Then, we describe the split strategy and training procedure. Finally, we describe the evaluation protocol. Detailed hyperparameter grids and final model configurations are provided in Appendix~\ref{sec:app_exp_hyperparams}.

\noindent \textbf{Model libraries.}
All neural probes were implemented using PyTorch~\cite{paszke_pytorch_2019}. This includes both response-level and line-level variants of the MLP and LSTM models. XGBoost models were implemented using the XGBoost~\cite{xgboost_introduction_2026} library. Using these standard libraries facilitates the reproducibility of this work.

\noindent \textbf{Data splitting and stratification.}
For each dataset variant and LLM, we performed a conservative $60 / 20 / 20$ split into train, validation, and test partitions. The splitting was performed at the problem level, meaning that all of the sampled candidate programs from the same benchmark problem were assigned to the same partition. This prevents leakage of similar candidates across splits.

Given dataset sizes of 10{,}550 samples per model for LiveCodeBench (LCB) and 11{,}400 samples per model for each BigCodeBench (BCB) prompt variant, this strategy still leaves sufficient training data while maintaining substantial validation and test sets. Additionally, we applied label-aware stratification to keep the distributions of \textit{correct} and \textit{tests\_failed} samples consistent across splits. This is important for datasets with severe class imbalances, as it ensures that models see a representative distribution of both classes during training and evaluation.

\noindent \textbf{Training setup and hyperparameters.}
To keep the main text focused, we report only the high-level procedure here and defer exact details on model hyperparameters to Appendix~\ref{sec:app_exp_hyperparams}. Across all models, we used early stopping on validation AUROC, with patience values selected from preliminary runs. For optimization of neural probes, we used Adam~\cite{kingma_adam_2017} with a binary cross-entropy loss (\textit{BCEWithLogitsLoss}). For XGBoost, we used the \textit{binary:logistic} objective with the \textit{log-loss} evaluation metric. We performed hyperparameter tuning via grid search for XGBoost and via Bayesian search for neural probes. The hyperparameter ranges were informed by prior work and preliminary experiments.

For XGBoost, we began with default settings and trained up to 500 boosting rounds with early stopping. The learning rate was the only parameter tuned explicitly for the final protocol, using multiplicative factors around a base value of $0.05$. For line-level experiments, we deviated from this base configuration by training for additional rounds and utilizing the \textit{scale\_pos\_weight} parameter to address severe label imbalances. 

For MLP probes, we implemented two architectures that align with prior introspective work~\cite{snyder_early_2024, bui_correctness_2025}. One variant consisted of a single hidden layer followed by a ReLU activation function. The other consisted of two hidden layers, each followed by a ReLU activation function. We performed a hyperparameter sweep with 10 training runs over learning rate, weight decay, batch size, and hidden sizes of all available layers. This design preserves comparability to prior studies while ensuring that our probes are well-tuned for our specific tasks.

For sequential probes, we used an LSTM encoder with an MLP classification head over the final hidden state. We explored variations in learning rate, weight decay, LSTM hidden size and depth, head model width and depth, dropout between the encoder and head model, and batch size, again executing a sweep with 10 training runs. Following common practice (see Subsection~\ref{subsec:lstm}), we applied gradient clipping with a maximum norm of $1.0$ to improve optimization stability. Additionally, for line-level experiments, we adapted the sequential probe to incorporate a positive class weight based on the training data ratio, using it as a parameter for the binary cross-entropy loss to address the severe class imbalances.

\noindent \textbf{Evaluation protocol.}
During training, XGBoost was evaluated on the validation split at every boosting iteration, while MLP and LSTM models were evaluated every 10 training iterations. For model selection, we used validation AUROC as the primary early-stopping signal and to select the best hyperparameter configurations. The final evaluation protocols report performance on the held-out test split, using AUROC as the primary metric and F1 and BSS as complementary metrics (see Subsection~\ref{subsec:metrics}). Higher values indicate better performance for each of these metrics.

\section{Response-Level} \label{sec:response_level}

The response-level experiments evaluate whether hidden-state features extracted from generated programs can predict functional correctness at the granularity of full responses. This stage serves two purposes. First, strong baselines are established before introducing more complex model designs. Second, it establishes a reference point for later analyses of IUE on code.

The section is organized as follows. Subsection~\ref{subsec:res_baselines} introduces static-feature baselines. Subsection~\ref{subsec:seq_models} then studies dynamic features using sequential models. Subsection~\ref{subsec:calibration_eval} evaluates confidence calibration, and Subsection~\ref{subsec:generalization} analyzes response-level generalization capabilities.

\subsection{Baselines} \label{subsec:res_baselines}

This subsection establishes static-feature baselines that are directly connected to the methodological choices in Chapter~\ref{cha:methodology}. We evaluate four open-weight LLMs (Qwen3-Coder, GPT-OSS-20B, NVIDIA-Nemotron-3-Nano, and Olmo 3.1 Instruct (32B)) across three dataset variants (LCB, BCB-Instruct, and BCB-Fusion).

\noindent \textbf{MLP probes vs. XGBoost on single-token features.}
We begin our evaluation by examining the static, single-token features established in prior introspective studies: first token, first code token, last code token, and last token~\cite{bui_correctness_2025}. To determine whether the choice of probe architecture significantly impacts the predictive performance of these features, we compare our default XGBoost implementation with the two MLP probe variants (featuring one and two hidden layers), with the results on the LCB dataset for Qwen3-Coder presented in Table~\ref{tab:res_baseline_qwen_lcb_mlp_vs_xgb}.

\begin{table}[htpb]
\centering
\footnotesize
\setlength{\tabcolsep}{5pt}
\begin{tabular}{lcccc}
\toprule
\textbf{Probe Model} & \textbf{Last Code} & \textbf{Last Token} & \textbf{First Code} & \textbf{First Token} \\
\midrule
XGBoost & 0.9026 & \textbf{0.9002} & 0.8804 & 0.8626 \\
MLP (1 hidden layer) & 0.8967 & 0.8945 & 0.8775 & 0.8619 \\
MLP (2 hidden layers) & \textbf{0.9057} & 0.8962 & \textbf{0.8827} & \textbf{0.8638} \\
\bottomrule
\end{tabular}
\caption{Preliminary model comparison using single-token features. Results are AUROC($\uparrow$) on the LCB dataset for Qwen3-Coder. The best result for each token position is in bold.}
\label{tab:res_baseline_qwen_lcb_mlp_vs_xgb}
\end{table}

The first comparison suggests near-parity between probe families, as absolute differences are small across all four token positions. To determine whether this pattern extends beyond a single model-dataset pair, Table~\ref{tab:res_baseline_lastcode_all_models} compares two-layer MLP probes against XGBoost using the \textit{last code token} feature for all four LLMs on both the LCB and BCB-Instruct datasets.

\begin{table}[htpb]
\centering
\footnotesize
\setlength{\tabcolsep}{6pt}
\begin{tabular}{lcc}
\toprule
\textbf{LLM} & \textbf{LCB (XGBoost / MLP-2)} & \textbf{BCB-Instruct (XGBoost / MLP-2)} \\
\midrule
Qwen3-Coder & 0.9026 / \textbf{0.9057} & 0.6841 / \textbf{0.6947} \\
GPT-OSS-20B & \textbf{0.8610} / 0.8519 & 0.6374 / \textbf{0.6571} \\
NVIDIA-Nemotron-3-Nano & \textbf{0.8911} / 0.8841 & \textbf{0.6419} / 0.6381 \\
Olmo 3.1 Instruct (32B) & 0.8542 / \textbf{0.8556} & 0.7309 / \textbf{0.7407} \\
\bottomrule
\end{tabular}
\caption{Two-layer MLP probe vs. XGBoost on the \textit{last code token} feature of LCB and BCB-Instruct. Results are AUROC($\uparrow$) with the better result for each cell in bold.}
\label{tab:res_baseline_lastcode_all_models}
\end{table}

The primary takeaway from both tables is that there is no significant difference in performance between MLP probes and XGBoost when evaluating static, single-token features. The consistency observed across all tested LLMs and datasets indicates that both model families can effectively leverage these features for correctness estimation. Importantly, this implies that the choice between neural probes and ensemble decision trees is not strictly bound by predictive power, allowing researchers to prioritize practical factors such as computational efficiency and ease of implementation.

For our subsequent experiments, we selected XGBoost as the default baseline. Its high training efficiency allowed us to explore a wide array of training tasks and configurations while maintaining comparable performance without significant computational overhead.

These findings are reinforced by similar observations in the literature. For example, Snyder et al.~\cite{snyder_early_2024} reported that scaling MLP probes up to 8 layers with widths of 256 provided virtually no benefit over a single-layer model, observing only a marginal AUROC increase from $0.71$ to $0.72$. This reaffirms the premise that increasing probe complexity does not necessarily lead to better performance in IUE. Ultimately, because we do not expect significant improvements from more complex MLPs within the single-token paradigm, utilizing a highly efficient model such as XGBoost becomes justified.

\noindent \textbf{Static single-token baselines.}
Table~\ref{tab:res_single_token_all} summarizes the complete single-token baseline results for all model-dataset pairs. Each cell reports AUROC/F1/BSS, and bold entries mark the best value for each metric (AUROC, F1, BSS) within every model-dataset cell.

\begin{table}[htpb]
\centering
\footnotesize
\setlength{\tabcolsep}{4pt}
\renewcommand{\arraystretch}{1.05}
\resizebox{\textwidth}{!}{%
\begin{tabular}{clcccc}
\toprule
\textbf{Dataset} & \textbf{Feature} & \textbf{Qwen3-Coder} & \textbf{GPT-OSS-20B} & \textbf{NVIDIA-Nemotron-3-Nano} & \textbf{Olmo 3.1 Instruct (32B)} \\
\midrule
\multirow{4}{*}{LCB} & LC & \textbf{0.9026}/\textbf{0.8484}/0.4576 & \textbf{0.8610}/0.9238/0.3369 & \textbf{0.8911}/\textbf{0.9581}/\textbf{0.4415} & \textbf{0.8542}/0.8047/0.3287 \\
& L & 0.9002/0.8426/\textbf{0.4814} & 0.8570/0.9213/0.3466 & 0.8621/0.9536/0.3766 & 0.8454/0.8015/\textbf{0.3522} \\
& FC & 0.8804/0.8282/0.4318 & 0.8598/\textbf{0.9287}/\textbf{0.3957} & 0.8292/0.9398/0.2869 & 0.8429/\textbf{0.8091}/0.3478 \\
& F & 0.8626/0.8344/0.3445 & 0.7800/0.8912/0.1139 & 0.7981/0.9235/0.1192 & 0.8172/0.7918/0.2762 \\
\midrule
\multirow{4}{*}{BCB-Instruct} & LC & \textbf{0.6841}/\textbf{0.5753}/\textbf{0.0957} & 0.6374/0.5541/-0.0280 & 0.6381/0.5168/\textbf{0.0423} & \textbf{0.7309}/\textbf{0.5026}/0.0802 \\
& L & 0.6745/0.5293/0.0720 & \textbf{0.6532}/\textbf{0.5601}/\textbf{0.0596} & \textbf{0.6411}/\textbf{0.5294}/0.0326 & 0.7251/0.4804/\textbf{0.1100} \\
& FC & 0.5925/0.4491/0.0229 & 0.6046/0.4131/0.0333 & 0.6262/0.5169/0.0142 & 0.6524/0.3017/0.0689 \\
& F & 0.6463/0.5703/0.0425 & 0.6376/0.5313/0.0442 & 0.6083/0.5047/-0.0344 & 0.6294/0.2881/0.0387 \\
\midrule
\multirow{4}{*}{BCB-Fusion} & LC & 0.6447/0.7092/0.0258 & 0.7089/0.7419/0.1272 & 0.6736/0.6749/0.0829 & 0.7094/0.6026/\textbf{0.1307} \\
& L & \textbf{0.6449}/\textbf{0.7109}/\textbf{0.0415} & \textbf{0.7245}/\textbf{0.7585}/\textbf{0.1377} & \textbf{0.6813}/0.6797/\textbf{0.0987} & \textbf{0.7137}/\textbf{0.6137}/0.1221 \\
& FC & 0.6175/0.6991/-0.0650 & 0.6483/0.7228/0.0645 & 0.6278/0.6442/0.0184 & 0.6585/0.5352/0.0750 \\
& F & 0.6142/0.7053/-0.0806 & 0.6799/0.7356/0.0964 & 0.6739/\textbf{0.6874}/0.0900 & 0.6552/0.5380/-0.0248 \\
\bottomrule
\end{tabular}%
}
\caption{Single-token XGBoost baselines across datasets and LLMs. Feature descriptions: L = Last token, LC = Last Code token, F = First token, FC = First Code token. Cells report AUROC($\uparrow$)/F1($\uparrow$)/BSS($\uparrow$). Bold marks the best value per metric (AUROC, F1, BSS) within each model-dataset pair.}
\label{tab:res_single_token_all}
\end{table}

Four trends emerge from these evaluations. First, performance consistently improves as feature extraction moves toward the end of the generated sequence. Bui et al.~\cite{bui_correctness_2025} also observed this effect, and our results strengthen their claim. This finding aligns with the intuition that late-generated hidden states inherently aggregate more context information crucial for robust correctness estimation.

Second, performance is generally strong on LCB across all LLMs, with AUROC values consistently exceeding $0.85$ for the best single-token features. This suggests that the extracted features are highly informative in the context of LCB, enabling XGBoost to effectively distinguish between correct and incorrect candidate programs. These results notably exceed prior benchmark thresholds from static, single-token features~\cite{bui_correctness_2025, snyder_early_2024}.

For example, our baseline achieves an AUROC of $0.90$ for Qwen3-Coder on LCB, which is a significant improvement over the maximum AUROC of $0.81$ reported by Snyder et al.~\cite{snyder_early_2024}. Similarly, our highest F1 score of $0.95$ (for NVIDIA-Nemotron-3-Nano on LCB) considerably outperforms the maximum of $0.83$ reported by Bui et al.~\cite{bui_correctness_2025}.

We acknowledge that these direct comparisons are complicated by variations in task, dataset, and model. Nevertheless, we would like to point out that our best results significantly improve upon previously reported performance levels of static, single-token probes. This improvement may be entirely due to the use of more recent and powerful LLMs, which can produce richer, more discriminative representations that enhance code correctness estimation.

Next, BCB-Instruct is substantially more challenging. Across most models and features, AUROC values drop significantly, often hovering around or below $0.65$, with Olmo as an exception at $0.73$. This decline aligns with the more challenging software engineering context and the shorter instruction format described in Section~\ref{sec:bcb}, where responses rely heavily on library usage and implicit assumptions.

Finally, BCB-Fusion generally outperforms BCB-Instruct for GPT-OSS-20B and NVIDIA-Nemotron-3-Nano, though not consistently across all models. This suggests that prompt engineering can change the separability of hidden-state features, yet the effect remains model-dependent.

\noindent \textbf{Static multi-token baselines.}
As a natural extension of the single-token experiments, we evaluate static, multi-token features, as described in Subsection~\ref{subsec:feat_select}. This experiment directly tests whether concatenating multiple token representations provides complementary information that improves correctness estimation beyond what single-token features capture. Specifically, we evaluate two multi-token features: the first and last tokens (F+L) and the first and last code tokens (FC+LC). As in our previous experiments, we train XGBoost probes and report AUROC/F1/BSS scores for each model-dataset pair, with results summarized in Table~\ref{tab:res_multi_token_all}.

\begin{table}[htpb]
\centering
\footnotesize
\setlength{\tabcolsep}{4pt}
\renewcommand{\arraystretch}{1.05}
\resizebox{\textwidth}{!}{%
\begin{tabular}{clcccc}
\toprule
\textbf{Dataset} & \textbf{Feature} & \textbf{Qwen3-Coder} & \textbf{GPT-OSS-20B} & \textbf{NVIDIA-Nemotron-3-Nano} & \textbf{Olmo 3.1 Instruct (32B)} \\
\midrule
\multirow{2}{*}{LCB} & F+L & \textbf{0.9034}/\textbf{0.8533}/\textbf{0.4860} & 0.8348/0.9115/0.2955 & \textbf{0.8993}/0.9522/0.3928 & 0.8559/0.8123/\textbf{0.3275} \\
& FC+LC & 0.9012/0.8435/0.4507 & \textbf{0.8672}/\textbf{0.9309}/\textbf{0.3972} & 0.8856/\textbf{0.9562}/\textbf{0.4226} & \textbf{0.8586}/\textbf{0.8168}/0.3200 \\
\midrule
\multirow{2}{*}{BCB-Instruct} & F+L & \textbf{0.6839}/\textbf{0.5808}/\textbf{0.0948} & \textbf{0.6569}/0.5638/\textbf{0.0756} & 0.6335/0.5313/-0.0132 & 0.6893/0.3971/0.0714 \\
& FC+LC & 0.6768/0.5801/0.0787 & 0.6549/\textbf{0.5699}/-0.0422 & \textbf{0.6406}/\textbf{0.5325}/\textbf{0.0345} & \textbf{0.7264}/\textbf{0.4868}/\textbf{0.0730} \\
\midrule
\multirow{2}{*}{BCB-Fusion} & F+L & 0.6209/0.7184/0.0356 & \textbf{0.7076}/\textbf{0.7601}/\textbf{0.1247} & \textbf{0.6790}/\textbf{0.6792}/\textbf{0.0345} & \textbf{0.7284}/\textbf{0.6124}/\textbf{0.0810} \\
& FC+LC & \textbf{0.6680}/\textbf{0.7332}/\textbf{0.0848} & 0.7011/0.7440/0.1128 & 0.6512/0.6547/-0.0033 & 0.7137/0.5993/0.0602 \\
\bottomrule
\end{tabular}%
}
\caption{Multi-token XGBoost baselines across datasets and LLMs. Feature descriptions: F+L = First token + Last token, FC+LC = First Code token + Last Code token. Cells report AUROC($\uparrow$)/F1($\uparrow$)/BSS($\uparrow$). Bold marks the best value per metric (AUROC, F1, BSS) within each model-dataset pair.}
\label{tab:res_multi_token_all}
\end{table}

The central finding is that multi-token features do not consistently outperform single-token baselines. While some model-dataset pairs show slight performance improvements (e.g., Qwen3-Coder on BCB-Fusion), many other configurations' metrics either remain flat or degrade relative to their best single-token counterpart (e.g., GPT-OSS-20B on BCB-Fusion).

This behavior suggests that concatenating static token representations fails to extract complementary correctness information beyond what single-token features already capture. Because static concatenation demonstrates limited utility in improving prediction performance, it directly motivates the following subsection, which explores dynamic, sequential feature representations.

\subsection{Sequential Models} \label{subsec:seq_models}
We propose sequential models based on the hypothesis that correctness signals may emerge at various points throughout the code generation process and not only at fixed token positions. To test this hypothesis, we employ an LSTM encoder over token-wise hidden states, followed by an MLP classification head on the final recurrent state. Unlike static probes, this setup can, in principle, model dependencies across the generated sequence and capture how uncertainty evolves during decoding.

\noindent \textbf{Full and newline-filtered sequences.}
In addition to using the entire sequence from the first code token to the end of the response, we evaluate two sequence-filtering strategies. The first strategy is a ``newline'' filter that retains only the hidden states aligned with the newline tokens (\texttt{\textbackslash n}), which are located at the end of each line of code. This filter also appends the final token hidden state to preserve the context of the end of the response. This design is motivated by prior findings that uncertainty at line boundaries can be especially informative for code risk estimation~\cite{zhu_uncertainty-guided_2025}. Therefore, our approach directly adapts the newline filter method introduced in \cite{huang_risk_2025}, although in this context, it is used for response-level predictions.

We compare the results of these two dynamic, sequential approaches with the best static, single-token XGBoost baseline in Table~\ref{tab:res_single_token_all} for each respective dataset-model pair. We report AUROC values for each combination and absolute changes relative to the XGBoost baseline on LCB and BCB-Instruct.

\begin{table}[htpb]
\centering
\footnotesize
\setlength{\tabcolsep}{4pt}
\renewcommand{\arraystretch}{1.05}
\resizebox{\textwidth}{!}{%
\begin{tabular}{llccc}
\toprule
\textbf{Dataset} & \textbf{LLM} & \textbf{XGBoost (best single-token)} & \textbf{LSTM (full sequence)} & \textbf{LSTM (newline filter)} \\
\midrule
\multirow{4}{*}{LCB}
& Qwen3-Coder & \textbf{0.9026} & 0.8897 (-0.0129) & 0.8824 (-0.0202) \\
& GPT-OSS-20B & \textbf{0.8610} & 0.8508 (-0.0102) & 0.8504 (-0.0106) \\
& NVIDIA-Nemotron-3-Nano & \textbf{0.8911} & 0.7851 (-0.1060) & 0.7778 (-0.1133) \\
& Olmo 3.1 Instruct (32B) & \textbf{0.8542} & 0.8135 (-0.0407) & 0.8008 (-0.0534) \\
\midrule
\multirow{4}{*}{BCB-Instruct}
& Qwen3-Coder & \textbf{0.6841} & 0.6523 (-0.0318) & 0.6707 (-0.0134) \\
& GPT-OSS-20B & 0.6532 & 0.6524 (-0.0008) & \textbf{0.6585 (+0.0053)} \\
& NVIDIA-Nemotron-3-Nano & 0.6411 & 0.6424 (+0.0013) & \textbf{0.6685 (+0.0274)} \\
& Olmo 3.1 Instruct (32B) & 0.7309 & 0.7239 (-0.0070) & \textbf{0.7452 (+0.0143)} \\
\bottomrule
\end{tabular}%
}
\caption{Comparison of sequential model variants against the best single-token XGBoost baseline. Cells report AUROC($\uparrow$), and values in parentheses denote absolute AUROC change relative to the XGBoost baseline.}
\label{tab:res_seq_lstm_vs_xgb}
\end{table}

Table~\ref{tab:res_seq_lstm_vs_xgb} shows that sequential models do not provide a consistent gain over single-token baselines. Full sequences improve over XGBoost in only one of eight settings (NVIDIA-Nemotron-3-Nano on BCB-Instruct), while newline-filtered sequences outperform in three settings, all of which are on BCB-Instruct. For every model on LCB, both sequential variants under-perform the single-token baseline, with the largest drop observed for NVIDIA-Nemotron-3-Nano.

Two additional patterns are noteworthy. First, on LCB, full sequences are consistently better than newline-filtered sequences, suggesting that removing intra-line token states discards useful information for this task. Second, the opposite trend appears on BCB-Instruct: newline filtering consistently outperforms full sequences. This indicates that line-boundary states offer a cleaner signal in library-heavy code generation tasks.

Overall, these results suggest that much of the response-level correctness signal is already captured by late single-token representations. A plausible explanation is that these signals in neighboring token states are highly correlated, so training on dynamic, longer sequences mainly introduces redundant features and optimization noise. In this setting, the added sequential capacity of the LSTM does not reliably translate into better discrimination between correct and incorrect programs.

\noindent \textbf{Entropy-filtered sequences.}
We also evaluate entropy-based sequence filtering as a second selection strategy. The underlying intuition is that tokens with higher uncertainty may provide more information about correctness. By filtering full hidden-state sequences based on token entropy, we aim to focus the model's attention on the most critical parts of the generation process where the LLM is less confident and more susceptible to errors. This reliance on entropy as an auxiliary filtering signal is motivated by recent findings on uncertainty dynamics in code generation~\cite{akyash_decortl_2025}. In their work, the authors demonstrate that, although synthesized code generally has low entropy, sharp spikes occur at critical syntactic and semantic positions. Consequently, these occurrences of high uncertainty can highlight fragile code regions.

As introduced in Subsection~\ref{subsec:feat_select}, we filter using normalized token entropy~\cite{akyash_decortl_2025}. This approach enables thresholding based on relative rather than absolute entropy, effectively accounting for inherent variations in model confidence across different generations. Due to computational constraints, this ablation was performed only for Qwen3-Coder on LCB, with results summarized in Table~\ref{tab:res_seq_entropy_qwen_lcb}.

\begin{table}[htpb]
\centering
\footnotesize
\setlength{\tabcolsep}{5pt}
\begin{tabular}{lcc}
\toprule
\textbf{Model, Filter Setting} & \textbf{AUROC} & \textbf{$\Delta$ vs. XGBoost} \\
\midrule
XGBoost (best single-token baseline) & \textbf{0.9026} & 0.0000 \\
LSTM, default entropy filter ($H \geq 0.20$) & 0.8665 & -0.0361 \\
LSTM, top-entropy ($H \geq 0.66$) & 0.8770 & -0.0256 \\
LSTM, middle-entropy band ($0.33 \leq H \leq 0.66$) & 0.8779 & -0.0247 \\
LSTM, low-entropy ($H \leq 0.33$) & 0.8960 & -0.0066 \\
LSTM, low-high entropy band ($0.15 \leq H \leq 0.33$) & 0.8775 & -0.0251 \\
LSTM, lower-entropy ($H \leq 0.15$) & 0.8944 & -0.0083 \\
\bottomrule
\end{tabular}
\caption{Entropy-filtered sequential model variants against the best
single-token XGBoost baseline for Qwen3-Coder on LCB. $H$ denotes normalized token entropy. Default entropy filter is based on~\cite{zhu_uncertainty-guided_2025}, who found $0.2$ to be an effective threshold for identifying high-risk tokens.}
\label{tab:res_seq_entropy_qwen_lcb}
\end{table}

The entropy ablation results reinforce the previous trend, showing that none of the entropy-filtered sequential models surpass the best single-token baseline. Interestingly, the strongest entropy-filtered result is achieved by retaining low-entropy tokens ($H \leq 0.33$), whereas filters that prioritize high-entropy regions perform considerably worse. This suggests that high entropy alone is not a sufficient proxy for positions critical to correctness at the response level. In practice, high-entropy tokens may capture ambiguous but harmless decisions. Meanwhile, low-entropy regions can carry the stable structural signals required for reliable correctness prediction.

In summary, across full, newline-filtered, and entropy-filtered sequences, sequential probes do not consistently improve over static single-token baselines. The central lesson for response-level IUE with regard to code generation is that additional sequential context is not automatically beneficial. Without better selection or aggregation mechanisms, features that cover the entire generation sequence seem to introduce more redundancy and noise than learnable signals.

\subsection{Comparison with Post-hoc Calibration} \label{subsec:calibration_eval}
The previous experiments focused primarily on discriminatory performance. However, for real-world correctness estimation, calibration quality is equally important. To address this, we conducted a dedicated calibration evaluation using BSS as our primary metric (see Subsection~\ref{subsec:metrics}). We compare the calibration of our static, single-token baselines with that of existing post-hoc calibration methods~\cite{campos_multicalibration_2025} for code correctness estimation.

To ensure a direct and fair comparison, we evaluate on the \href{https://huggingface.co/datasets/lavis-nlp/CALIBRI}{CALIBRI} dataset, which was introduced by \cite{campos_multicalibration_2025} and serves as their primary benchmark. This benchmark was not curated as part of this thesis and contains Qwen3-Coder generations on LCB with response-level correctness labels. Importantly, unlike our datasets, CALIBRI includes samples without code segments that would be labeled \textit{no\_program} in our pipeline (see Section~\ref{sec:datasets}). We intentionally retain these samples to strictly match their evaluation protocol.

For this evaluation, we adopt our strongest single-token baseline for Qwen3-Coder on LCB: an XGBoost probe trained on the last-token hidden state. This baseline achieved the highest BSS in Table~\ref{tab:res_single_token_all}. We then trained this probe configuration on the CALIBRI dataset and evaluated it on the exact test split used in the original study. Table~\ref{tab:res_calib_calibri} presents the results of our probe alongside the baselines and multicalibration approaches reported by \cite{campos_multicalibration_2025}.

\begin{table}[htpb]
\centering
\footnotesize
\setlength{\tabcolsep}{6pt}
\begin{tabular}{lcc}
\toprule
\textbf{Method} & \textbf{BSS ($\uparrow$)} & \textbf{Accuracy ($\uparrow$)} \\
\midrule
IUE probe (last-token XGBoost, ours) & \textbf{0.5888} & \textbf{0.8534} \\
IGLB~\cite{campos_multicalibration_2025} & 0.4680 & 0.8239 \\
LINR~\cite{campos_multicalibration_2025} & 0.4631 & 0.8216 \\
HB~\cite{campos_multicalibration_2025} & 0.3832 & 0.7973 \\
Platt~\cite{campos_multicalibration_2025} & 0.3769 & 0.7985 \\
LOGR~\cite{campos_multicalibration_2025} & 0.3026 & 0.8269 \\
IGHB~\cite{campos_multicalibration_2025} & 0.2238 & 0.7284 \\
\bottomrule
\end{tabular}
\caption{Calibration evaluation on \href{https://huggingface.co/datasets/lavis-nlp/CALIBRI}{CALIBRI} (Qwen3-Coder, LCB). BSS($\uparrow$) and Accuracy($\uparrow$) are reported, with the best result for each metric in bold. IUE probe reports our last-token XGBoost baseline. All other methods are from~\cite{campos_multicalibration_2025}.}
\label{tab:res_calib_calibri}
\end{table}

Two key observations emerge from these results. First, our single-token probe achieves the highest BSS, significantly outperforming the strongest multicalibration baseline (IGLB). Second, this improvement is not confined to calibration alone with the IUE probe simultaneously delivering the highest accuracy. In this setting, hidden-state-based code correctness estimation appears to provide confidence scores that are significantly more aligned with empirical correctness than post-hoc calibration methods.

While this comparison serves as a strong proof of concept, its scope is inherently constrained to a single model (Qwen3-Coder) and a specific external benchmark. Nevertheless, the result provides strong evidence that introspective features can outperform dedicated calibration pipelines for code generation. This motivates the comprehensive generalization analyses for response-level IUE probes presented in the following subsection.

\subsection{Generalization Capabilities} \label{subsec:generalization}
While probe-based introspective methods in NLG often struggle with generalization under domain and task shifts~\cite{ch-wang_androids_2024, preis_hallucination_2025}, their robustness in code generation tasks remains underexplored. To address \hyperref[item:rq2]{RQ2}, we conduct a targeted evaluation of out-of-distribution (OOD) behavior. To prioritize observing the effects of data distribution shifts, we fix our probing architecture to the computationally efficient static XGBoost baselines established in Subsection~\ref{subsec:res_baselines}.

We systematically evaluate all six static probe configurations (four single-token, two multi-token) under three complementary generalization settings: (1) \textit{cross-task generalization} between the LCB and BCB benchmarks, (2) \textit{cross-domain generalization} via a held-out domain approach on BCB, and (3) \textit{cross-token-position generalization} using randomized extraction windows to test whether probes learn robust global correctness signals or overfit to specific generation positions.

\noindent \textbf{Cross-task (cross-benchmark) generalization.}
We evaluate transfer between LCB and both BCB variants (BCB-Instruct and BCB-Fusion). For each direction, probes are trained using the source benchmark's train and validation splits and evaluated on the target benchmark's test split. We do not report an evaluation between the two BCB variants because they share the same underlying benchmark and only differ in prompt format, resulting in only a minor distribution shift.

Table~\ref{tab:res_gen_cross_lcb_bcb} summarizes the best AUROC per model and direction for the cross-task transfers. We report the best feature in parentheses and the absolute difference to the best in-distribution (ID) AUROC on the target benchmark for the corresponding dataset-LLM pair. Feature abbreviations follow earlier sections: L = last token, LC = last code token, F+L = first+last token, FC+LC = first+last code token.

\begin{table}[htpb]
\centering
\footnotesize
\setlength{\tabcolsep}{4pt}
\renewcommand{\arraystretch}{1.05}
\resizebox{\textwidth}{!}{%
\begin{tabular}{lcccc}
\toprule
\textbf{Train $\rightarrow$ Test} & \textbf{Qwen3-Coder} & \textbf{GPT-OSS-20B} & \textbf{NVIDIA-Nemotron-3-Nano} & \textbf{Olmo 3.1 Instruct (32B)} \\
\midrule
LCB $\rightarrow$ BCB-Instruct & 0.6483 (LC, $\Delta=\mathbf{-0.0358}$) & 0.5970 (L, $\Delta=-0.0562$) & 0.5838 (F+L, $\Delta=-0.0573$) & 0.6149 (FC+LC, $\Delta=-0.1160$) \\
LCB $\rightarrow$ BCB-Fusion & 0.6509 (LC, $\Delta=\mathbf{+0.0060}$) & 0.6875 (L, $\Delta=-0.0370$) & 0.6236 (L, $\Delta=-0.0577$) & 0.6713 (L, $\Delta=-0.0424$) \\
BCB-Instruct $\rightarrow$ LCB & 0.8240 (LC, $\Delta=\mathbf{-0.0786}$) & 0.7342 (LC, $\Delta=-0.1268$) & 0.7771 (L, $\Delta=-0.1140$) & 0.7410 (LC, $\Delta=-0.1132$) \\
BCB-Fusion $\rightarrow$ LCB & 0.8164 (LC, $\Delta=-0.0862$) & 0.7415 (F+L, $\Delta=-0.1195$) & 0.8426 (L, $\Delta=-0.0485$) & 0.8149 (L, $\Delta=\mathbf{-0.0393}$) \\
\bottomrule
\end{tabular}%
}
\caption{Cross-benchmark transfer between LCB and both BCB variants. Cells report best AUROC($\uparrow$), best feature, and absolute difference to the best in-distribution AUROC on the target benchmark. Best delta ($\Delta=\text{OOD}-\text{ID}$) per transfer direction is in bold.}
\label{tab:res_gen_cross_lcb_bcb}
\end{table}

Two observations stand out. First, testing on LCB yields strong AUROC results when models are trained on the more challenging BCB datasets. Transferring from BCB-Fusion to LCB is particularly effective across models, which is expected since BCB-Fusion closely mirrors the LCB prompt structure (see Section~\ref{sec:bcb}). Notably, the transfer from BCB-Instruct to LCB also demonstrates strong performance, exceeding that of BCB-Fusion in one instance (Qwen3-Coder).

These results are significant because they show that probes trained on complex datasets can detect correctness signals that generalize well to easier evaluation settings despite the expected decrease in performance compared to ID training. This suggests that the probes identify aspects of code correctness that are not entirely task-specific and provides encouraging evidence for the broader applicability of introspective methods to code generation.

Second, the transfers from LCB to BCB tasks also show promise. However, testing on the more challenging BCB datasets leads to lower AUROC values. This is expected given the increased difficulty of the test splits, yet the performance drop relative to ID training remains modest for all LLMs, particularly for the LCB to BCB-Fusion transfer. In fact, there is one instance of a slight performance gain relative to the BCB-Fusion ID baseline, which is a strong indication of cross-task robustness.

Overall, these findings demonstrate a robustness of response-level IUE models across tasks. Even under data distribution shifts, probe performance remains substantially above random and can, occasionally, surpass ID performance. At the same time, a persistent OOD-ID gap confirms that benchmark-specific effects are still present. Consequently, while it is feasible to train a resilient generalist probe that can perform well across different tasks, benchmark-specific training is still necessary to achieve optimal performance.

\noindent \textbf{Cross-domain generalization.}
To test domain-level generalization of IUE on code, we perform experiments using a held-out domain approach on the BCB-Instruct dataset. Following the BCB domain taxonomy (see Figure~\ref{fig:bcb_domains}), we trained probes on all problems except those belonging to a specific held-out domain (e.g., ``System''). The non-held-out data is split into a $80 / 20$ stratified training and validation set. The remaining samples, which all use function calls from the held-out domain, serve as the remaining test split. Each of these domain test splits has a different structure, with varying sample sizes and label balances. However, it was verified that each test split contains both positive and negative labels (see Appendix~\ref{fig:bcb_domains_models_combined}).

This cross-domain experiment evaluates whether introspective probes can identify domain-invariant correctness features that generalize to unseen programming contexts. This ability is essential for practical applications where the distribution of code may shift across domains (e.g., from data science to web  engineering).

Table~\ref{tab:res_gen_bcb_domain_detail} reports detailed results for three representative domains with different test split sizes: System ($30\%$ of BCB), Time ($10\%$), and Cryptography ($5\%$). We report the best OOD AUROC, best feature, and $\Delta$ relative to the best ID AUROC for each model on BCB-Instruct.

\begin{table}[htpb]
\centering
\footnotesize
\setlength{\tabcolsep}{4pt}
\renewcommand{\arraystretch}{1.05}
\resizebox{\textwidth}{!}{%
\begin{tabular}{lcccc}
\toprule
\textbf{Model} & \textbf{ID Best} & \textbf{System} & \textbf{Time} & \textbf{Cryptography} \\
\midrule
Qwen3-Coder & 0.6841 & 0.6643 (LC, $\Delta=\mathbf{-0.0198}$) & 0.6402 (L, $\Delta=-0.0439$) & 0.6353 (L, $\Delta=-0.0488$) \\
GPT-OSS-20B & 0.6569 & 0.6502 (F+L, $\Delta=\mathbf{-0.0066}$) & 0.6105 (L, $\Delta=-0.0464$) & 0.6281 (L, $\Delta=-0.0288$) \\
NVIDIA-Nemotron-3-Nano & 0.6411 & 0.6568 (F+L, $\Delta=+0.0157$) & 0.6508 (LC, $\Delta=+0.0097$) & 0.6746 (FC+LC, $\Delta=\mathbf{+0.0335}$) \\
Olmo 3.1 Instruct (32B) & 0.7309 & 0.7107 (LC, $\Delta=-0.0202$) & 0.7570 (LC, $\Delta=\mathbf{+0.0261}$) & 0.7147 (L, $\Delta=-0.0162$) \\
\bottomrule
\end{tabular}%
}
\caption{Representative cross-domain generalization on BCB-Instruct. Cells report best AUROC($\uparrow$) on the held-out domain, best feature, and absolute difference to each model's best ID AUROC on BCB-Instruct. Best delta ($\Delta$) per model is in bold.}
\label{tab:res_gen_bcb_domain_detail}
\end{table}

Domain-wise tables for the remaining held-out domain configurations (Computation, General, Network, and Visualization) are provided in Appendix~\ref{sec:app_exp_domain_tables}.

Across these representative held-out domains, the cross-domain generalization pattern is comparatively stable. All absolute OOD-ID deltas are below $0.05$, which is notably smaller than the cross-task gaps reported in Table~\ref{tab:res_gen_cross_lcb_bcb}. This is consistent with the weaker shift induced by holding out a BCB domain relative to transferring between different benchmarks. Importantly, the signed differences are mixed: several settings show small degradations, while others improve, indicating no systematic OOD collapse within this domain-transfer setup.

At the domain level, the \textit{System} split (the largest among the three shown domains) yields the most consistent behavior, with only modest deviations from ID performance for all models. The \textit{Time} and \textit{Cryptography} splits show larger dispersion, which is plausible given their smaller test shares ($10\%$ and $5\%$). Even in these settings, improvements still appear for specific model-domain pairs (e.g., Olmo 3.1 Instruct on \textit{Time}), reinforcing that domain shift effects are heterogeneous rather than uniformly negative. Finally, NVIDIA-Nemotron-3-Nano stands out as the only model that improves over the best ID performance across all three domains, suggesting that it may have learned more robust, domain-invariant correctness patterns than the other models.

To avoid over-interpreting individual domains, we evaluate all seven BCB domains using the shortfall metric
$$
\text{shortfall}=\max(0,\text{AUROC}_{\text{ID-best}}-\text{AUROC}_{\text{OOD-best}}).
$$
Here, $\text{AUROC}_{\text{ID-best}}$ denotes each model's best ID AUROC on BCB-Instruct, and $\text{AUROC}_{\text{OOD-best}}$ denotes the best AUROC of each model on one of the held-out domain splits. By definition, shortfall only penalizes degradation and assigns a value of zero to OOD improvements. This makes the aggregate statistic more robust to outlier effects from smaller domain splits, where a positive OOD-ID delta may occur due to smaller sample size. To preserve directional information, we additionally report the mean signed difference ($\Delta=\text{AUROC}_{\text{OOD-best}}-\text{AUROC}_{\text{ID-best}}$) across all seven domains. Table~\ref{tab:res_gen_bcb_domain_summary} summarizes these aggregate results.

\begin{table}[htpb]
\centering
\footnotesize
\setlength{\tabcolsep}{6pt}
\begin{tabular}{lcc}
\toprule
\textbf{Model} & \textbf{Shortfall $\mu$ ($\downarrow$)} & \textbf{$\Delta=\text{OOD}-\text{ID}$ $\mu$ ($\uparrow$)} \\
\midrule
Qwen3-Coder & 0.0217 & -0.0153 \\
GPT-OSS-20B & 0.0219 & -0.0200 \\
NVIDIA-Nemotron-3-Nano & \textbf{0.0012} & \textbf{+0.0218} \\
Olmo 3.1 Instruct (32B) & 0.0178 & -0.0141 \\
\bottomrule
\end{tabular}
\caption{Aggregate cross-domain generalization on BCB-Instruct. Shortfall $\mu$ reports the mean shortfall across all seven held-out domains for each LLM, and $\Delta$ $\mu$ reports the mean signed difference across all seven held-out domains for each LLM. \\ Best value per metric is in bold.}
\label{tab:res_gen_bcb_domain_summary}
\end{table}

The aggregate results are promising for all models. The mean shortfall remains low ($< 2.2\%$), indicating that domain shifts within BCB have a more subtle impact on response-level IUE performance compared to cross-task transfers (see Table~\ref{tab:res_gen_cross_lcb_bcb}). Only having to expect a small performance degradation for a domain shift on code generation tasks is a highly encouraging result. NVIDIA-Nemotron-3-Nano is the most robust model, with a near-zero mean shortfall and a positive mean signed difference, suggesting that its OOD domain performance is, on average, at least as strong as its ID performance.

Overall, this experiment shows that the domain-level OOD performance on BCB-Instruct remains relatively stable across all four models, with only minor degradations compared to ID results. This supports the conclusion that domain shifts within a consistent code generation task do not severely disrupt response-level IUE, and that probes can learn correctness-relevant patterns that largely transfer across different programming domains.

\noindent \textbf{Cross-token-position generalization.}
The final generalization experiment at the response level evaluates the extent to which probes overfit to their static token positions. Unlike previous baselines, which used hidden states from specific positions in the response, such as the \textit{last code token}, this experiment investigates whether models can learn features relevant to correctness estimation that are robust to positional variance. To test this, we implemented a ``random tail'' (RT) dataset. Rather than using a static position, the RT dataset uniformly samples a single token representation per program from a predefined suffix window of the generation sequence.

The size of this window is controlled by a parameter, $\text{RT} \in \{0.001, 0.05, \dots, 1.0\}$, representing the fraction of the total sequence length considered for sampling. For example, $\text{RT} = 1.0$ uniformly samples the token position between the start of the generated code segment and the end of the response sequence. Conversely, $\text{RT} = 0.001$ restricts sampling to the final $0.1\%$ of the response tokens, closely approximating a \textit{last token} extraction. During evaluation, we sample from an identical tail range with the same RT, but employ a different random seed. This ensures that the concrete token positions observed during testing differ from those seen during training, thereby testing the model's generalization capabilities across token positions. Table~\ref{tab:res_gen_rt} reports AUROC across LCB and BCB-Instruct for all four models and a range of RT values.

\begin{table}[htpb]
\centering
\footnotesize
\setlength{\tabcolsep}{5pt}
\resizebox{\textwidth}{!}{%
\begin{tabular}{llccccccc}
\toprule
\multirow{2}{*}{\textbf{Dataset}} & \multirow{2}{*}{\textbf{Model}} & \multicolumn{7}{c}{\textbf{Random Tail (RT) Fraction}} \\
\cmidrule(lr){3-9}
& & \textbf{0.001} & \textbf{0.05} & \textbf{0.25} & \textbf{0.5} & \textbf{0.75} & \textbf{0.95} & \textbf{1.0} \\
\midrule
\multirow{4}{*}{LCB} & Qwen3-Coder & 0.9045 & 0.8515 & 0.8351 & 0.8363 & 0.8279 & 0.8326 & 0.8202 \\
& GPT-OSS-20B & 0.8551 & 0.7983 & 0.7426 & 0.7364 & 0.7261 & 0.7232 & 0.7212 \\
& NVIDIA-Nemotron-3-Nano & 0.8617 & 0.8128 & 0.7444 & 0.7393 & 0.7808 & 0.7667 & 0.7437 \\
& Olmo 3.1 Instruct (32B) & 0.8440 & 0.7974 & 0.7687 & 0.7563 & 0.7713 & 0.7587 & 0.7708 \\
\midrule
\multirow{4}{*}{BCB-Instruct} & Qwen3-Coder & 0.6754 & 0.6302 & 0.5961 & 0.6037 & 0.6027 & 0.5677 & 0.5827 \\
& GPT-OSS-20B & 0.6500 & 0.5585 & 0.5665 & 0.5729 & 0.5492 & 0.5408 & 0.5322 \\
& NVIDIA-Nemotron-3-Nano & 0.6469 & 0.5739 & 0.5702 & 0.5792 & 0.5757 & 0.5556 & 0.5615 \\
& Olmo 3.1 Instruct (32B) & 0.7176 & 0.6337 & 0.6476 & 0.6208 & 0.6221 & 0.6189 & 0.6116 \\
\bottomrule
\end{tabular}
}%
\caption{``Random tail'' (RT) cross-token-position generalization for each LLM across LCB and BCB-Instruct. Each cell reports AUROC($\uparrow$). Larger RT fractions represent a more challenging generalization scenario.}
\label{tab:res_gen_rt}
\end{table}

As expected, $\text{RT}=0.001$ comes close to the best static, single-token baselines from Table~\ref{tab:res_single_token_all}. Increasing the RT fraction introduces a clear trade-off between performance and robustness: AUROC decreases when moving from very late tokens to broader suffix windows. However, the decline is not unbounded. For most model-dataset pairs, the largest drop occurs up to $\text{RT}=0.25$, after which performance tends to stabilize around a task-specific floor. This observation supports the hypothesis that response-level correctness information is distributed across many token positions, even though late positions remain the most informative.

Taken together, the cross-token-position results indicate that position-robust code correctness estimation is possible despite moderate performance loss. This is a relevant insight for practical monitoring scenarios using IUE in which static token positions cannot be guaranteed. For example, one could monitor the code generation of autonomous agents using their hidden states at random token positions.

However, while robust response-level predictions provide a useful overall confidence score, practical developer assistance demands a more precise localization of errors for generated code. Addressing the need to identify the exact source of a failure motivates the transition to fine-grained hallucination detection.

\section{Line-Level} \label{sec:line_level}

The line-level experiments evaluate whether introspective features can do more than provide a response-level confidence score and instead identify likely fault locations in generated code. This represents a substantially more challenging task than estimating response-level correctness because the model must identify specific lines of code that are likely incorrect by distinguishing a small number of faulty lines from a large number of correct ones, often within partially correct programs.

This fine-grained setting is particularly relevant for practical development workflows, such as debugging and reviewing generated code. In these scenarios, developers benefit more from detailed, localized feedback than from a single binary prediction for the entire response. Furthermore, it directly addresses \hyperref[item:rq3]{RQ3}, which investigates whether IUE can support fine-grained hallucination detection on realistic code generation tasks.

The section is organized as follows. Subsection~\ref{subsec:line_baselines} presents line-level baselines in a mixed setting that includes both correct and incorrect programs. Then, Subsection~\ref{subsec:fault_loc} studies a conditional setting in which programs are assumed to be incorrect beforehand (e.g., after response-level filtering). This shifts the focus of the task to fault localization within a program that is known to be faulty.

\subsection{Baselines} \label{subsec:line_baselines}
The baseline experiments are conducted on the augmented LCB datasets introduced in Section~\ref{sec:augmentation}, using Qwen3-Coder and GPT-OSS-20B. We evaluate two dataset variants. \textbf{V1} is a filtered setting that retains only samples with small, localized faults (fewer than four incorrect lines), while \textbf{V2} uses the full augmented dataset without this additional filter. For all line-level experiments, whitespace-only and comment-only lines are removed from the datasets prior to training and evaluation.

As the augmentation statistics in Table~\ref{tab:aug_summary_combined} show, both datasets have severe class imbalances at the line level, a problem that is particularly pronounced in \textbf{V1}. Specifically, the proportion of faulty lines in \textbf{V1} is merely $1.21\%$ for Qwen3-Coder and $0.27\%$ for GPT-OSS-20B. In contrast, \textbf{V2} presents a relatively more balanced, though still highly skewed, distribution with corresponding shares of $13.52\%$ and $1.40\%$. It is important to acknowledge these proportions when interpreting subsequent experimental results, especially for threshold-sensitive metrics like the F1 score.

Following the methodology outlined in Section~\ref{subsec:train_obj}, we invert the labels for \hyperref[item:rq3]{RQ3} at both the token and line levels, designating incorrect code as the positive class. This aligns optimization with the practically relevant minority class.

\noindent \textbf{Static token-to-line vs. line-level baselines.}
We first evaluate two static XGBoost baselines with class weighting (\textit{scale\_pos\_weight}) to compensate for label imbalance. The first model is trained at the token level and performs line-level predictions via max-pooling over token uncertainty scores within each line. This follows the intuition that a line should be flagged as faulty if at least one token is highly suspicious, similar to the span-level aggregation in prior fine-grained probing work~\cite{ch-wang_androids_2024} for NLG. The second model predicts line labels directly from a fixed line representation, in this case the \textit{last code token} of the line, excluding whitespace and newline tokens. Table~\ref{tab:line_xgb_baselines} summarizes the results of these two static baselines across both dataset variants and LLMs, reporting AUROC and F1 scores.

\begin{table}[htpb]
\centering
\footnotesize
\setlength{\tabcolsep}{5pt}
\renewcommand{\arraystretch}{1.05}
\begin{tabular}{llcc}
\toprule
\textbf{Dataset} & \textbf{Baseline} & \textbf{Qwen3-Coder} & \textbf{GPT-OSS-20B} \\
\midrule
\multirow{2}{*}{V1} & Token $\rightarrow$ Line & 0.7258 / 0.0390 & \textbf{0.7431} / \textbf{0.0533} \\
& Line-level (last code token) & \textbf{0.7291} / \textbf{0.0601} & 0.7318 / 0.0127 \\
\multirow{2}{*}{V2} & Token $\rightarrow$ Line & 0.7575 / 0.2308 & \textbf{0.7294} / \textbf{0.0862} \\
& Line-level (last code token) & \textbf{0.7639} / \textbf{0.2593} & 0.7210 / 0.0651 \\
\bottomrule
\end{tabular}
\caption{Line-level XGBoost baselines on augmented dataset variants. Cells report AUROC($\uparrow$) / F1($\uparrow$). The token $\rightarrow$ line baseline applies max-pooling over token-level predictions, while the line-level baseline uses the last code token of each line. Bold denotes the better approach per metric within each dataset-model pair.}
\label{tab:line_xgb_baselines}
\end{table}

Two findings emerge from these baseline evaluations. First, the AUROC values achieved on these highly imbalanced datasets are moderate for both baselines and datasets. Although these scores fall short of the robust response-level performance (see Subsection~\ref{subsec:res_baselines}), they remain considerably above random predictions. This is a promising finding given the inherent difficulty of fine-grained fault localization. The comparative performance of the two baselines is mixed. For Qwen3-Coder, the direct line-level baseline slightly outperforms the token-to-line approach. However, the reverse is true for GPT-OSS-20B.

Second, the reduced class imbalance in \textbf{V2} demonstrably improves performance across both metrics for Qwen3-Coder. Despite this improvement, the F1 scores remain fundamentally low across all configurations, even when the corresponding AUROC indicates reasonably strong discrimination capability. This disparity highlights that strong ranking quality does not automatically translate into well-calibrated binary classifications.

Overall, the static baselines suggest that neither approach consistently outperforms the other across different models. Yet, the AUROC performance reinforces the broader conclusion that hidden states encode features relevant to correctness at a localized level. Consequently, the primary challenge in line-level IUE on code is not the absence of a learning signal, but rather the difficulty of translating that signal into well-calibrated confidence scores.

\noindent \textbf{Sequential line-level baselines.}
To test whether a richer local context can improve fine-grained fault localization, we trained sequential baselines that encode each line as a full sequence of token hidden states and predict one line label per sequence. Treating the line as a sequence allows the model to learn the sequential adaptation of local code correctness patterns as each line unfolds. Specifically, we train an LSTM encoder, followed by an MLP classification head, optimized with a positive-class weighted binary cross-entropy loss. Table~\ref{tab:line_lstm_baselines} reports the sequential baselines AUROC / F1, with absolute differences to the static, line-level XGBoost baselines from Table~\ref{tab:line_xgb_baselines} in parentheses.

\begin{table}[htpb]
\centering
\footnotesize
\setlength{\tabcolsep}{6pt}
\renewcommand{\arraystretch}{1.05}
\begin{tabular}{lcc}
\toprule
\textbf{Dataset} & \textbf{Qwen3-Coder} & \textbf{GPT-OSS-20B} \\
\midrule
V1 & 0.7273 / 0.0891 ($\Delta=-0.0018 / +0.0290$) & 0.7557 / 0.0010 ($\Delta=+0.0239 / -0.0117$) \\
V2 & 0.7402 / 0.2778 ($\Delta=-0.0237 / +0.0185$) & 0.7268 / 0.2015 ($\Delta=+0.0058 / +0.1364$) \\
\bottomrule
\end{tabular}
\caption{Sequential line-level LSTM baselines on augmented dataset variants. Cells report LSTM AUROC($\uparrow$) / F1($\uparrow$) with metric-wise deltas in parentheses, where $\Delta_M = M_{\text{LSTM}} - M_{\text{XGBoost}}$ and $M \in \{\text{AUROC}, \text{F1}\}$ (refer to Table~\ref{tab:line_xgb_baselines} for XGBoost).}
\label{tab:line_lstm_baselines}
\end{table}

The sequential baselines improved the F1 score for three out of four of the evaluated model-dataset pairs. The most substantial gain was observed for GPT-OSS-20B on \textbf{V2}. However, similar to the results with response-level sequential models (see Subsection~\ref{subsec:seq_models}), the models do not outperform the static, line-level XGBoost approach consistently across both evaluation metrics. Specifically, the LSTM achieves slightly higher AUROC scores  for GPT-OSS-20B only, while underperforming on Qwen3-Coder across both dataset variants. This mixed behavior suggests that adding within-line sequence context does not necessarily lead to superior fine-grained performance.

\noindent \textbf{Threshold optimization.}
Since AUROC is threshold-independent and the F1 score is not, we performed an additional threshold sweep using $\text{t} \in \{0.1, 0.2, \dots, 0.9\}$ for select line-level baselines. The results are summarized in Table~\ref{tab:line_threshold_optimization}, which reports the default F1 score at the standard $0.5$ threshold and the best F1 score across all tested thresholds.

\begin{table}[htpb]
\centering
\footnotesize
\setlength{\tabcolsep}{6pt}
\begin{tabular}{lccc}
\toprule
\textbf{LLM, Line-level baseline (dataset)} & \textbf{Default F1 ($t{=}0.5$)} & \textbf{Best F1} & \textbf{Best Threshold ($t$)} \\
\midrule
Qwen3-Coder, XGBoost Token$\rightarrow$Line (V1) & 0.0390 & 0.0660 & 0.8 \\
Qwen3-Coder, XGBoost Line-level (V2) & 0.2593 & \textbf{0.3156} & 0.8 \\
Qwen3-Coder, LSTM Line-level (V2) & 0.2778 & 0.2886 & 0.7 \\
\bottomrule
\end{tabular}
\caption{Post-hoc threshold optimization for select line-level baselines. Default F1 is reported at the standard $0.5$ threshold, while Best F1 is the maximum F1 achieved across the tested thresholds $\text{t} \in \{0.1, 0.2, \dots, 0.9\}$. Best overall F1 is in bold.}
\label{tab:line_threshold_optimization}
\end{table}

Threshold tuning confirms that calibration choices matter, but also that the attainable gains are limited. The most significant effect is observed with Qwen3-Coder on \textbf{V2}, where the line-level XGBoost baseline improves from approximately $0.26$ to $0.32$ F1 after optimizing the decision threshold. Other settings improve as well, but remain lower in absolute value. This pattern is consistent with a core limitation of the mixed line-level setting. Positive and negative line-score distributions overlap significantly, so adjusting the threshold, $t$, primarily trades precision against recall without revealing a stable operating point. In other words, the probes recover a useful ranking signal, as reflected by AUROC, but converting that signal into reliable line scores remains difficult.

Taken together, these baselines provide evidence that correctness-relevant information on the generated code exists at a line level, while also clarifying why thresholded F1 scores remain low in practice. The next subsection therefore evaluates a conditional fault localization setup in which all program samples are assumed to already be known as incorrect, reducing the task complexity to prioritizing the most likely faulty lines within an incorrect program.

\subsection{Fault Localization Models}  \label{subsec:fault_loc}
This experiment evaluates a conditional, line-level fault localization setup that assumes each analyzed program is already known to be incorrect at the response level. In practical terms, this corresponds to a two-stage workflow. First, a response-level IUE model filters for likely incorrect code generations. Then, a line-level model prioritizes likely fault regions within those candidates. To emulate this scenario, we trained and evaluated on dataset variant \textbf{V3} (see Table~\ref{tab:aug_summary_combined}), which only contains incorrect programs.

We use the same probe architectures as in Subsection~\ref{subsec:line_baselines}: a static token-to-line XGBoost probe, a static line-level XGBoost probe using the \textit{last code token}, and a sequential line-level LSTM. Table~\ref{tab:fault_loc_v3_models} reports AUROC / F1 across all three probes for each LLM. Since F1 is threshold-sensitive, we report the best value from a post-hoc sweep over $\text{t} \in \{0.1, 0.2, \dots, 0.9\}$.

\begin{table}[htpb]
\centering
\footnotesize
\setlength{\tabcolsep}{5pt}
\renewcommand{\arraystretch}{1.05}
\begin{tabular}{lcc}
\toprule
\textbf{Baseline} & \textbf{Qwen3-Coder} & \textbf{GPT-OSS-20B} \\
\midrule
XGBoost Token$\rightarrow$Line & 0.7267 / \textbf{0.6675} & 0.7015 / 0.3970 \\
XGBoost Line-level (last code token) & 0.7407 / 0.6560 & 0.7270 / 0.3905 \\
LSTM Line-level & \textbf{0.7410} / 0.6456 & \textbf{0.7294} / \textbf{0.4043} \\
\bottomrule
\end{tabular}
\caption{Line-level fault localization evaluation on V3 (incorrect programs only) for each LLM. Cells report AUROC($\uparrow$) / F1($\uparrow$). F1 values are computed at each baseline's best threshold from a sweep over $\text{t} \in \{0.1, 0.2, \dots, 0.9\}$. Best value per metric and LLM is in bold.}
\label{tab:fault_loc_v3_models}
\end{table}

Compared to the mixed-program setting (\textbf{V1/V2}), AUROC remains in a similar range, indicating that ranking quality is broadly stable under this conditional setup. The main change appears in the F1 scores. For Qwen3-Coder, F1 increases substantially for all baselines, and for GPT-OSS-20B, the score also improves considerably, though not as much. This is consistent with the class-balance shift in \textbf{V3}, where the proportion of incorrect lines is notably higher than in \textbf{V1/V2}, particularly for Qwen3-Coder (see Table~\ref{tab:aug_summary_combined}). Consequently, this conditional fault localization setting helps the probes translate ranking signal into thresholded classifications more effectively.

\noindent \textbf{Top-$K$ point-of-failure ranking.}
Finally, to specifically evaluate practical debugging utility, we utilize the Top-$K$ hit rate metric from Subsection~\ref{subsec:metrics}. For each incorrect program, we rank the lines by scores predicted by the models and determine if at least one incorrectly labeled line appears in the top $K$. We focus on the two static XGBoost baselines, token-to-line and line-level (last code token), and compare them to a random baseline that assigns independent random scores ($x_n \overset{\text{i.i.d.}}{\sim} \mathrm{Uniform}(0,1)$) per line. The random baseline is averaged over 1000 trials. Table~\ref{tab:fault_loc_topk} reports Top-$K$ hit rates with $K \in \{1, 2, 3\}$ for each baseline and LLM.

\begin{table}[htpb]
\centering
\footnotesize
\setlength{\tabcolsep}{3.6pt}
\renewcommand{\arraystretch}{1.05}
\resizebox{\textwidth}{!}{%
\begin{tabular}{lcccccc}
\toprule
\multirow{2}{*}{\textbf{Baseline}} & \multicolumn{3}{c}{\textbf{Qwen3-Coder}} & \multicolumn{3}{c}{\textbf{GPT-OSS-20B}} \\
\cmidrule(lr){2-4}\cmidrule(lr){5-7}
& \textbf{Hit@1} & \textbf{Hit@2} & \textbf{Hit@3} & \textbf{Hit@1} & \textbf{Hit@2} & \textbf{Hit@3} \\
\midrule
Token$\rightarrow$Line & \textbf{0.6163} & \textbf{0.7209} & \textbf{0.8052} & \textbf{0.4937} & \textbf{0.6076} & \textbf{0.7089} \\
Line-level (last code token) & 0.5669 & 0.6483 & 0.7355 & 0.3418 & 0.5316 & 0.6203 \\
Random (1000 trials) & 0.3766 $\pm$ 0.0191 & 0.5102 $\pm$ 0.0182 & 0.5883 $\pm$ 0.0183 & 0.1940 $\pm$ 0.0376 & 0.3054 $\pm$ 0.0416 & 0.3836 $\pm$ 0.0439 \\
\bottomrule
\end{tabular}
}
\caption{Top-$K$ point-of-failure ranking evaluation on V3 (incorrect programs only) for each LLM. Cells report Top-$K$ hit rate with $K \in \{1, 2, 3\}$ (Hit@K, $\uparrow$) as defined in Subsection~\ref{subsec:metrics}. Random baseline reports mean $\pm$ standard deviation over 1000 trials. Best value per metric and LLM is in bold.}
\label{tab:fault_loc_topk}
\end{table}

Three observations stand out. First, in this conditional ranking scenario, the token-to-line probe consistently outperforms the direct line-level probe for both LLMs. This contrasts with the mixed setting, where both static baselines often had similar line-level AUROC scores, and suggests that token-level context aggregation is particularly useful when the task is to prioritize fault locations within an already faulty program.

Second, both probes clearly outperform the random baseline for all values of $K$, confirming that they detect meaningful localized code correctness signals rather than random patterns. Third, compared to previous work~\cite{huang_risk_2025}, which reports a best Hit@3 of $0.641$ on EditEval (a code-editing benchmark), our token-to-line probe achieves substantially higher Hit@3 values on LCB (a code-synthesis benchmark): $0.8052$ for Qwen3-Coder and $0.7089$ for GPT-OSS-20B. While direct comparison should be interpreted cautiously due to differences in benchmarks and evaluation protocols, these results still provide encouraging evidence that IUE can support practical fine-grained fault localization in realistic code generation tasks.

Finally, the evaluation shows that, once incorrect programs have been pre-selected, introspective probes can provide meaningful, line-level fault localization for downstream debugging support.

%% file: chapters/discussion.tex
\chapter{Discussion} \label{cha:discussion}

This chapter interprets the empirical results in direct relation to the research questions defined in Section~\ref{sec:rqs}. Our focus is on the implications of the observed patterns for the practical application of Introspective Uncertainty Estimation (IUE) in code generation tasks. We first discuss the response-level findings for \hyperref[item:rq1]{RQ1}, then analyze generalization capabilities for \hyperref[item:rq2]{RQ2}, and finally examine fine-grained hallucination detection for \hyperref[item:rq3]{RQ3}. The chapter concludes with threats to validity that impact the interpretation of our findings.

\section{RQ1: To what extent can internal artifacts produced by LLMs during the forward pass of code generation tasks be used to estimate code correctness?} \label{sec:disk_rq1}

This research question asks whether artifacts produced during decoding carry meaningful information about correctness for code generation tasks. Building on prior code-focused introspective work~\cite{bui_correctness_2025, ribeiro_llms_2025}, we explicitly tested whether richer representations of the generation process improve response-level prediction beyond static, single-token baselines. Our key finding is that hidden states are clearly predictive of correctness, but additional representational complexity does not necessarily lead to better performance.

\noindent \textbf{Probe architectures.}
The first finding is that the choice of model family among lightweight probes is not decisive in our settings. Table~\ref{tab:res_baseline_qwen_lcb_mlp_vs_xgb} and Table~\ref{tab:res_baseline_lastcode_all_models} show near-parity between MLP probes and XGBoost on static single-token features, with only small absolute differences across token positions, models, and datasets. This suggests that the main performance indicator lies in the extracted hidden-state features rather than in increasingly complex downstream classifiers. In practice, this justifies using XGBoost as an efficient baseline without compromising predictive quality.

\noindent \textbf{Code generation tasks.}
The second finding is a clear task-level performance gap. On LCB, the best single-token AUROC values are consistently high (above $0.85$), while BCB-Instruct is substantially more difficult (see Table~\ref{tab:res_single_token_all}). This pattern is consistent with the benchmark characteristics introduced in Section~\ref{sec:bcb}. BCB tasks require reasoning that relies heavily on libraries and implicit assumptions, and therefore create a more difficult correctness prediction problem. In other words, hidden states remain informative, but their separability is limited by the complexity of the task.

At the same time, the formulation of the prompt alters this separability. Although not uniformly across all models, BCB-Fusion improves over BCB-Instruct for GPT-OSS-20B and NVIDIA-Nemotron-3-Nano (see Table~\ref{tab:res_single_token_all}). This model-dependent behaviour provides important insight into RQ1: not only does prompt engineering affect the quality of the generated text itself, it also changes the quality of the introspective features used to estimate correctness. Notably, these results also show that improved benchmark performance does not necessarily lead to improved IUE performance for all LLMs, suggesting that generation capability and introspective separability are related but distinct properties.

\noindent \textbf{Static multi-token probes.}
The multi-token ablation further refines this conclusion. As shown in Table~\ref{tab:res_multi_token_all}, concatenating the hidden states of static tokens positions (F+L or FC+LC) does not consistently outperform the best single-token features. While some combinations improve marginally, many remain unchanged or degrade. This challenges the assumption that adding more token positions can increase predictive power. In our setting, static concatenation appears to introduce redundant information more frequently than it provides complementary correctness signals.

\noindent \textbf{Sequential models.}
The same pattern extends to dynamic sequence modeling. Table~\ref{tab:res_seq_lstm_vs_xgb} shows that LSTM-based sequential probes do not produce a consistent gain over static, single-token XGBoost baselines. Improvements occur only in a minority of settings and are concentrated on BCB-Instruct. On LCB, sequential variants underperform across all models. Entropy-filtered ablations (see Table~\ref{tab:res_seq_entropy_qwen_lcb}) reinforce this result: no entropy-filtered variant surpasses the best single-token baseline. Interestingly, low-entropy filtering performs better than high-entropy filtering, implying that high-uncertainty tokens are not necessarily the most discriminative for response-level correctness.

A plausible interpretation is that much of the response-level correctness signal is already concentrated in late token states, while adjacent sequence states contain strongly correlated correctness patterns. In this case, longer sequential inputs may introduce optimization noise and increase sample complexity without providing sufficient novel information. Consequently, sequential models alone are not a reliable path to better response-level IUE for code synthesis tasks.

\noindent \textbf{Comparison with post-hoc calibration.}
Beyond discrimination, calibration outcomes provide additional evidence for the utility of hidden-state probing. On CALIBRI, our single-token XGBoost probe achieves both the highest BSS and the highest accuracy compared to established post-hoc calibration methods (see Table~\ref{tab:res_calib_calibri}). This is important for practical deployment scenarios, as confidence scores derived from IUE probes appear to be more aligned with empirical correctness than scores produced by post-hoc calibration of output-level model probabilities alone.

Overall, the evidence for \hyperref[item:rq1]{RQ1} is clearly positive, yet context-dependent. Internal artifacts, especially hidden states, can be used effectively to estimate code correctness, and they provide both strong discrimination and calibration in favorable settings. However, the strongest evidence supports relatively simple feature designs, particularly single-token representations towards the end of the response. More complex feature constructions, including static concatenation, sequential modeling, and entropy-based token filtering, did not produce consistent improvements in our experiments. Therefore, for response-level code correctness estimation, careful feature-position selection appears to be more important than increasing probe complexity.

\section{RQ2: To what extent do introspective uncertainty estimation methods generalize across code generation tasks?} \label{sec:disc_rq2}

Generalization is a known weakness of probe-based introspective methods in NLG, where performance often degrades significantly under task and domain shifts~\cite{ch-wang_androids_2024, preis_hallucination_2025}. For code generation tasks, evidence has remained limited. To answer \hyperref[item:rq2]{RQ2}, we evaluated out-of-distribution (OOD) behavior using three experiments that provide a comprehensive view of generalization capabilities: cross-task transfer, cross-domain transfer within BCB, and cross-token-position transfer via randomized extraction windows.

\noindent \textbf{Cross-task (cross-benchmark) generalization.}
As shown in Table~\ref{tab:res_gen_cross_lcb_bcb}, the cross-benchmark results indicate that response-level IUE can transfer across tasks, though performance generally degrades. The strongest performance is observed when training on BCB datasets and testing on LCB, particularly with BCB-Fusion. This is consistent with the methodological setup in Section~\ref{sec:bcb}, where BCB-Fusion was explicitly designed to resemble the LCB prompt structure. Meanwhile, transfers from BCB-Instruct to LCB also perform well, exceeding BCB-Fusion's performance for Qwen3-Coder. This finding suggests that cross-task robustness cannot be reduced to prompt similarity alone and that these probes are demonstrating some degree of generalization.

As expected, the opposite direction (LCB to BCB datasets) is more challenging due to the higher complexity and stronger library dependence of BCB tasks. Nevertheless, the degradation relative to in-distribution (ID) performance is generally moderate. In fact, one transfer setting (Qwen3-Coder on BCB-Fusion) even achieves a small positive OOD-ID difference. This suggests that the probes are not merely memorizing benchmark-specific artifacts, but rather, are capturing task-agnostic correctness patterns. These transferable correctness cues enable robust cross-task performance while often retaining a measurable task-specific OOD-ID performance gap that should not be ignored for practical deployments.

\noindent \textbf{Cross-domain generalization within BCB.}
Cross-domain generalization experiments, with a specific held-out domain of programming tasks on BCB-Instruct, further strengthen this interpretation. As shown in Table~\ref{tab:res_gen_bcb_domain_detail}, all representative OOD-ID differences remain below $0.05$, which is substantially smaller than the cross-task transfer gaps. The signed differences are mixed, with both small degradations and small improvements, and therefore show no evidence of systematic OOD collapse under domain shifts.

The aggregate summary in Table~\ref{tab:res_gen_bcb_domain_summary} confirms this stability, showing a mean shortfall below $2.2\%$ for all models. This indicates that shifts between software domains inside the same task family perturb response-level IUE significantly less than shifts between distinct code generation tasks. NVIDIA-Nemotron-3-Nano, with a near-zero shortfall across all domains, even illustrates that domain-robust, response-level IUE is achievable for some models.

\noindent \textbf{Cross-token-position generalization.}
The ``random tail'' (RT) experiments in Table~\ref{tab:res_gen_rt} add a complementary perspective, showing that generalization across token positions is also feasible. The results indicate that performance scales with token position: late positions in the model response remain the most informative, approaching the performance of the best static single-token baselines for response-level IUE. As RT increases, performance decreases, reflecting the expected trade-off between positional robustness and peak predictive capability.

Importantly, widening the sampling window does not cause performance to collapse. Instead, performance drops quickly at first and then plateaus around a task-specific floor. This finding indicates that correctness-relevant information is distributed across many token positions, even though late tokens are the strongest carriers. For practical systems, this is encouraging because it enables uncertainty monitoring under positional variance, such as in streaming autonomous agent generations, where fixed extraction positions are not always available. Consequently, IUE probes remain useful for code correctness estimation even when token positions are not fixed, albeit with expected performance degradation.

In summary, \hyperref[item:rq2]{RQ2} can be answered with optimism. Response-level IUE models demonstrably generalize across tasks, domains, and token positions, which is encouraging for their practical use beyond narrow ID settings. However, performance degradation varies by transfer setting and model, and a persistent OOD-ID gap remains in the most challenging cross-task scenarios. Therefore, although training a robust ``generalist'' probe is feasible for broad uncertainty monitoring, task-specific adaptation remains the most reliable way to achieve optimal IUE performance.

\section{RQ3: Is fine-grained hallucination detection on code feasible using introspective uncertainty estimation methods?} \label{sec:disc_rq3}

This research question explores whether introspective signals can advance beyond coarse, response-level detection and support fault localization in generated code. In practical terms, this implies determining whether IUE can assist developers in not only identifying whether a program is likely incorrect, but also locating the most likely source of the error. This problem is substantially harder than response-level classification because probes must distinguish a small number of incorrect lines from a large number of correct lines. Since prior code-focused introspective studies are primarily response-level, the line-level analysis in this thesis offers direct evidence of the feasibility of fine-grained hallucination detection for code synthesis tasks.

\noindent \textbf{Baselines.}
The mixed dataset settings in Tables~\ref{tab:line_xgb_baselines} and~\ref{tab:line_lstm_baselines}, which include both correct and incorrect programs, show that line-level hidden-state features are informative. The ability to discriminate between positive and negative samples remains significantly above random, demonstrating that the probes learn meaningful, localized fault signals. At the same time, the performance is far below the response-level baselines (see Table~\ref{tab:res_single_token_all}) for LCB, confirming the expected increase in difficulty for fine-grained fault localization.

Another finding is that there is no universally dominant architecture in this context. The static XGBoost baselines exchange advantages across models, and the sequential LSTM models improve across some evaluated settings and metrics, but also do not consistently outperform the static baselines. Similar to the response-level findings, additional sequential feature complexity does not reliably result in better discrimination capability. These findings suggest that the primary obstacle for line-level IUE is not the choice of probe architecture or feature selection, but rather the inherent difficulty of the task itself, which is compounded by class imbalance and calibration.

\noindent \textbf{Threshold optimization.}
In the mixed setting, probes can rank lines reasonably well, but converting that ranking into robust binary decisions is difficult (see Table~\ref{tab:line_xgb_baselines}). To investigate the discrepancy between good discriminatory performance and low overall F1 scores, we evaluated threshold optimization. As shown in Table~\ref{tab:line_threshold_optimization}, post-hoc threshold tuning improves F1 scores, yet these performance gains remain moderate and do not resolve the broader calibration problem in this setting.

This behavior is consistent with the properties of the datasets in Table~\ref{tab:aug_summary_combined}. When faulty lines are rare, threshold choice primarily trades precision for recall without exposing a stable operating point. Therefore, the limiting factor is not the absence of correctness information in hidden states but rather the difficulty of mapping that signal to well-calibrated, line-level confidence scores under severe imbalance.

\noindent \textbf{Fault localization models.}
By focusing exclusively on incorrect programs, the conditional setup achieves a more favorable class balance (see \textbf{V3} in Table~\ref{tab:aug_summary_combined}). As shown in Table~\ref{tab:fault_loc_v3_models}, AUROC remains broadly comparable to that of the mixed setting, while the F1 scores improve substantially for both LLMs. This finding reinforces the interpretation that the underlying fault localization signal exists in both settings. However, the mixed setting's class imbalance hinders the conversion of that signal into reliable binary predictions. In conclusion, while the probes can still identify likely faulty lines, the thresholding step is more effective when the search space is limited to incorrect programs, thereby reducing task complexity and balancing the classes.

From a practical application perspective, this finding is crucial because it advocates for a two-stage IUE pipeline instead of a single line-level classifier. The first stage can provide a response-level rejection of likely incorrect outputs, while the second stage focuses entirely on localizing faults within these candidate generations. With this approach, line-level probes bypass the most severe calibration bottlenecks, making them considerably more practical for direct developer support.

\noindent \textbf{Top-$K$ point-of-failure ranking.}
For debugging workflows, ranked suggestions are often more useful than strict line-wise binary labels. Therefore, we use the Top-$K$ hit rate metric to evaluate the practical utility of our models for fault localization. This metric assesses whether at least one incorrectly labeled line appears in the top $K$ ranked lines for each program. The results in
Table~\ref{tab:fault_loc_topk} show that the token-to-line probe consistently outperforms the direct line-level baseline for both LLMs and across $K$ values. Both approaches also clearly exceed the random baseline for all $K$ values, confirming that the probes identify non-random, localized correctness cues.

The best Hit@3 results are particularly encouraging. Compared to prior fine-grained introspective research on code editing~\cite{huang_risk_2025}, these values suggest that strong point-of-failure ranking is also achievable on code synthesis tasks. The broader implication is that token-level context aggregation is especially effective when the objective is to prioritize likely incorrect lines inside already-faulty programs. Thus, it is necessary to extract full token-level hidden states for line-level IUE to achieve optimal performance rather than using single-token features per line, which we have demonstrated to be less effective in this setting.

Overall, the answer to \hyperref[item:rq3]{RQ3} is affirmative. Fine-grained hallucination detection for code synthesis tasks is feasible with introspective methods, particularly when framed as ranking-based fault localization in a two-stage pipeline. In contrast, direct thresholded line classification on mixed programs remains challenging due to class imbalance and calibration limitations. Consequently, the strongest practical application of line-level IUE is not as a standalone model but rather as targeted debugging guidance that directs developers' attention toward high-risk lines in known incorrect generations.

\section{Threats to Validity} \label{sec:threats}

This thesis has several limitations that are important to acknowledge when interpreting the empirical results. We highlight the most relevant threats in terms of label quality, feature approximation, and fine-grained label generation.

\noindent \textbf{Reliance on benchmark test suites for response-level labels.}
Our response-level labels are derived from benchmark test outcomes. This method is scalable and reproducible, but it is not a perfect oracle for code correctness~\cite{spiess_calibration_2024, ribeiro_llms_2025}. Incomplete or unreliable test suites can produce false positives (incorrect programs labeled as correct), while passing all tests does not guarantee broader software quality properties such as reliability, readability, or maintainability. Consequently, the response-level IUE probes in this thesis are trained to predict benchmark-measured correctness rather than universal software quality. This limitation is common in uncertainty estimation (UE) research, but it should be kept in mind when transferring conclusions to real-world deployment settings, where code correctness criteria are often more nuanced~\cite{sharma_assessing_2025}.

\noindent \textbf{Feature approximation introduced by prefilling.}
Hidden-state features are extracted by replaying saved prompt-response pairs with Transformers~\cite{wolf_huggingfaces_2020} instead of recording representations directly during generation with vLLM~\cite{kwon_efficient_2023}. This design choice is practical and widely used in IUE research~\cite{sriramanan_llm-check_2024, ch-wang_androids_2024, ribeiro_llms_2025, huang_risk_2025}. However, it introduces a feature approximation: the extracted representations may differ slightly from those produced in the original decoding environment. Our analysis of token probabilities indicates that these differences are small but statistically significant. Therefore, the extracted features are close proxies for the original generation artifacts, but they are not exact replicas. While this approximation may affect the absolute performance of the probes, it is unlikely to introduce systematic bias in comparisons across tasks or probe architectures since all experiments use the same prefilling procedure.

\noindent \textbf{Fine-grained labels from automated repair and diff-based alignment.}
Token- and line-level labels are generated through an automated ``fix-then-diff'' pipeline. This procedure assumes that the repaired, test-passing programs isolate the original faults with minimal edits. In practice, however, an LLM prompted to repair code may introduce broader rewrites than necessary, and alignment-based differencing may misattribute the location of faults. Even with careful post-processing, the resulting labels are only approximations rather than ground truth bug annotations. This limitation affects the interpretation of our line-level results because our experiments reflect the ability to identify fault locations tagged by the augmentation pipeline rather than the ability to reconstruct human-annotated bugs.

\noindent \textbf{Representativeness constraints.}
Although response-level experiments span multiple models and two benchmarks, fine-grained augmentation is restricted to LCB and two LLMs due to computational and time constraints. Consequently, line-level findings are strongest in this setting and may not transfer to other code generation tasks.

To improve transparency despite these limitations, we release code, generation outputs, and labels for both response-level and line-level experiments. This enables external validation of our findings and facilitates future work on better oracles and higher-quality fine-grained labels.

%% file: chapters/conclusion.tex
\chapter{Conclusion} \label{cha:conclusion}

This chapter concludes the thesis by providing a brief summary of the central insights and offering an outlook on future research directions.

\section{Summary}

This thesis examined whether hidden-state-based Introspective Uncertainty Estimation (IUE) can improve trust in LLM-based code generation. The central finding is that internal representations can provide robust signals for code correctness at different levels of granularity, with applications in practical uncertainty estimation workflows.

At the response level, the most effective configurations were often the simplest, particularly those involving representations from the final token positions of the response. More complex feature constructions and sequential variants did not consistently provide additional benefits, indicating that practical IUE should retain lightweight features and simple models.

Generalization beyond training settings is feasible across multiple different scenarios, making robust deployments promising. However, to achieve the best performance, they should still include task-aware adaptation. At a fine level of granularity, line-level detection is most practical within a two-stage workflow: response-level risk screening followed by ranking-based fault localization within likely incorrect programs.

Overall, this thesis positions IUE as a practical complement to post-hoc verification of generated code. It is not a replacement for testing and other methods of ensuring code quality, but rather a way to identify risks early on in an LLM-assisted development process and guide debugging efforts more effectively.

\section{Future Work}

There are several directions in which future work could build upon the findings of this thesis and address some of its limitations.

\noindent \textbf{Improving label quality.}
As discussed in Section~\ref{sec:threats}, benchmark-derived labels and automated repair-based augmentation are scalable but not perfect. A natural next step is to incorporate higher-quality human annotations for code correctness and fault localization, for example through curated feedback workflows or review protocols. Better line-level labels would not only improve fault localization models directly, but could also facilitate stronger response-level labels, thereby enhancing the entire workflow.

\noindent \textbf{Generalization across programming languages.}
This thesis focuses on Python-centric code synthesis benchmarks. Expanding the analysis to include additional programming languages would enable new generalization experiments. These evaluations would reveal whether the learned introspective signals are language-agnostic or tied to specific syntax and library ecosystems, as well as the effectiveness of model transfers to less common programming languages.

\noindent \textbf{Uncertainty estimation for non-functional code qualities.}
Future work should also move beyond functional correctness to investigate uncertainty estimation for non-functional requirements such as security, performance, reliability, readability, and maintainability. Specifically, efforts to determine whether internal artifacts can predict instruction following for verifiable non-functional constraints, such as adherence to specific style or safety rules, which can be labeled automatically through post-hoc verification methods. This would expand IUE from code correctness estimation to a more comprehensive approach to assess code quality, further improving trust in LLM-assisted software development.

\noindent \textbf{Native feature extraction during inference.}
The prefilling workflow used in this thesis is practical but introduces approximation and additional feature extraction overhead. A key direction for future research should be to extract hidden states directly during generation at selected layers and token positions, ideally using efficient LLM inference engines. Enabling native feature extraction during inference for IUE would improve feature accuracy, reduce redundant computational overhead, and facilitate real-time uncertainty monitoring for practical applications.

In conclusion, this thesis demonstrated that introspective signals from hidden states provide a reliable foundation for uncertainty estimation on code generation tasks. Although crucial challenges remain, the findings indicate that IUE is a promising approach toward achieving more reliable, trustworthy, and practical LLM-assisted software development.

%% file: chapters/appendix.tex
\chapter{Prompts}
\label{cha:appendix_prompts}

\section{LiveCodeBench (LCB)}

\subsection{Example Prompt} \label{sec:app_prompt_lcb}
\begin{verbatim}
System:
You are an expert Python programmer. You will be given a 
question (problem specification) and will generate a 
correct Python program that matches the specification
and passes all tests.

User:
### Question:
There are three cards with letters `a`, `b`, `c` placed in 
a row in some order. You can do the following operation at
most once: 

- Pick two cards, and swap them. Is it possible that the 
row becomes `abc` after the operation? Output "YES" if it
is possible, and "NO" otherwise.

Input
The first line contains a single integer t (1 <= t <= 6) 
— the number of test cases.
The only line of each test case contains a single string 
consisting of each of the three characters `a`, `b`, and 
`c` exactly once, representing the cards.

Output
For each test case, output "YES" if you can make the row
`abc` with at most one operation, or "NO" otherwise.

Sample Input 1:
6
abc
acb
bac
bca
cab
cba

Sample Output 1:
YES
YES
YES
NO
NO
YES

Note
In the first test case, we don't need to do any operations, 
since the row is already `abc`.
In the second test case, we can swap `c` and `b`: `acb` -> 
`abc`.

### Format: Read the inputs from stdin, solve the problem, 
and write the answer to stdout. Enclose your code within 
delimiters as follows.
```python
# YOUR CODE HERE
```
### Answer: (use the provided format with backticks)
\end{verbatim}

\subsection{Augmentation Prompt}
\label{sec:app_augmentation_prompt}
\begin{verbatim}
System:
You are a code repair assistant. Your task is to fix buggy 
programs by changing the minimum number of tokens. Only 
return the corrected code in a single fenced code block.

User:
Below is a programming problem followed by a buggy solution.

{problem_prompt}

### Buggy solution:
{program}

Fix the solution by editing the **minimum** number of tokens 
needed to make the program pass all tests. Preserve existing 
comments when they remain correct; update or remove comments 
if their context is wrong. Only rename identifiers if they 
are actually wrong or misleading. Only return the corrected 
code in a single fenced code block.
\end{verbatim}

\section{BigCodeBench (BCB)}

\subsection{Instruct Example Prompt}
\label{sec:app_prompt_bcb_instruct}
\begin{verbatim}
System:
Please provide a self-contained Python script that solves 
the following problem in a markdown code block:

User:
Calculates the average of the sums of absolute differences 
between each pair of consecutive numbers for all 
permutations  of a given list. Each permutation is shuffled 
before calculating the differences.

Args:
- numbers (list): A list of numbers. Default is numbers 
from 1 to 10.

The function should output with:
    float: The average of the sums of absolute differences 
    for each shuffled permutation of the list.

You should write self-contained code starting with:
```
import itertools
from random import shuffle
def task_func(numbers=list(range(1, 3))):
```
\end{verbatim}

\subsection{Fusion Example Prompt}
\label{sec:app_prompt_bcb_fusion}
\begin{verbatim}
System:
You are an expert Python programmer. You will be given a 
question (problem specification) and will generate a 
correct Python program that matches the specification 
and passes all tests.

User:
### Question:
Write a self-contained Python function that solves the 
following problem:

Calculates the average of the sums of absolute differences 
between each pair of consecutive numbers for all 
permutations of a given list. Each permutation is shuffled 
before calculating the differences.

Args:
- numbers (list): A list of numbers. Default is numbers 
from 1 to 10.

Returns:
float: The average of the sums of absolute differences for 
each shuffled permutation of the list.

Requirements:
- itertools
- random.shuffle

Example:
>>> result = task_func([1, 2, 3])
>>> isinstance(result, float)
True

You should write self-contained code starting with:
```
import itertools
from random import shuffle
def task_func(numbers=list(range(1, 3))):
```
### Format: Enclose your code within delimiters as follows.
```python
# YOUR CODE HERE
```
### Answer: (use the provided format with backticks)
\end{verbatim}

\chapter{Datasets}
\label{cha:appendix_datasets}
\captionsetup{list=no}

\section{LiveCodeBench (LCB)}

\subsection{t-SNE Feature Visualization} \label{sec:app_feature_vis}

\begin{figure}[H]
    \centering
    \includegraphics[width=0.8\textwidth]{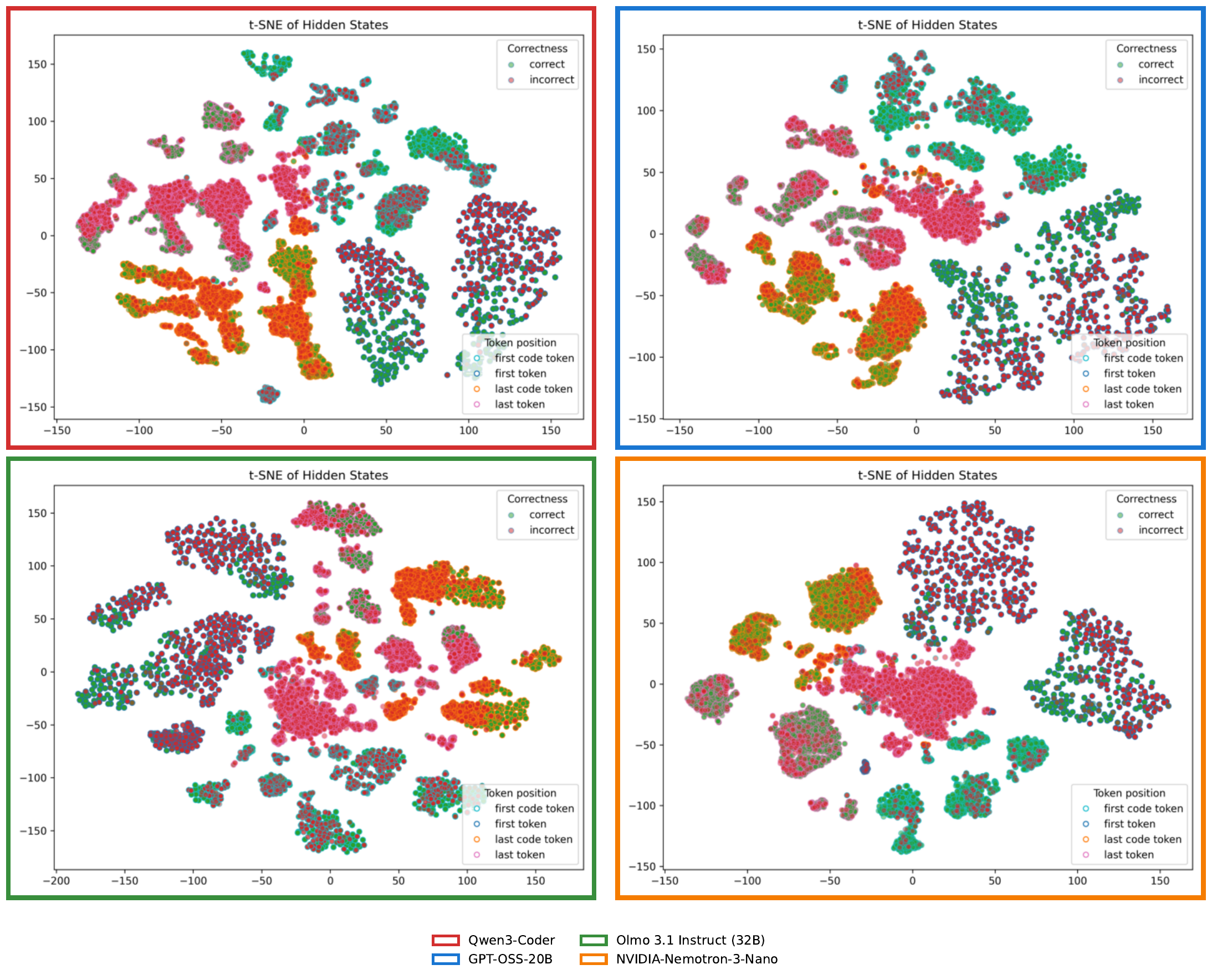}
    \caption{Combined 2D t-SNE distributions of select hidden-state features extracted from the last layer of the models on LCB. The distributions differ between correct v.s. incorrect code generations and show varying degrees of separation across the models and token positions.}
    \label{fig:lcb_tsne_models_combined}
\end{figure}

\subsection{Augmentation Examples with Insertions} \label{sec:app_augmentation_insertions}

\begin{figure}[H]
    \centering
    \includegraphics[width=0.8\textwidth]{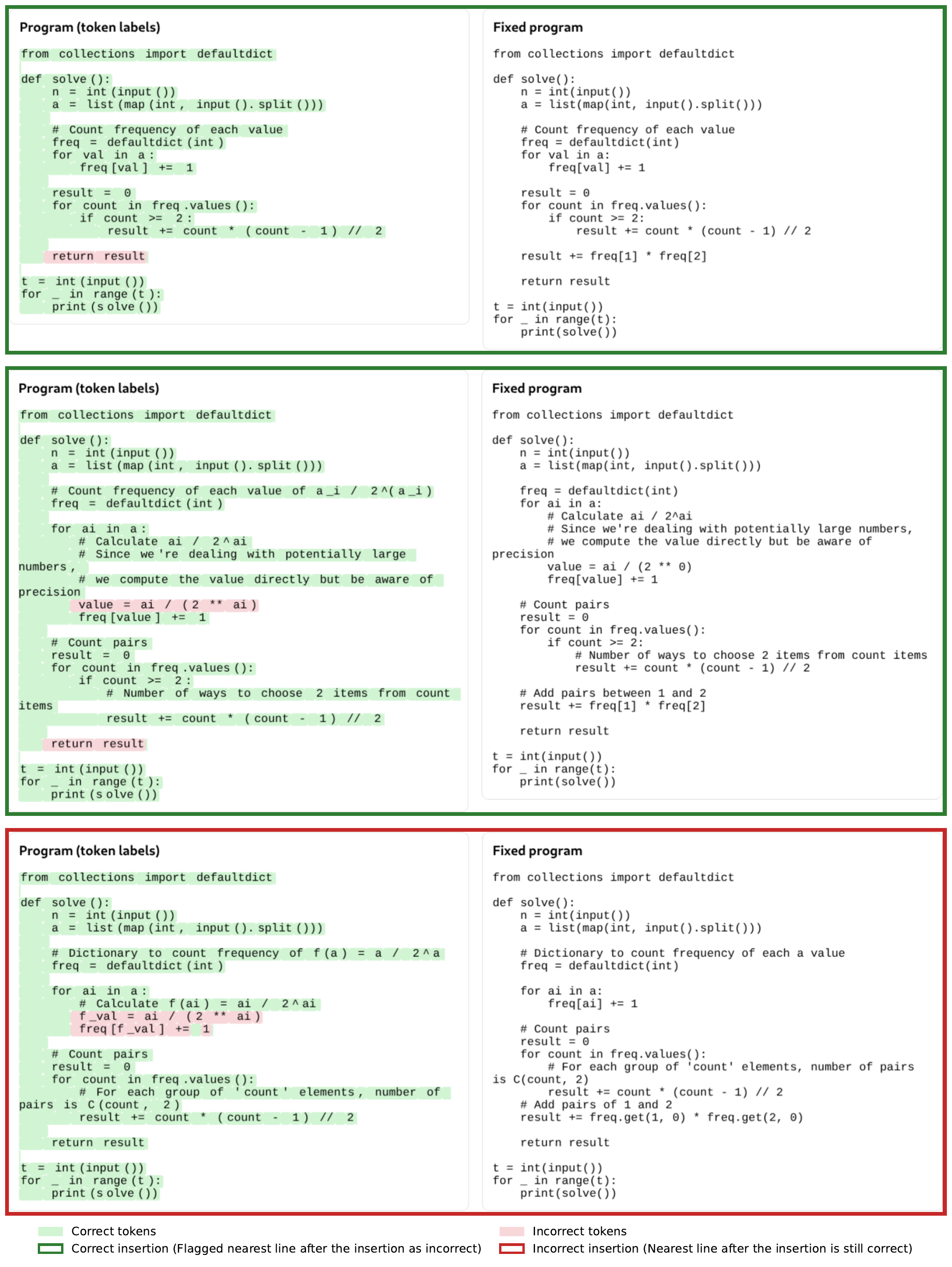}
    \caption{Combined visualization of three augmented programs generated by the minimal-edit fix prompt for LCB. Correct and incorrect insertion handling is visualized with green and red borders. In the program code, line labels are highlighted in light green and light red. A correctly flagged insertion is one where the nearest line after the insertion in the original code is flagged as incorrect. An incorrectly flagged insertion is one where the nearest line after the insertion in the original code is not flagged as incorrect.}
    \label{fig:lcb_aug_samples_combined}
\end{figure}

\section{BigCodeBench (BCB)}

\subsection{Domain Label Distributions} \label{sec:app_bcb_domains}

\begin{figure}[H]
    \centering
    \includegraphics[width=0.8\textwidth]{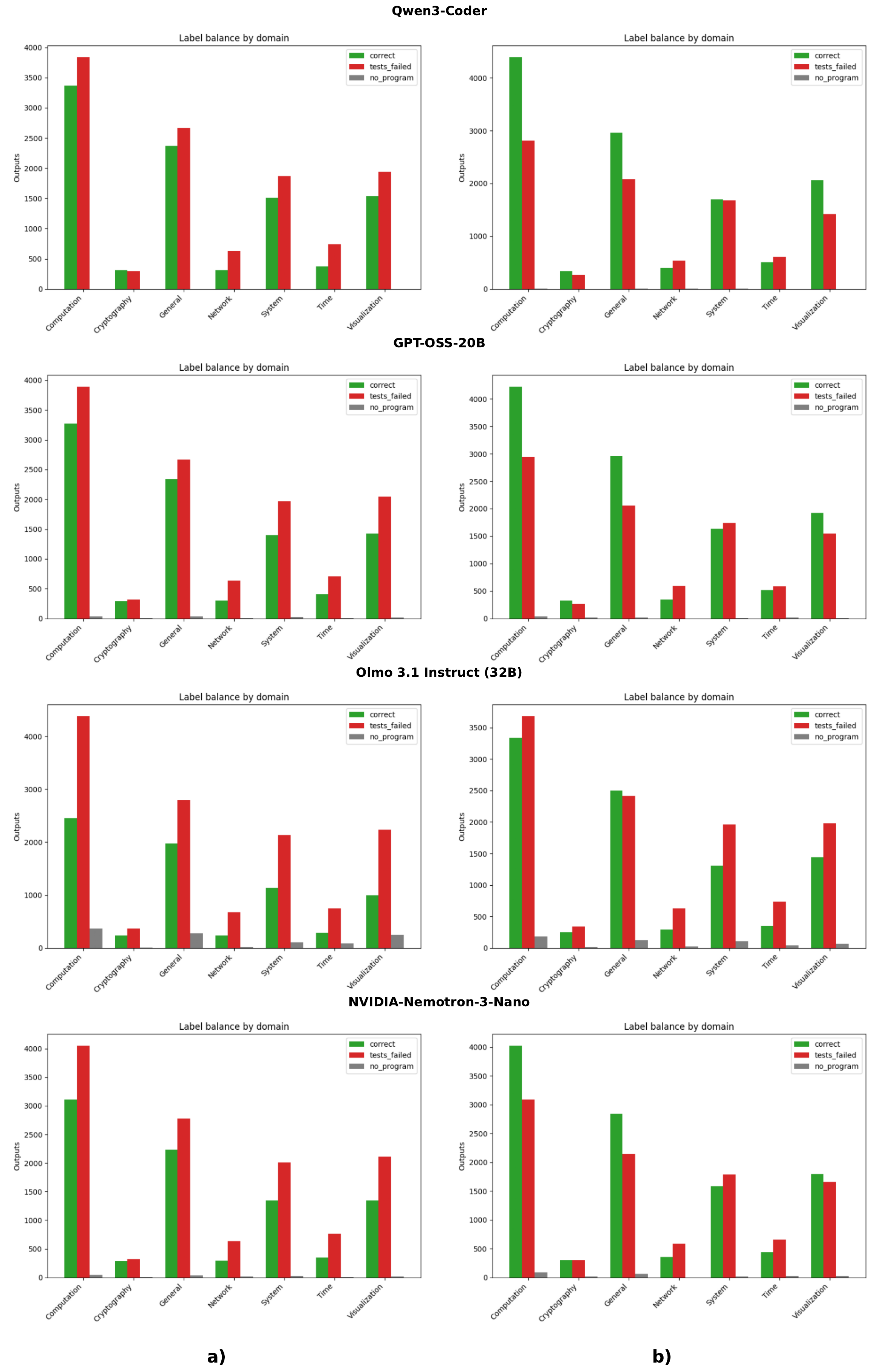}
    \caption{Combined visualization of domain label distributions for BCB across a) Instruct and b) Fusion prompt versions. The Fusion prompt increases the number of correct programs across all domains and models.}
    \label{fig:bcb_domains_models_combined}
\end{figure}

\chapter{Experiments}
\label{cha:appendix_experiments}

\section{Hyperparameter Configuration Overview}
\label{sec:app_exp_hyperparams}

This appendix summarizes the concrete hyperparameter choices and search spaces referenced in Section~\ref{sec:setup}. Unless noted otherwise, the same protocol was used for both response-level and line-level experiments.

\subsection{XGBoost Probes}
\label{sec:app_exp_xgboost}

\noindent \textbf{Base configuration.}
\begin{itemize}
    \item max\_depth: 6
    \item min\_child\_weight: 1.0
    \item subsample: 0.8
    \item colsample\_bytree: 0.8
    \item reg\_lambda: 1.0
    \item reg\_alpha: 0.0
    \item tree\_method: auto
\end{itemize}

\noindent \textbf{Line-level configuration.}
\begin{itemize}
    \item max\_depth: 4
    \item min\_child\_weight: 10.0
    \item subsample: 0.8
    \item colsample\_bytree: 0.8
    \item reg\_lambda: 2.0
    \item reg\_alpha: 0.1
    \item tree\_method: auto
    \item scale\_pos\_weight: set to the ratio of negative to positive samples in the training data to address class imbalance
\end{itemize}

\noindent \textbf{Base training setup.}
\begin{itemize}
    \item num\_boost\_round: 500
    \item early\_stopping\_rounds: 50
    \item Validation metrics: log-loss and AUROC
\end{itemize}

\noindent \textbf{Line-level training setup.}
\begin{itemize}
    \item num\_boost\_round: 3000
    \item early\_stopping\_rounds: 300
    \item Validation metrics: log-loss and AUROC
\end{itemize}

\noindent \textbf{Base tuning.}
Only the learning rate (eta) was tuned, using multiplicative factors \\ $[0.25, 0.5, 0.75, 1.0, 1.25, 1.5, 2.0]$ around a base value of $0.05$.

\noindent \textbf{Line-level tuning.}
Only the learning rate (eta) was tuned, using multiplicative factors $[0.8, 0.9, 1.0, 1.1, 1.2]$ around a base value of $0.02$.

\subsection{MLP Probes}
\label{sec:app_exp_mlp}

Two MLP families were implemented to mirror prior work~\cite{snyder_early_2024, bui_correctness_2025}.

\noindent \textbf{Snyder-style MLP.}
\begin{itemize}
    \item Architecture: single hidden layer (default width 256)
    \item Optimizer: Adam
    \item Default training horizon: 1000 iterations
    \item Default batch size: 128
    \item Default learning rate: $10^{-4}$
    \item Default weight decay: $10^{-2}$
    \item Activation function: ReLU
    \item Loss: \texttt{BCEWithLogitsLoss}
    \item Early stopping patience: 150 validation checks
\end{itemize}

\noindent \textbf{Sweep space (10 trials).}
\begin{itemize}
    \item Learning rate: log-uniform in $[10^{-6}, 10^{-3}]$
    \item Weight decay: $\{10^{-1}, 10^{-2}, 10^{-3}\}$
    \item Hidden width: $\{128, 256, 512\}$
    \item Batch size: $\{128, 256, 512\}$
\end{itemize}

\noindent \textbf{Bui-style MLP.}
\begin{itemize}
    \item Architecture: two hidden layers (default widths 128 and 64)
    \item Optimizer: Adam
    \item Default training horizon: 50 epochs
    \item Default batch size: 32
    \item Default learning rate: $10^{-3}$
    \item Activation functions: ReLU
    \item Loss: \texttt{BCEWithLogitsLoss}
    \item Early stopping patience: 10 epochs
\end{itemize}

\noindent \textbf{Sweep space (10 trials).}
\begin{itemize}
    \item Learning rate: log-uniform in $[10^{-5}, 10^{-2}]$
    \item Weight decay: $\{0, 10^{-1}, 10^{-2}, 10^{-3}\}$
    \item Hidden widths: $\{[128,64], [256,128], [512,256]\}$
    \item Batch size: $\{64, 128, 256\}$
\end{itemize}

\subsection{LSTM Probes}
\label{sec:app_exp_lstm}

\noindent \textbf{Base architecture.}
\begin{itemize}
    \item Single-layer LSTM encoder (default hidden size 256)
    \item MLP head over final hidden state (default one hidden layer)
    \item Loss: \texttt{BCEWithLogitsLoss}
\end{itemize}

\noindent \textbf{Training setup.}
\begin{itemize}
    \item Optimizer: Adam
    \item Training horizon: 1000 iterations
    \item Batch size: 128
    \item Learning rate: $10^{-4}$
    \item Weight decay: 0
    \item Early stopping patience: 150 validation checks
    \item Gradient clipping: max norm 1.0
\end{itemize}

\noindent \textbf{Line-level training setup.}
\begin{itemize}
    \item Optimizer: Adam
    \item Training horizon: 1500 iterations
    \item Batch size: 128
    \item Learning rate: $2 \times 10^{-4}$
    \item Weight decay: 0
    \item Early stopping patience: 300 validation checks
    \item Gradient clipping: max norm 1.0
    \item Dropout between encoder and head: 0.1
    \item Positive class weighting: set to the ratio of negative to positive samples in the training data to address class imbalance
\end{itemize}

\noindent \textbf{Base sweep space (10 trials).}
\begin{itemize}
    \item Learning rate: log-uniform in $[10^{-5}, 10^{-3}]$
    \item Weight decay: $\{0, 10^{-2}, 10^{-4}\}$
    \item LSTM hidden size: $\{128, 256, 512\}$
    \item Head hidden size: $\{64, 128, 256\}$
    \item LSTM depth: $\{1, 2\}$ layers
    \item Head depth: $\{1, 2\}$ layers
    \item Batch size: $\{32, 64, 128\}$
    \item Dropout (between LSTM and head): $\{0, 0.1, 0.2\}$
\end{itemize}

\noindent \textbf{Line-level sweep space (10 trials).}
\begin{itemize}
    \item Learning rate: log-uniform in $[10^{-5}, 5 \times 10^{-4}]$
    \item LSTM hidden size: $\{128, 256, 512\}$
    \item LSTM depth: $\{1, 2\}$ layers
\end{itemize}

\section{Supplementary Generalization Tables}
\label{sec:app_exp_domain_tables}

This section reports additional domain generalization results for BCB-Instruct that are referenced in Section~\ref{subsec:generalization}. Each table reports the best in-distribution (ID) AUROC from BCB-Instruct and the best out-of-distribution (OOD) AUROC on the held-out domain, including the feature that achieved the best OOD score and the signed difference $\Delta=\text{OOD}-\text{ID}$.

\begin{table}[H]
\centering
\footnotesize
\setlength{\tabcolsep}{4pt}
\renewcommand{\arraystretch}{1.05}
\begin{tabular}{lcc}
\toprule
\textbf{Model} & \textbf{ID Best} & \textbf{Computation} \\
\midrule
Qwen3-Coder & 0.6841 & 0.6570 (L, $\Delta=-0.0271$) \\
GPT-OSS-20B & 0.6569 & 0.6213 (F+L, $\Delta=-0.0356$) \\
NVIDIA-Nemotron-3-Nano & 0.6411 & 0.6466 (L, $\Delta=+0.0054$) \\
Olmo 3.1 Instruct (32B) & 0.7309 & 0.7023 (L, $\Delta=-0.0286$) \\
\bottomrule
\end{tabular}
\caption{Cross-domain generalization results on BCB-Instruct with \textit{Computation} as held-out domain.}
\label{tab:app_bcb_domain_computation}
\end{table}

\begin{table}[H]
\centering
\footnotesize
\setlength{\tabcolsep}{4pt}
\renewcommand{\arraystretch}{1.05}
\begin{tabular}{lcc}
\toprule
\textbf{Model} & \textbf{ID Best} & \textbf{General} \\
\midrule
Qwen3-Coder & 0.6841 & 0.6743 (L, $\Delta=-0.0098$) \\
GPT-OSS-20B & 0.6569 & 0.6402 (L, $\Delta=-0.0166$) \\
NVIDIA-Nemotron-3-Nano & 0.6411 & 0.6528 (L, $\Delta=+0.0117$) \\
Olmo 3.1 Instruct (32B) & 0.7309 & 0.7090 (LC, $\Delta=-0.0219$) \\
\bottomrule
\end{tabular}
\caption{Cross-domain generalization results on BCB-Instruct with \textit{General} as held-out domain.}
\label{tab:app_bcb_domain_general}
\end{table}

\begin{table}[H]
\centering
\footnotesize
\setlength{\tabcolsep}{4pt}
\renewcommand{\arraystretch}{1.05}
\begin{tabular}{lcc}
\toprule
\textbf{Model} & \textbf{ID Best} & \textbf{Network} \\
\midrule
Qwen3-Coder & 0.6841 & 0.7294 (F, $\Delta=+0.0453$) \\
GPT-OSS-20B & 0.6569 & 0.6701 (F, $\Delta=+0.0133$) \\
NVIDIA-Nemotron-3-Nano & 0.6411 & 0.7259 (F+L, $\Delta=+0.0848$) \\
Olmo 3.1 Instruct (32B) & 0.7309 & 0.7275 (L, $\Delta=-0.0034$) \\
\bottomrule
\end{tabular}
\caption{Cross-domain generalization results on BCB-Instruct with \textit{Network} as held-out domain.}
\label{tab:app_bcb_domain_network}
\end{table}

\begin{table}[H]
\centering
\footnotesize
\setlength{\tabcolsep}{4pt}
\renewcommand{\arraystretch}{1.05}
\begin{tabular}{lcc}
\toprule
\textbf{Model} & \textbf{ID Best} & \textbf{Visualization} \\
\midrule
Qwen3-Coder & 0.6841 & 0.6814 (FC+LC, $\Delta=-0.0027$) \\
GPT-OSS-20B & 0.6569 & 0.6379 (L, $\Delta=-0.0190$) \\
NVIDIA-Nemotron-3-Nano & 0.6411 & 0.6328 (LC, $\Delta=-0.0084$) \\
Olmo 3.1 Instruct (32B) & 0.7309 & 0.6967 (FC+LC, $\Delta=-0.0342$) \\
\bottomrule
\end{tabular}
\caption{Cross-domain generalization results on BCB-Instruct with \textit{Visualization} as held-out domain.}
\label{tab:app_bcb_domain_visualization}
\end{table}